\documentclass[twocolumn,showpacs,aps,prd,floatfix,preprintnumbers]{revtex4}
\usepackage{graphicx} 
\usepackage{dcolumn}
\usepackage{amsmath}
\usepackage{color}

\input babarsym   

  \renewcommand{\pipi}{\ensuremath{\pip\pim}\xspace}

\renewcommand{\mm}{\ensuremath{M_{\mathrm{\rm rec}}}\xspace}

\def\gevccc{\mbox{${\mathrm{GeV^2}}/c^4$}}

\def\calR         {{\ensuremath{\cal R}\xspace}}

\def\calL         {{\ensuremath{\cal L}\xspace}}

\def\etal {{\it et al.}}

\newcommand{\etactwo}{\ensuremath{\eta_{c}(2S)}\xspace}

\renewcommand{\mm}{\ensuremath{M^2_{\mathrm{\rm rec}}}\xspace}

\newcommand{\etactoetapkk}{\ensuremath{\etac \to \etapr \Kp \Km}\xspace}
\newcommand{\etactoetappipi}{\ensuremath{\etac \to \etapr \pip \pim}\xspace}

\newcommand{\etaprkk}{\ensuremath{\etapr \Kp \Km}\xspace}

\newcommand{\etaprpipi}{\ensuremath{\etapr \pip \pim}\xspace}
\newcommand{\etapipi}{\ensuremath{\eta \pip \pim}\xspace}

\def\Kstarz   {\ensuremath{K^*_0(1430)^0}\xspace}

\newcommand{\al}{\ensuremath{\kern 0.5em }}
\newcommand{\all}{\ensuremath{\kern 0.25em }}
\newcommand{\alm}{\ensuremath{\kern -0.50em }}
\newcommand{\almm}{\ensuremath{\kern -1.00em }}
\newcommand{\aln}{\ensuremath{\kern -0.25em }}

\renewcommand{\gg}{\ensuremath{\gamma\gamma}}
\mathchardef\myhyphen="2D

\begin{document}
\begin{flushleft}
  \mbox{\large {\babar-PUB-21/001}}\\
  \mbox{\large {SLAC-PUB-17606}}
\end{flushleft}  
\title{
 \large \bf\boldmath Light meson spectroscopy from Dalitz plot analyses of \etac decays to $\etapr \Kp \Km$, $\etapr \pip \pim$, and $\eta \pip \pim$ produced in two-photon interactions
}

\author{J.~P.~Lees}
\author{V.~Poireau}
\author{V.~Tisserand}
\affiliation{Laboratoire d'Annecy-le-Vieux de Physique des Particules (LAPP), Universit\'e de Savoie, CNRS/IN2P3,  F-74941 Annecy-Le-Vieux, France}
\author{E.~Grauges}
\affiliation{Universitat de Barcelona, Facultat de Fisica, Departament ECM, E-08028 Barcelona, Spain }
\author{A.~Palano}
\affiliation{INFN Sezione di Bari, I-70126 Bari, Italy}
\author{G.~Eigen}
\affiliation{University of Bergen, Institute of Physics, N-5007 Bergen, Norway }
\author{D.~N.~Brown}
\author{Yu.~G.~Kolomensky}
\affiliation{Lawrence Berkeley National Laboratory and University of California, Berkeley, California 94720, USA }
\author{M.~Fritsch}
\author{H.~Koch}
\author{T.~Schroeder}
\affiliation{Ruhr Universit\"at Bochum, Institut f\"ur Experimentalphysik 1, D-44780 Bochum, Germany }
% ...
\collaboration{The \babar\ Collaboration}
\noaffiliation

\begin{abstract}
We study the processes $\gg\to\etac\to\etaprkk$, $\etaprpipi$, and $\etapipi$ using a data sample
of 519~\invfb\ recorded with the \babar\ detector operating at the SLAC PEP-II
asymmetric-energy \epem\ collider at center-of-mass energies at and near the
$\Upsilon(nS)$ ($n = 2,3,4$) resonances.
This is the first observation of the decay $\etac \to \etapr \Kp \Km$ and we measure the branching fraction $\Gamma(\etactoetapkk)/(\Gamma(\etactoetappipi)=0.644\pm 0.039_{\rm stat}\pm 0.032_{\rm sys}$. 
Significant interference is observed between $\gg\to \etac\to \etapipi$ and the non-resonant two-photon process $\gg \to \etapipi$.
A Dalitz plot analysis is performed of \etac decays to $\etapr \Kp \Km$, $\etapr \pip \pim$, and $\eta \pip \pim$.
Combined with our previous analysis of $\etac \to K \bar K \pi$, we measure the $K^*_0(1430)$ parameters and the ratio between its $\etapr K$ and $\pi K$ couplings. 
The decay $\etac \to \etapr \pip \pim$ is dominated by the $f_0(2100)$ resonance, also observed in $J/\psi$ radiative decays.
A new $a_0(1700)\to \eta \pi$ resonance is observed in the $\etac \to \etapipi$
channel.
We also compare \etac decays to $\eta$ and \etapr final states in association with scalar mesons as they relate to the identification of the scalar glueball.
\end{abstract}
\pacs{13.25.Gv, 14.40.Pq, 14.40.Df, 14.40.Be}
\maketitle

\section{Introduction}

Scalar mesons remain a puzzle in light meson spectroscopy: they have complex structure,
and there are too many states to be accommodated within the quark model without difficulty~\cite{polosa}. 
In particular, the structure of the isospin I=$1\over 2$  $K \pi$
$S$-wave is still poorly understood, which limits the precision of measurements involving a $K \pi$ system in the final state,
including recent searches for \CP violation in $B$ meson decay~\cite{cp}, 
and studies of new exotic resonances~\cite{zs} and charmed mesons~\cite{bs}.

Decays of the \etac, the lightest pseudoscalar $c \bar{c}$ state, provide a window on light meson states.
The \babar\ experiment first performed a Dalitz plot analysis of $\etac \to \Kp \Km \piz$ and $\etac \to \Kp \Km \eta$ using an isobar model~\cite{Lees:2014iua}. 
The analysis reported the first observation of $K^*_0(1430) \to K \eta$, and observed that \etac decays into three pseudoscalars are dominated by intermediate scalar mesons.
This newly observed $K^*_0(1430)$ decay mode was expected to be small and in fact was not observed in the  study of $\Km p \to \Km \eta p$ interactions~\cite{lass_keta}.
More recently, the \babar\ experiment performed a measurement of the I=$1\over 2$ $K \pi$ $S$-wave amplitude from a Dalitz plot analyses of $\etac \to K \bar K \pi$~\cite{Lees:2015zzr}.
Further information on the properties of the $K^*_0(1430)$ resonance has been obtained by the CLEO experiment in an analysis of the $\Dp \to \Km \pip \pip$ decay~\cite{Bonvicini:2008jw},
and by the BESIII experiment, which observed its decay to $K \etapr$ using $\chi_{c1}$ decays to \etaprkk~\cite{Ablikim:2014tww}.

The existence of gluonium states is still an open issue for
Quantum Chromodynamics (QCD). Lattice QCD calculations predict the lightest gluonium states
to have quantum numbers $J^{PC}=0^{++}$ and $2^{++}$ and to be in
the mass region below 2.5 \gevcc~\cite{lattice}.
In particular, the $J^{PC}=0^{++}$ glueball is predicted to have a mass around 1.7 \gevcc.
Searches for these states have been performed using many supposed ``gluon rich'' reactions such as radiative decays of the heavy quarkonium states \jpsi~\cite{kopke,Dobbs} and \OneS~\cite{Lees:2018qrk}. 
However, despite intense experimental searches, there has been no conclusive experimental observation~\cite{klempt,ochs}.
The identification of the scalar glueball is further complicated by possible mixing with standard $q \bar q$ states.
The broad $f_0(500)$, $f_0(1370)$~\cite{mink}, $f_0(1500)$~\cite{amsler1,amsler2}, 
$f_0(1710)$~\cite{gg,Gui:2012gx} and possibly $f_0(2100)$~\cite{Ablikim:2013hq} have been suggested as scalar glueball candidates.
In the BESIII partial wave analysis of the radiative \jpsi decay to $\eta \eta$~\cite{Ablikim:2013hq}, the authors conclude that the production rates of $f_0(1710)$ and $f_0(2100)$ are both
about one order of magnitude larger than that of the $f_0(1500)$ and no clear evidence is found for $f_0(1370)$.
A feature of the scalar glueball is that its $s \bar s$ decay mode should be favored with respect to 
$u \bar u$ or $d \bar d$~\cite{chano,chao}.

In the present analysis, we consider the three-body \etac decays to \etaprkk, \etaprpipi, and \etapipi ,  
using two-photon interactions, $\epem\to\epem\gamma^*\gamma^*\to\epem\etac$.
If both of the virtual photons are quasi-real,
the allowed $J^{PC}$ values of any produced resonances are $0^{\pm+}$, $2^{\pm+}$, $4^{\pm+}$...~\cite{Yang}.       
Angular momentum conservation, parity conservation, and charge conjugation
invariance imply that these quantum numbers also apply to these final states.
The possible presence of a gluonic component of the \etapr meson, due to the so-called gluon anomaly, has been discussed in recent years~\cite{Harland-Lang:2013ncy,Bass:2018xmz}. 
A comparison of the $\eta$ and \etapr content of \etac decays might yield information on the possible gluonic content of resonances decaying to $\pip \pim$ or $\Kp \Km$.
The $\gg \to \etaprpipi$ process has been recently studied by the Belle experiment~\cite{Xu:2018uye} but no Dalitz plot analysis was performed.

This article is organized as follows. In Sec.\ II, a brief description of the
\babar\ detector is given. Section III is devoted to the event reconstruction and data selection. In Sec.\ IV, we describe the efficiency and resolution studies,
while in Sec.\ V we report the measurement of the \etac branching fraction. 
In Sec. VI we describe the Dalitz plot analysis methodology, and in Secs. VII, VIII, and IX we analyze \etac decays to \etaprkk, \etaprpipi, and \etapipi , respectively. 
The results are summarized in Sec.~X.

\section{The \babar\ detector and dataset}

The results presented here are based on the full data set collected
with the \babar\ detector 
at the PEP-II asymmetric-energy $e^+e^-$ collider
located at SLAC, and correspond 
to an integrated luminosity of 519~\invfb~\cite{BaBar:2013agn} recorded at
center-of-mass energies at and near the $\Upsilon (nS)$ ($n=2,3,4$)
resonances. 
The \babar\ detector is described in detail in ref.~\cite{BABARNIM}.
Charged particles are detected, and their
momenta are measured, by means of a five-layer, double-sided microstrip detector
and a 40-layer drift chamber, both operating  in the 1.5~T magnetic 
field of a superconducting
solenoid.
Photons are measured and electrons are identified in a CsI(Tl) crystal
electromagnetic calorimeter. Charged-particle
identification is provided by the measurement of specific energy loss in
the tracking devices, and by an internally reflecting, ring-imaging
Cherenkov detector.
The pions tracking efficiency increases from 98\% to 100\% in the momentum range 0.5-3 GeV/c while the average kaon identification efficiency is 84\%.
Muons and \KL\ mesons are detected in the
instrumented flux return  of the magnet.
Monte Carlo (MC) simulated events~\cite{geant}, with reconstructed sample sizes 
of the order $10^3$ times larger than the corresponding data samples, are
used to evaluate the signal efficiency and to determine background features. 
Two-photon events are simulated  using the GamGam MC
generator~\cite{BabarZ}.
In this article, the inclusion of charge-conjugate processes is implied, unless stated otherwise.

\section{ {\boldmath Event reconstruction and selection}}
\label{sec:reco}

\subsection{{\boldmath Reconstruction of the $\etapr h^+ h^-$ final state}}

We first study the reactions
\begin{equation}
  \gamma \gamma \to \etapr h^+ h^-,
  \label{eq:etaphh}
\end{equation}
where $h^+h^-$ indicates a $\pip \pim$ or $\Kp \Km$ system. The selection criteria are optimized for the \etac signal, as described below.
The \etapr is reconstructed in the two decay modes $\etapr \to \rho^0 \gamma$, $\rho^0 \to \pip \pim$, and $\etapr \to \eta \pip \pim$, $\eta \to \gamma \gamma$.
To reconstruct these final states we select events in which the $e^+$ and $e^-$ beam particles are scattered at small angles, and hence are undetected, ensuring that both virtual photons are quasi-real.
We consider photon candidates with reconstructed energy in the electromagnetic calorimeter greater than 100 MeV.
All pairs of photon candidates are combined, assuming they originate from the \epem interaction region, and pairs with invariant-mass within $\pm 20$ \mevcc\ ($\pm 150$ \mevcc) of the neutral pion ($\eta$ meson) mass are considered \piz\ ($\eta$) candidates.
We consider events with exactly 4 well-measured charged-particle tracks with transverse momentum greater than 0.1 GeV/c, and fit them to a common vertex, which must be within the $e^+ e^-$ interaction region and have a $\chi^2$ fit probability greater than 0.1\%. 
Tracks are identified as either charged kaons or pions using a high-efficiency algorithm
that rejects more than half the background with negligible signal loss.
A track can be identified as both kaon or pion (or neither) at this point. 
For the $\etapr \to \rho^0 \gamma$ selection, we allow the presence of only two $\gamma$ candidates, where \piz candidates are excluded.
For the $\etapr \to \eta \pip \pim$ we require exactly one  $\eta$ candidate, 
no more than three additional background photon candidates, and no \piz\ candidate in the event.
These selections are optimized on the data using as reference the \etac signal.

To reconstruct $\etapr \to \rho^0 \gamma$ decays,
we consider $\pip \pim$ pairs in the mass region $0.620 < m(\pip \pim)<0.875$ \gevcc. 
Each of these $\rho^0$ candidates is combined with all $\gamma$ candidates,
and any combination with invariant-mass in the range $0.935<m(\rho^0 \gamma)<0.975$ \gevcc\ is considered an  \etapr candidate. 
We compute the angle $\theta_{\gamma}$, 
defined as the angle between the \pip\ and the $\gamma$ in the $\pip \pim$ rest frame.
The distribution of $\theta_{\gamma}$ is expected to be proportional to $\sin^2\theta_{\gamma}$~\cite{Rosner}. We thus scan the $\rho^0 \gamma$ mass spectrum with varying selection on $|\cos\theta_{\gamma}|$ and obtain a small reduction of the combinatorial background by requiring $|\cos\theta_{\gamma}|<0.85$.
The above selection reduces the \etapr signal and background yields by 3\% and 17\%, respectively.

To improve the mass experimental resolution, 
the \etapr four-momentum is constructed by adding the momenta of the $\pip$, $\pim$, and $\gamma$, and computing the \etapr energy by assigning the Particle Data Group (PDG)~\cite{PDG} nominal mass. This method, tested on MC simulations, improves the resolution by $\approx 20\%$.

\begin{figure*}
\begin{center}
\includegraphics[width=8.5cm]{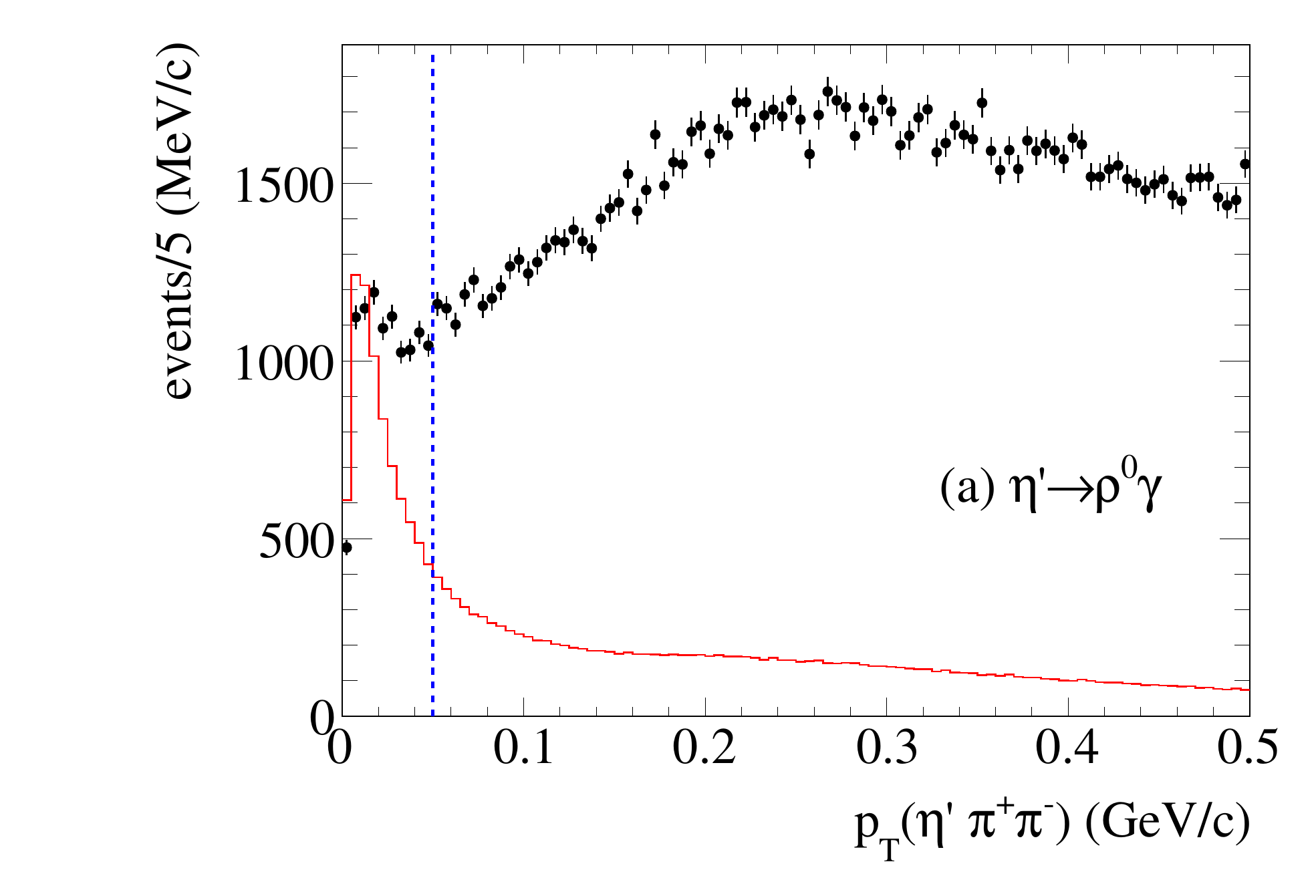}
\includegraphics[width=8.5cm]{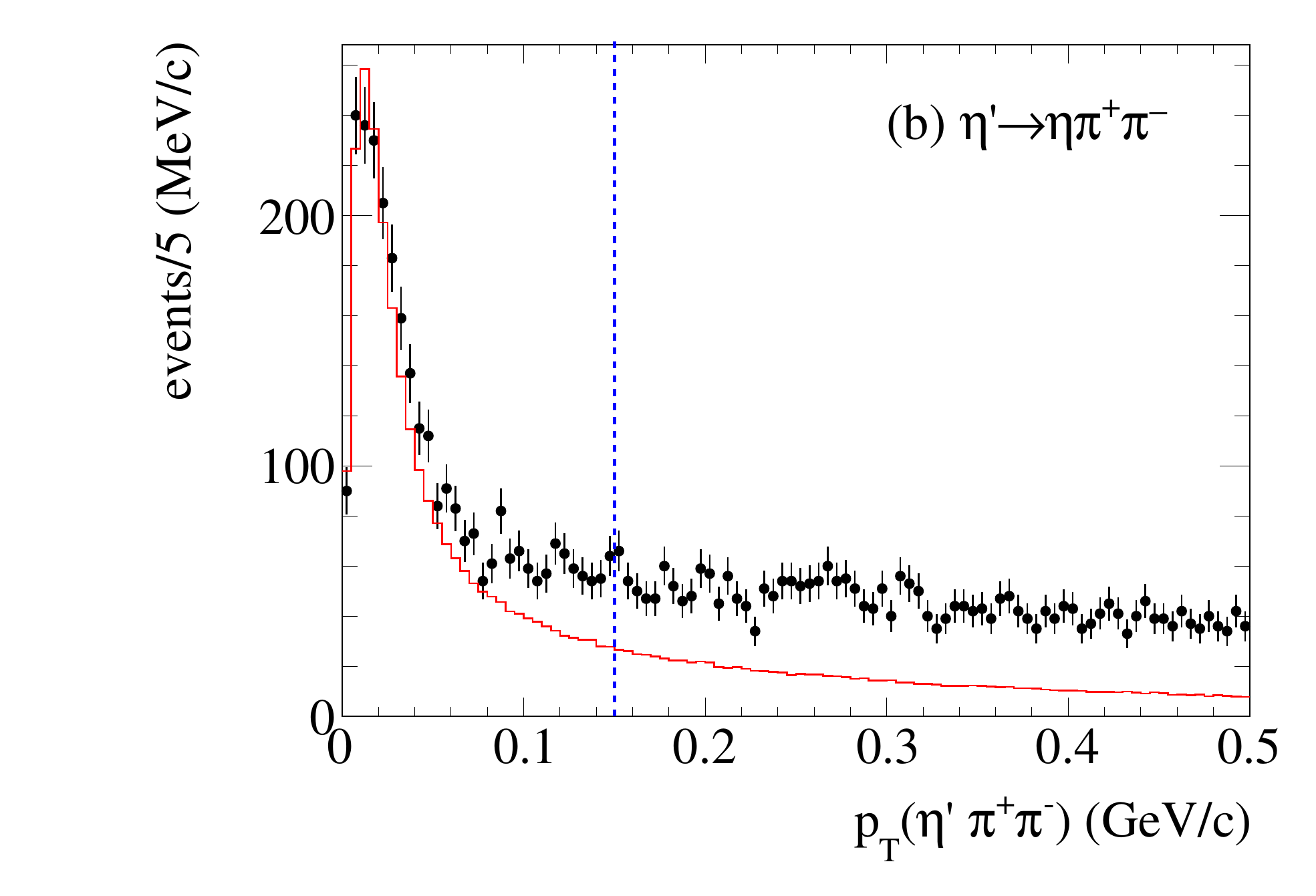}
\includegraphics[width=8.5cm]{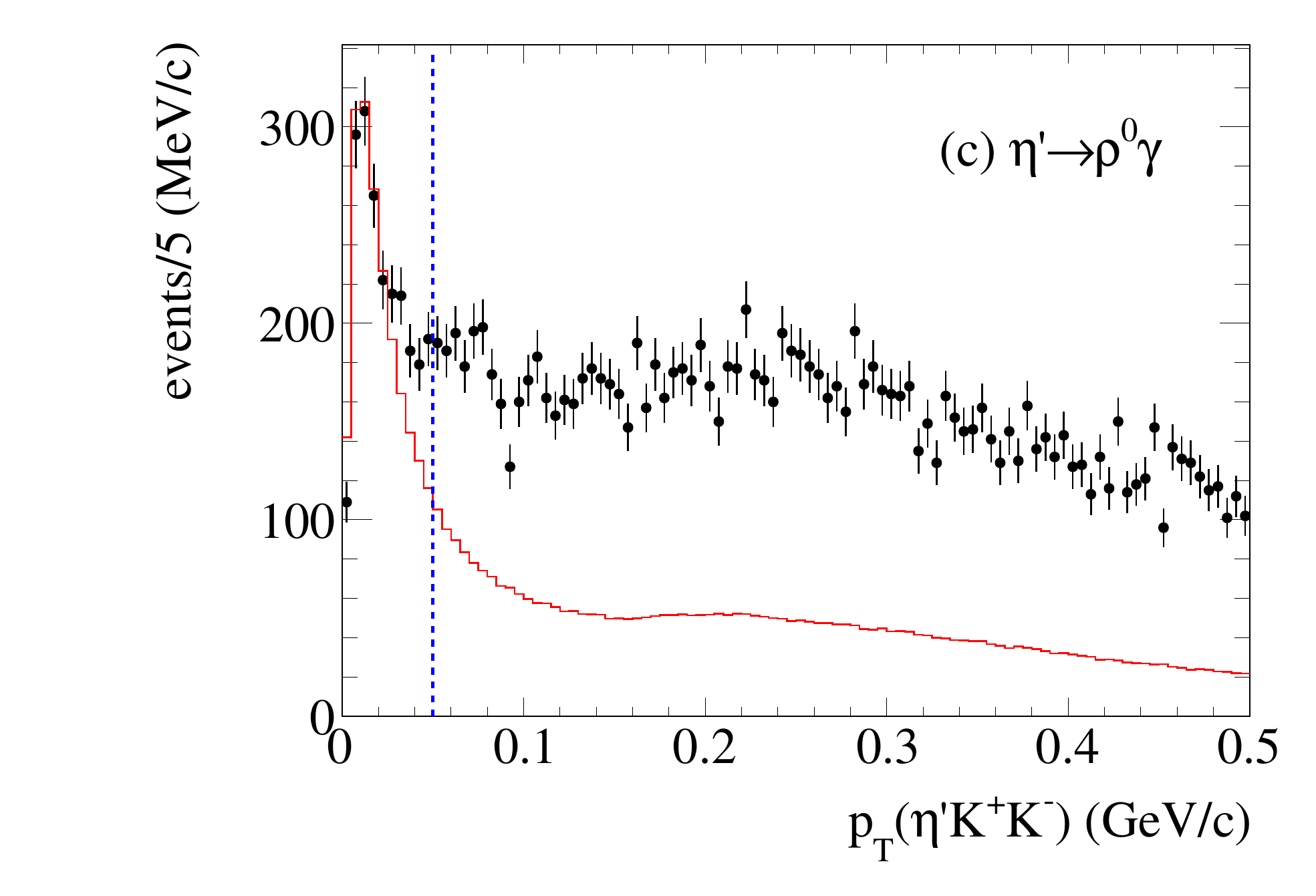}
\includegraphics[width=8.5cm]{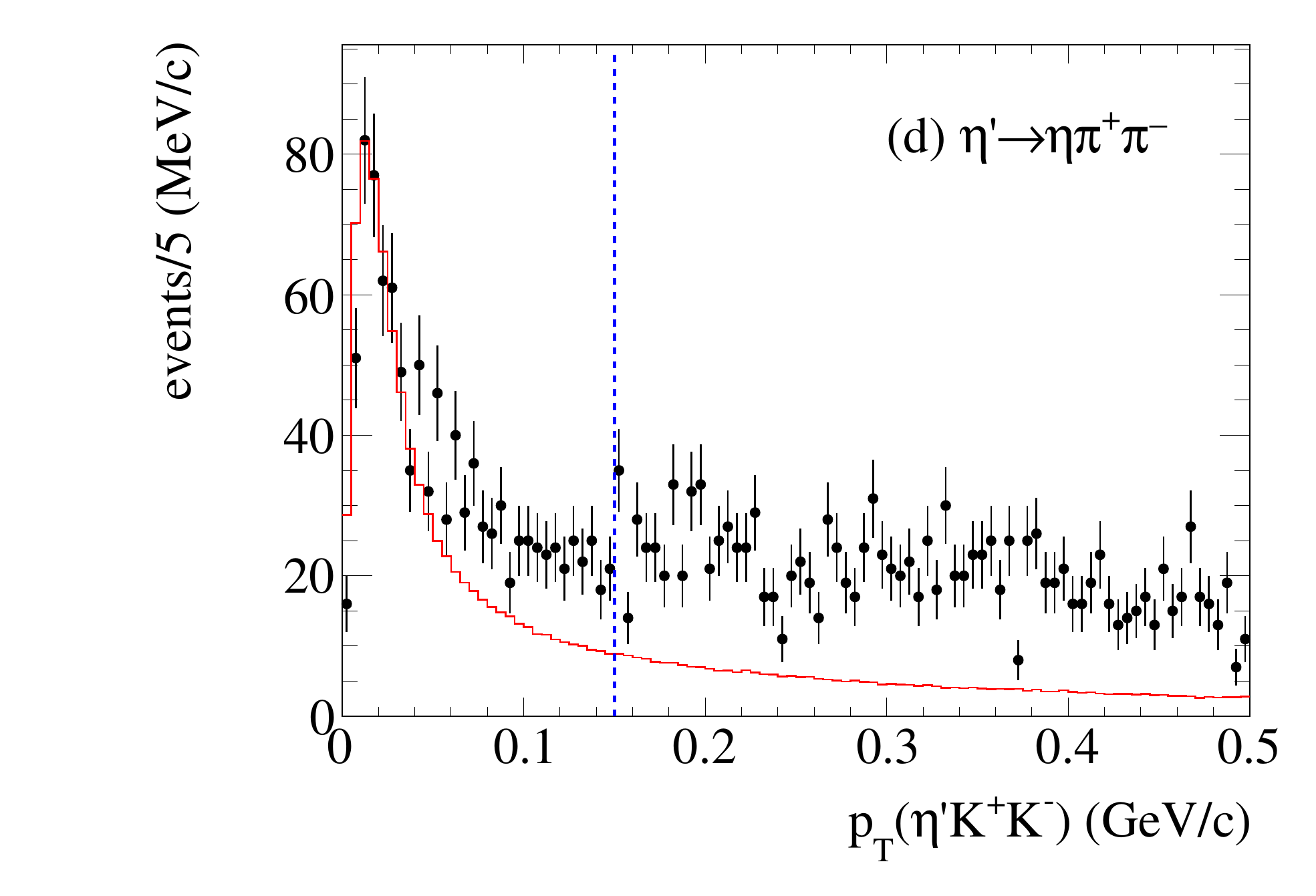}
\caption{Distributions of the transverse momenta of the 
(a,b) $\etapr \pip \pim$ and 
(c,d) $\etapr \Kp \Km$ systems for events satisfying all other selection criteria, in which the $\etapr$ is reconstructed in the 
(a,c) $\rho^0 \gamma$ and 
(b,d) $\eta \pip \pim$ decay modes.  
The data are represented by points with error bars, and the \etac MC simulation by solid (red) histograms with arbitrary normalization. 
The (blue) dashed lines indicate the selection used to isolate two-photon event candidates.}
\label{fig:fig1}
\end{center}
\end{figure*}

To reconstruct $\etapr \to \eta \pip \pim$ decays,
we perform a kinematic fit to the $\eta$ candidate, 
and require the $\eta \pip \pim$ mass to be within $\pm 2\sigma$ of the fitted \etapr mass ($956.8~\pm~0.5$)~\mevcc, where $\sigma=2.9$ \mevcc\ is the width of the resolution function describing the \etapr signal. 
Similarly, to improve the experimental resolution, the \etapr four momentum is constructed by adding the momenta of the $\pip$, $\pim$, and $\eta$, and computing the \etapr energy by assigning the PDG mass. 

Background arises mainly from random combinations of particles from \epem\ annihilation, 
from other two-photon processes, 
and from events with initial-state photon radiation (ISR). 
The ISR background is dominated by events with a single high-energy photon recoiling against the reconstructed hadronic system, 
which in the mass region of interest is typically a $J^{PC}=1^{--}$ resonance~\cite{isr}.
We discriminate against ISR events by requiring the recoil mass $\mm\equiv(p_{\epem}-p_{\rm rec})^2 > 2$~GeV$^2$/$c^4$,
where $p_{\epem}$ is the four-momentum of the initial state \epem\, 
and $p_{\rm rec}$ is the reconstructed four-momentum of the candidate $\etapr(\eta) h^+h^-$ system.

We define \pt\ as the magnitude of the transverse momentum of the $\etapr h^+h^-$ system, in the \epem\ rest frame, with respect to the beam axis.
Well reconstructed two-photon events with quasi-real photons are expected to have low values of \pt . 
Substantial background arises from $\gamma \gamma \to 2h^+ 2h^-$ events, combined with a background photon candidate.
These are removed by requiring $\pt(2h^+ 2h^-)>0.1$ \gevc.

We retain events with \pt below a maximum value that is optimized with respect to the \etac signal for each decay mode.
We produce $\etapr h^+ h^-$ invariant-mass spectra with different maximum \pt values, and fit them to extract the number of \etac signal events ($N_s$) (defined as the \mbox{2.93-3.03~\gevcc}\ interval) and the number of background events underneath the \etac signal ($N_b$). 
We then compute the purity, defined as $P = N_s/(N_s + N_b)$, the figure of merit $S = N_s/\sqrt{N_s + N_b}$, and their product, $PS$.

\subsubsection{Reconstruction of the \etaprpipi final state}

For the final selection of the $\etapr \pip \pim$ final state, we require all four charged tracks to be positively identified as pions, using an algorithm based on multivariate analysis~\cite{BDT} that is more than 98\% efficient for the tracks in the sample, while suppressing kaons by a factor of at least seven.

Figures~\ref{fig:fig1}(a) and~\ref{fig:fig1}(b) show the \pt distributions for selected events in the charmonium region. This region is defined as reconstructed invariant-mass \mbox{$m(\etapr (\eta) h^+ h^-)>2.7$~\gevcc}.
In the case of $\etapr \to \rho^0 \gamma$ an upper mass requirement $m(\etapr \pip \pim)<3.5$~\gevcc\ is applied
because of the large number of combinations produced by the presence of the $\gamma$.
The data are compared with expectations from \etac signal MC simulations; 
a signal from two-photon production is observed in the data in both cases, 
and is particularly clean for $\etapr \to \eta \pip \pim$. 
In a scan of the $S$, $P$, and $PS$ variables as functions of the maximum \pt value, 
we observe a broad maximum of $S$ starting at 0.05 \gevc\ for the $\etapr \to \rho^0 \gamma$ decay candidates, 
and a maximum of $PS$ at 0.15~\gevc\ for the $\etapr \to \eta \pip \pim$ candidates.
We require $\pt<0.05$~\gevc\ and $\pt<0.15$~\gevc, respectively, as indicated by the dashed lines in the figures.

\begin{figure*}
\begin{center}
\includegraphics[width=8.5cm]{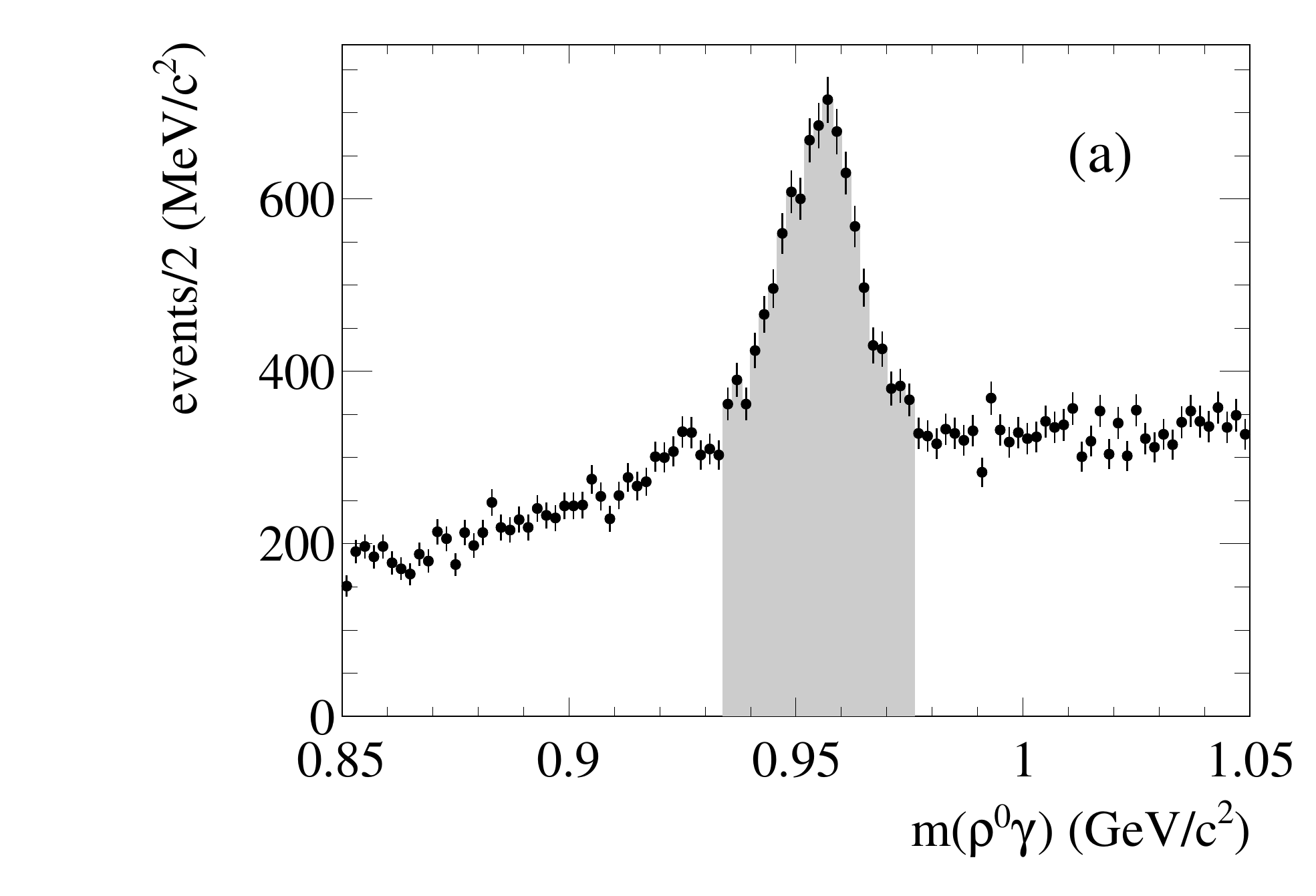}
\includegraphics[width=8.5cm]{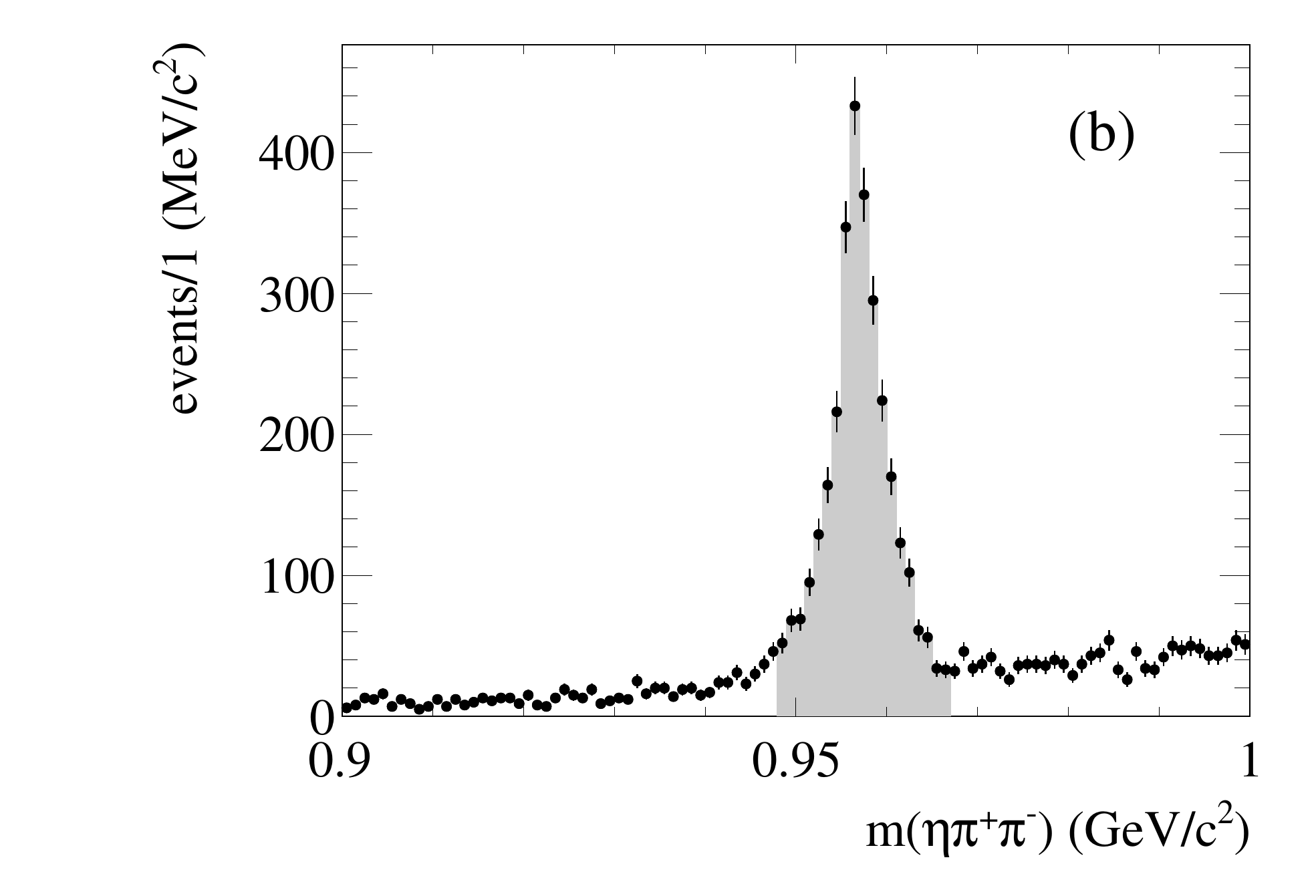}
\includegraphics[width=8.5cm]{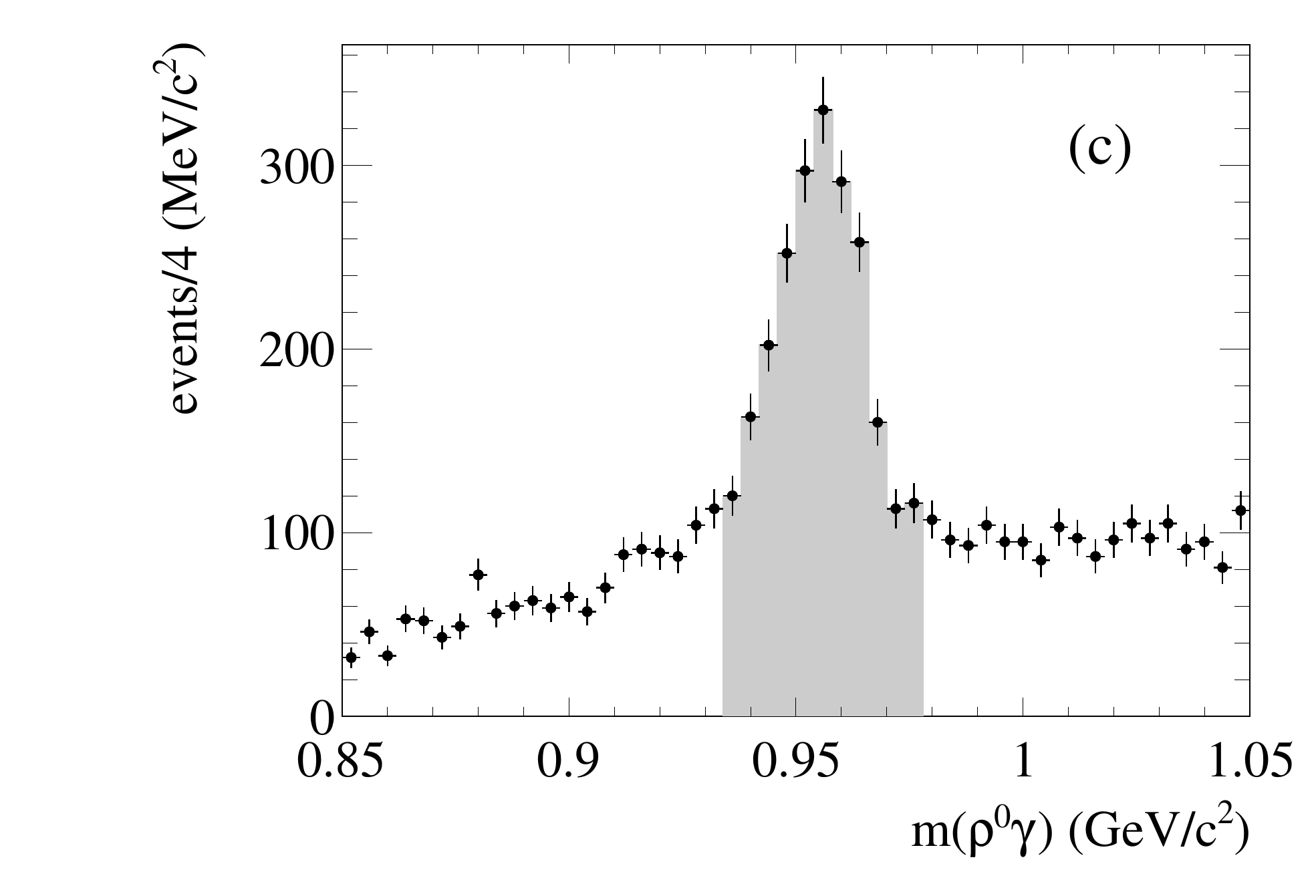}
\includegraphics[width=8.5cm]{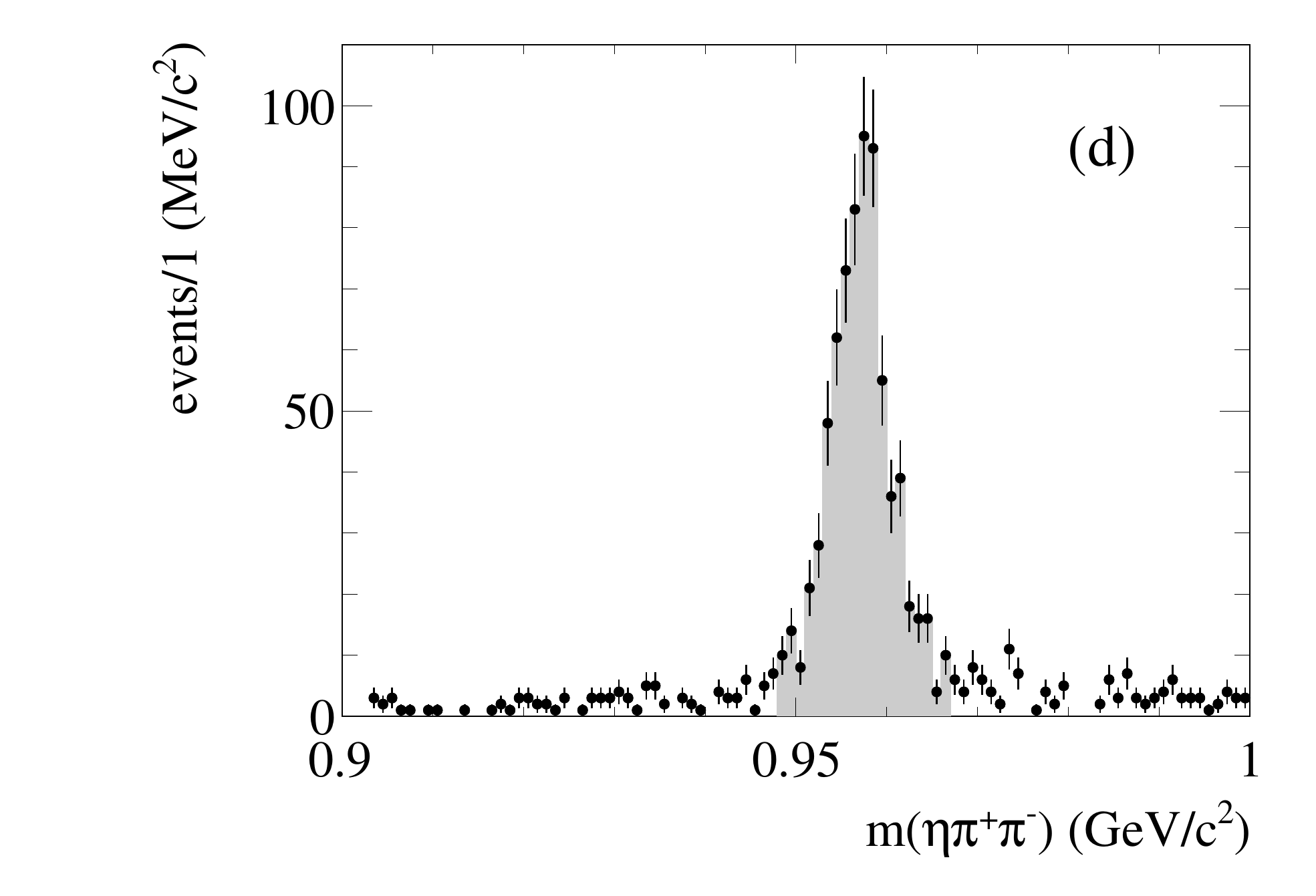}
\caption{Invariant-mass distributions of (a) $\rho^0 \gamma$ and (b) $\eta \pip \pim$ for $\gamma \gamma \to \etapr \pip \pim$ candidates satisfying all other selection criteria. 
Corresponding (c) $\rho^0 \gamma$ and (d) $\eta \pip \pim$ invariant-mass distributions for $\gamma \gamma \to \etapr \Kp \Km$ candidates.  
The shaded areas indicate the \etapr selections.}
\label{fig:fig2}
\end{center}
\end{figure*}

\begin{figure*}
\begin{center}
\includegraphics[width=8.5cm]{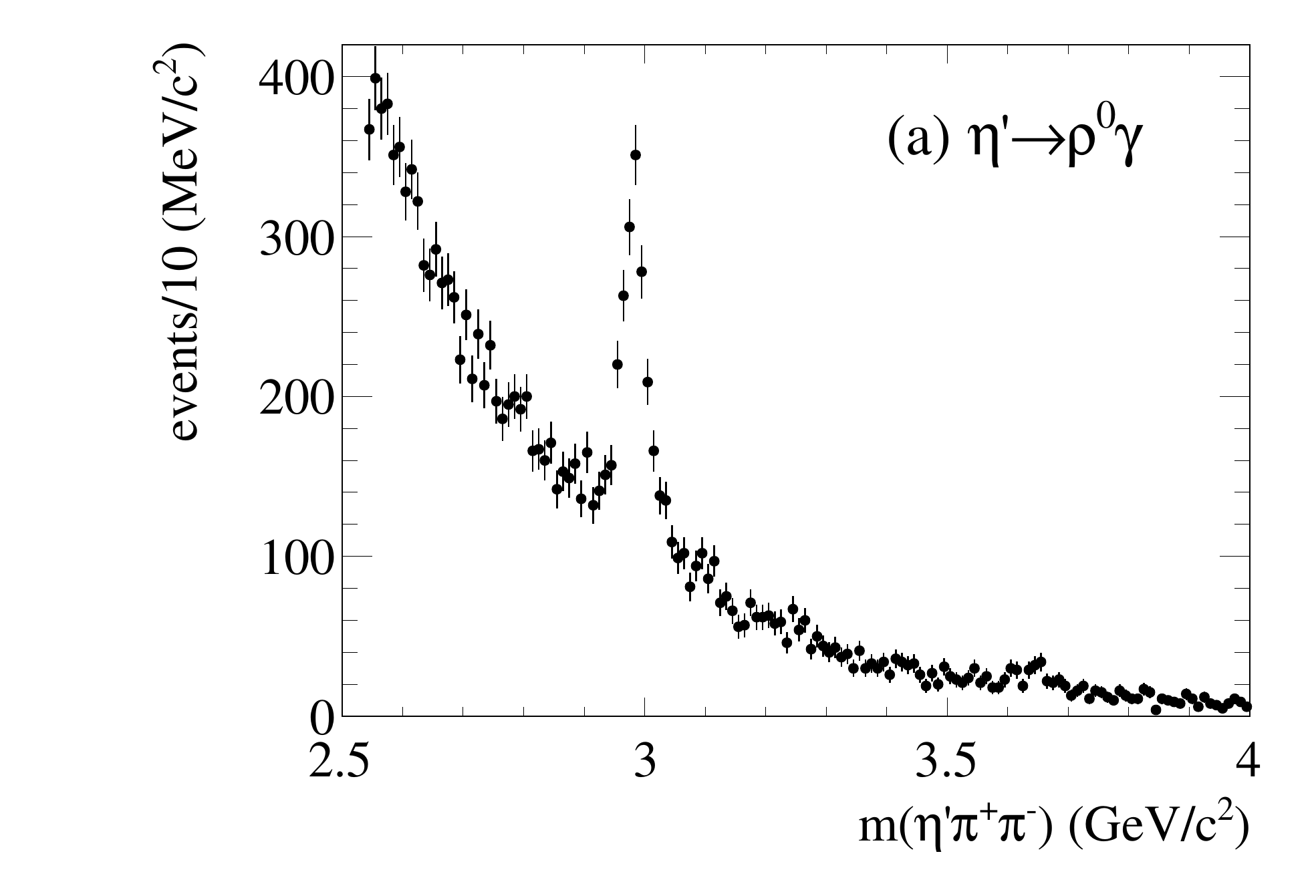}
\includegraphics[width=8.5cm]{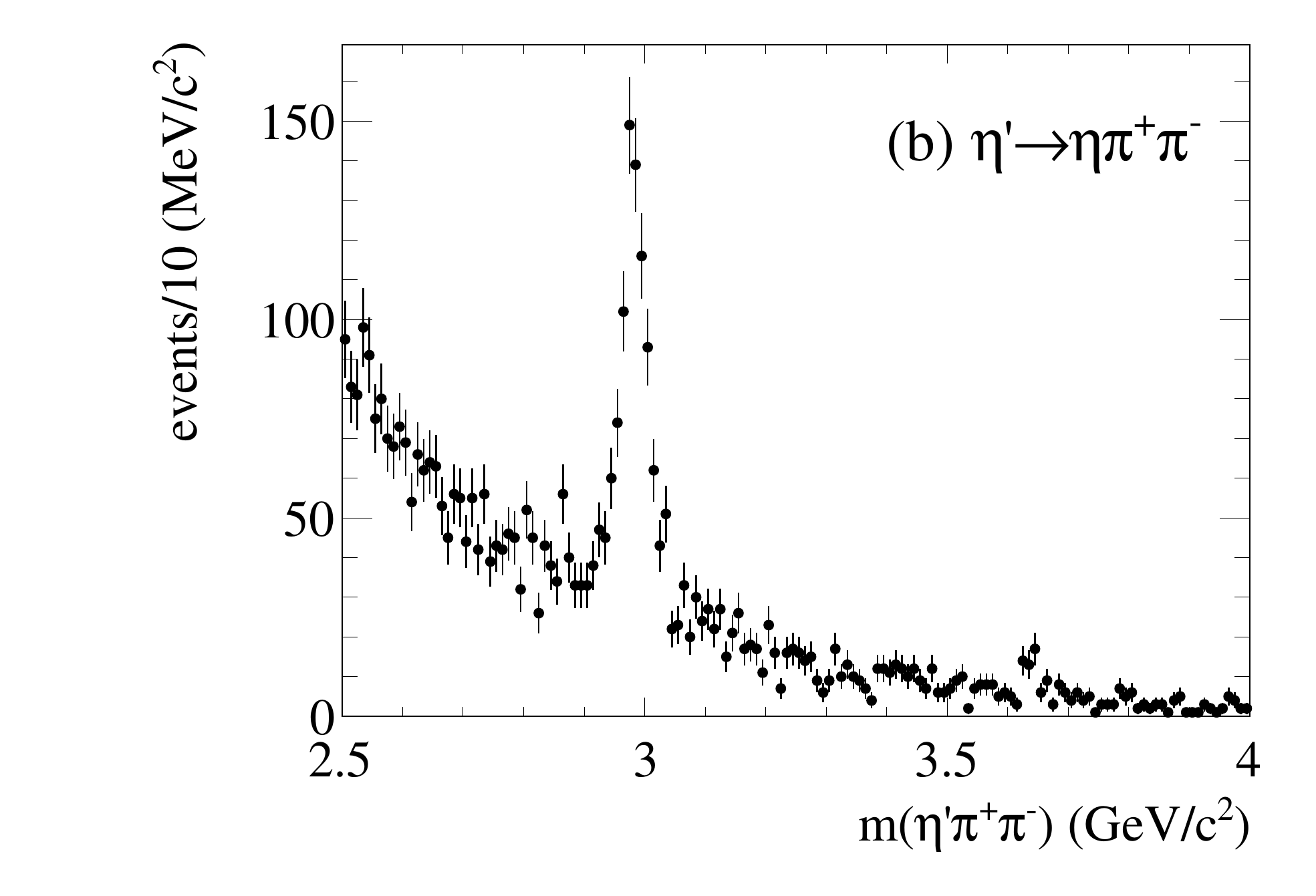}
\includegraphics[width=8.5cm]{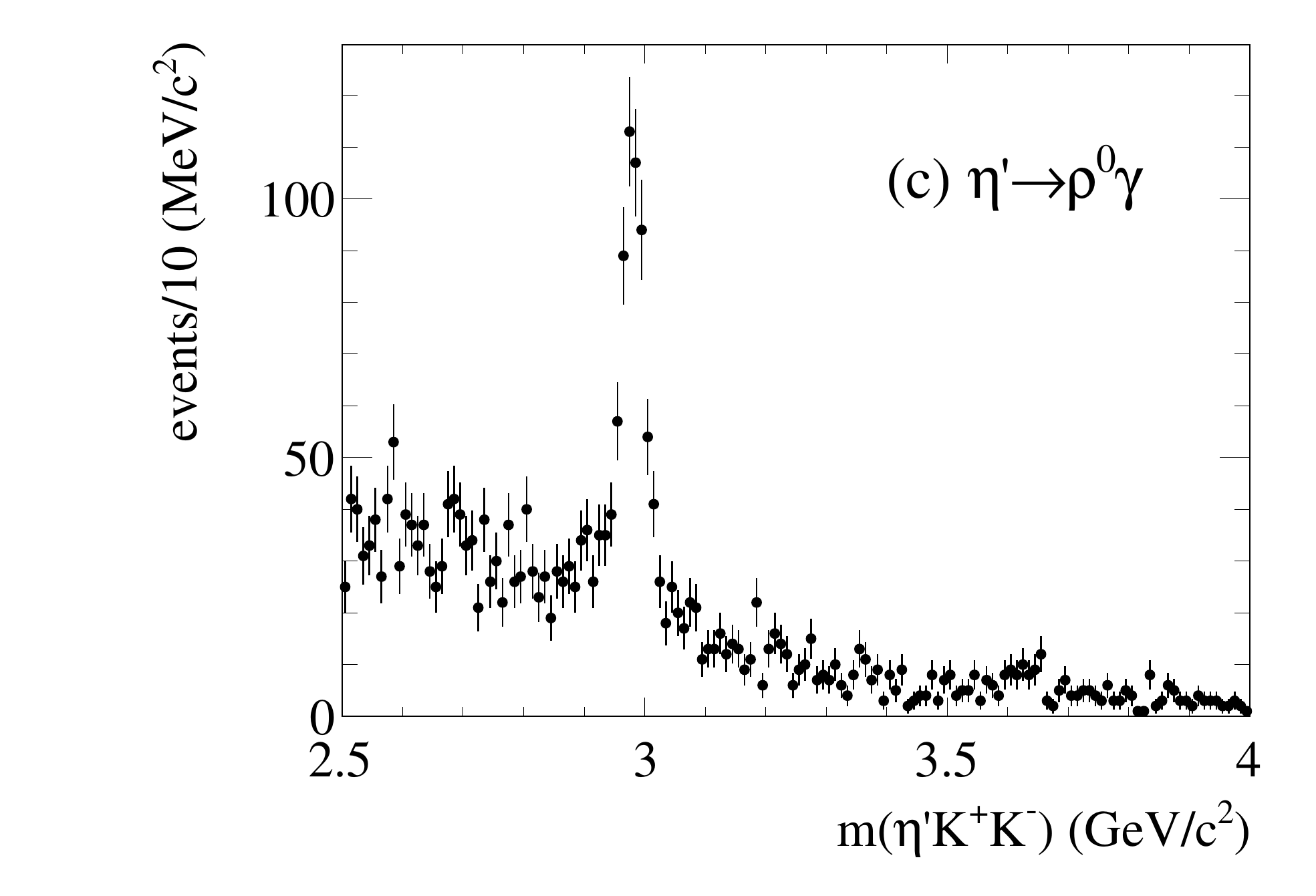}
\includegraphics[width=8.5cm]{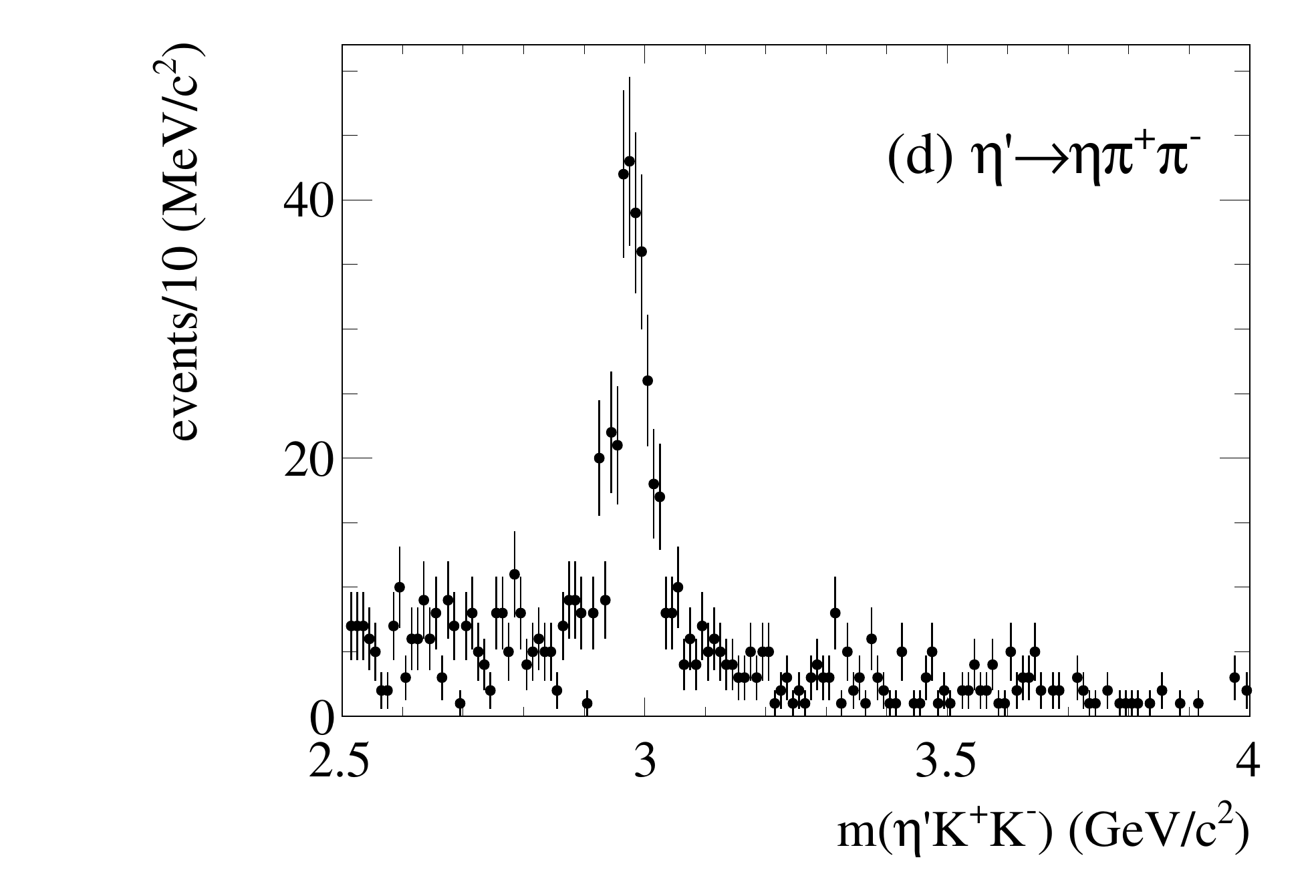}
\caption{The \etaprpipi invariant-mass spectra for selected events with (a) $\etapr\to\rho^0 \gamma$ and (b) $\etapr\to\eta \pip \pim$.
The \etaprkk invariant-mass spectra for selected events with (c) $\etapr\to\rho^0 \gamma$ and (d) $\etapr\to\eta \pip \pim$.}
\label{fig:fig3}
\end{center}
\end{figure*}

Figures~\ref{fig:fig2}(a) and ~\ref{fig:fig2}(b) show the 
$\rho^0 \gamma$ and $\eta \pip \pim$ invariant-mass distributions, respectively, for events satisfying all selection criteria except that on these masses.
Clear \etapr signals are visible, and the shaded regions indicate the selection windows, (0.935-0.975) \gevcc\ for $\etapr \to \rho^0\gamma$ and (0.948-0.966) \gevcc\ for $\etapr \to \eta \pip \pim$.
Figures~\ref{fig:fig3}(a,b) show the \etaprpipi invariant-mass spectra for the selected events in the data.  
Prominent \etac signals are observed, and there is some activity in the \etactwo mass region. 

\begin{figure*}
\begin{center}
\includegraphics[width=8.5cm]{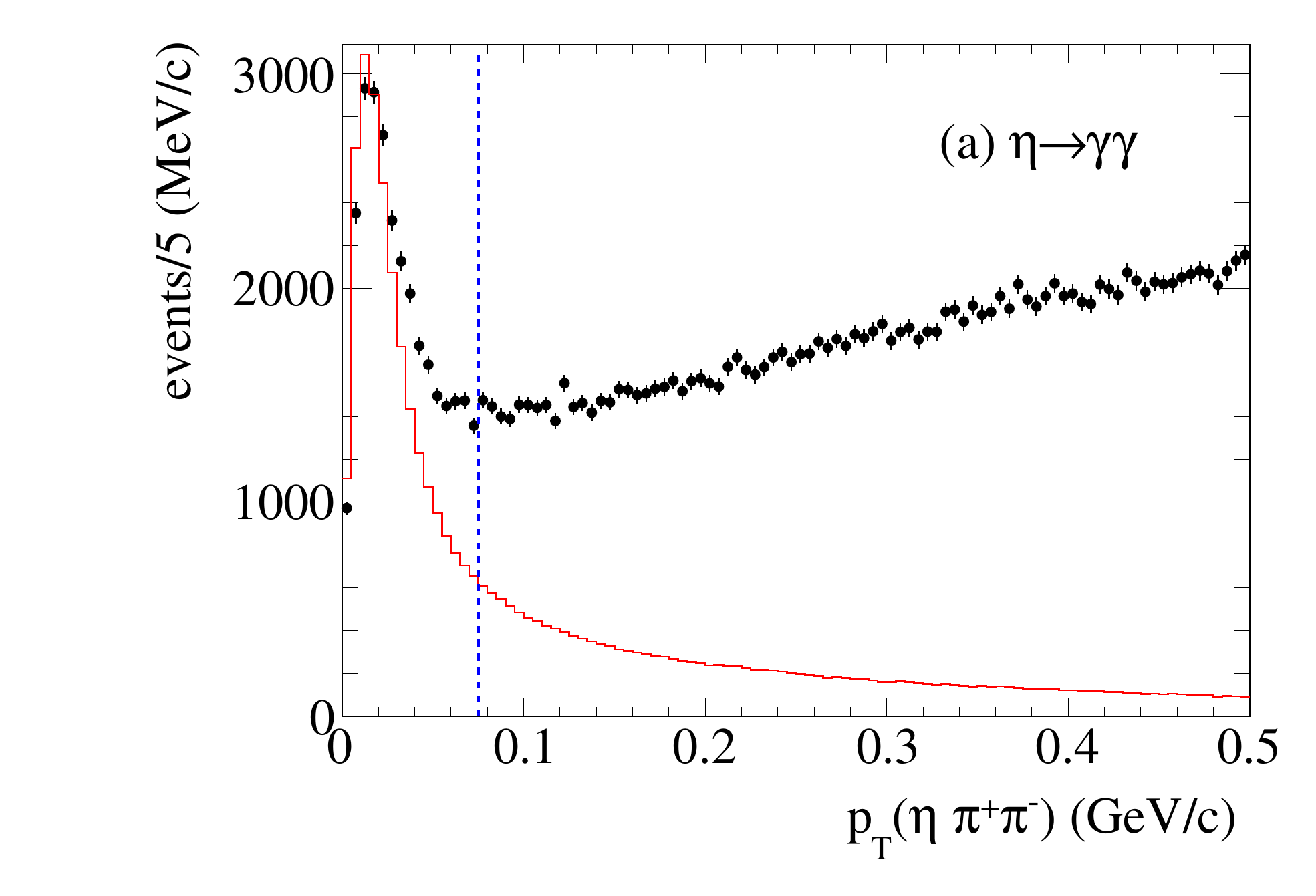}
\includegraphics[width=8.5cm]{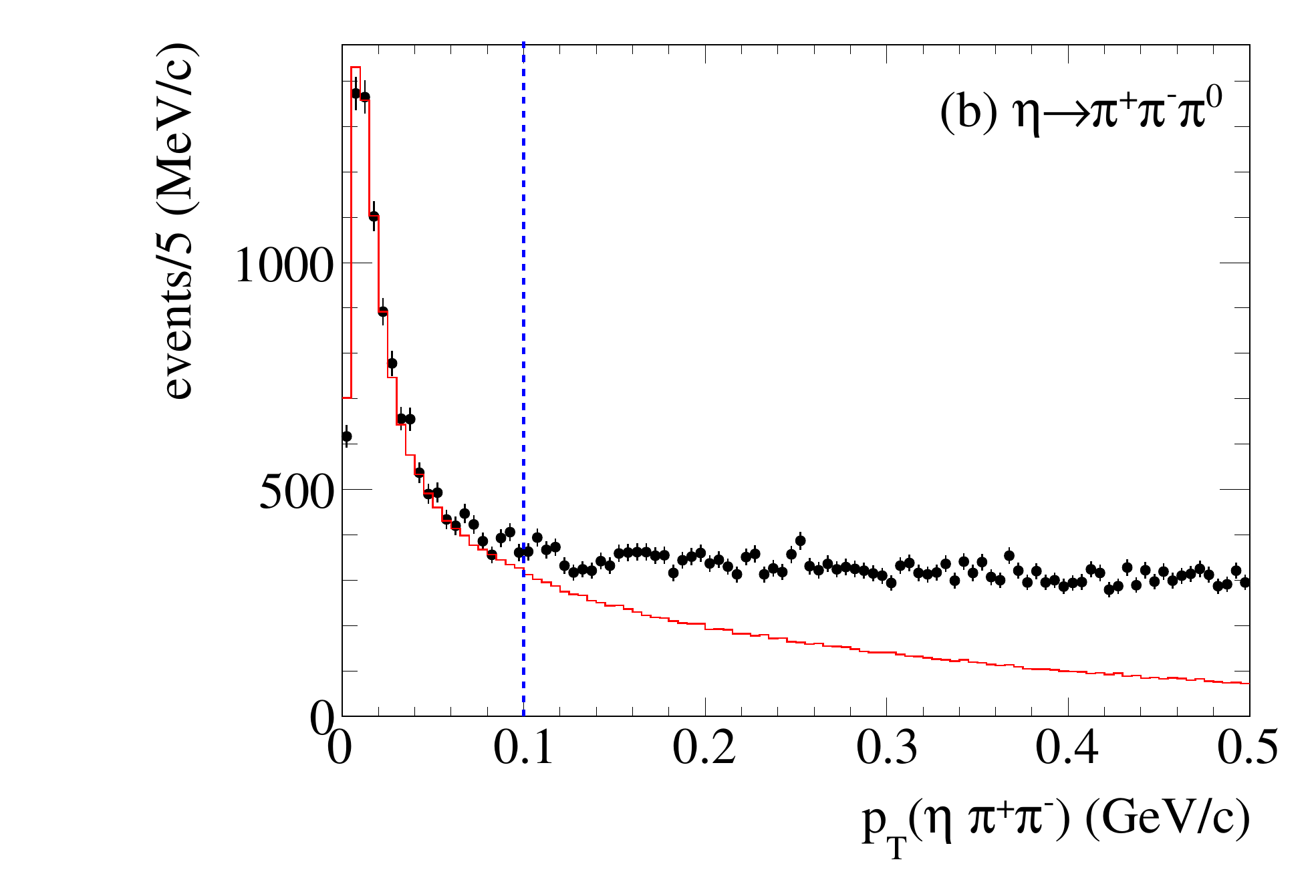}
\caption{Distributions of the transverse momentum $p_T(\eta \pip \pim)$ for selected $\gamma\gamma\to\eta\pip\pim$ candidates with (a) $\eta \to \gamma \gamma$ and (b) $\eta \to \pip \pim \piz$, in the charmonium mass region. 
The data are represented by the points with error bars, 
and the \etac MC simulation as solid (red) histograms with arbitrary normalizations. 
The dashed (blue) lines indicate the selection used to isolate two-photon event candidates.}
\label{fig:fig4}
\end{center}
\end{figure*}

If there are multiple candidates in the same event, we retain them all. 
The fraction of events having two combinations in the \etac mass region is 3\% (and 3.4\% in \etac signal MC simulations) for $\etac \to \etapr \pip \pim$ with $\etapr \to \rho^0 \gamma$.
No multiple candidates are found for $\etapr\to\pip\pim\piz$ or any of the other final states discussed below.

\subsubsection{Reconstruction of the \etaprkk final state}

For the \etaprkk final state, 
we require the two charged tracks assigned to the \etapr decay to be positively identified as pions and 
the other two to be positively identified as kaons.
The algorithm is more than 92\% efficient for kaon identification, while suppressing pions by a factor of at least five.
The \pt distributions for events in the charmonium region, compared with MC \etac signal simulations, are shown in Figs.~\ref{fig:fig1}(c)-(d), where signals of the two-photon reaction can be seen. 
To minimize  systematic uncertainties in the measurements of the branching fractions, the same \pt requirements as for the \etaprpipi final state are used, indicated by the dashed lines in the figures.

The corresponding \etapr signals for this final state are shown in Figs.~\ref{fig:fig2}(c)-(d), 
and the \etaprkk invariant-mass spectra are shown in Figs.~\ref{fig:fig3}(c)-(d).
Prominent \etac signals with low background are present in both invariant-mass spectra with possible weak activity in the \etactwo mass region. 
The decay $\etac \to \etaprkk$ is observed here for the first time.

\subsection{{\boldmath Reconstruction of the $\eta \pip \pim$ final state}}

We study the reaction
\begin{equation}
  \gamma \gamma \to \eta \pip \pim, 
  \label{eq:etapipi}
\end{equation}
where $\eta \to \gamma \gamma$ and $\eta \to \pip \pim \piz$.

\subsubsection{$\eta \to \gamma \gamma$}

For reaction~(\ref{eq:etapipi}), where $\eta \to \gamma \gamma$, 
we again consider well-measured charged-particle tracks with transverse momenta greater than 0.1~\gevc\ and photons with energy greater than 0.1~\gev, 
and each pair of $\gamma$'s is kinematically fitted to the \piz\ and $\eta$ hypotheses.
We require exactly two selected tracks, fit them to a common vertex, and require the fitted vertex to be within the interaction region and the $\chi^2$ probability of the fit to be greater than 0.1\%.
We retain events having exactly one $\eta$ candidate, no \piz\ candidates, and no more than three background $\gamma$'s.

The two charged tracks are required to be loosely identified as pions.
Most ISR events are removed by requiring $\mm\equiv(p_{\epem}-p_{\rm rec})^2>2$~GeV$^2$/$c^4$.
Further background is due to the presence of ISR events from $\psi(2S) \to \eta J/\psi \to \eta\mup\mun$, where the two muons are misidentified as pions. 
This background is efficiently removed by vetoing events having two loosely identified muons.
Background from the process $\gamma \gamma \to \pip \pim$ is removed by requiring $\pt(\pip \pim)>0.05$ \gevc.

The \pt distribution for such events in the charmonium mass region is compared with \etac signal MC simulation in Fig.~\ref{fig:fig4}(a), where a clear signal of the two-photon reaction is observed.
Optimizing the \etac figure of merit ($S$) and purity ($P$),
we require $\pt<0.1$ \gevc. The resulting $\eta \pip \pim$ invariant-mass spectrum is shown in Fig.~\ref{fig:fig6}(a), where
the \etac signal can be observed together with some weak activity in the \etactwo mass region.

\subsubsection{$\eta \to \pip \pim \piz$}

For reaction~(\ref{eq:etapipi}), where $\eta \to \pip \pim \piz$, we
require exactly four well-measured charged-particle tracks with the vertex $\chi^2$ fit probability greater than 0.1\%.
In order to have sensitivity to low momentum \piz\ mesons, we consider photons with energy greater than 30 \mevcc. 
We allow no more than two kinematically fitted \piz\ candidates and no more than five background $\gamma$'s.
Candidate $\gamma \gamma \to 2 \pip 2 \pim$ events are removed by requiring $\pt( 2 \pip 2 \pim)>0.05$~\gevc. 
Background ISR events are removed by requiring \mbox{$\mm\equiv(p_{\epem}-p_{\rm rec})^2>2$~GeV$^2$/$c^4$.} All four charged tracks are required to be loosely identified as pions.

\begin{figure}
\begin{center}
\includegraphics[width=8.5cm]{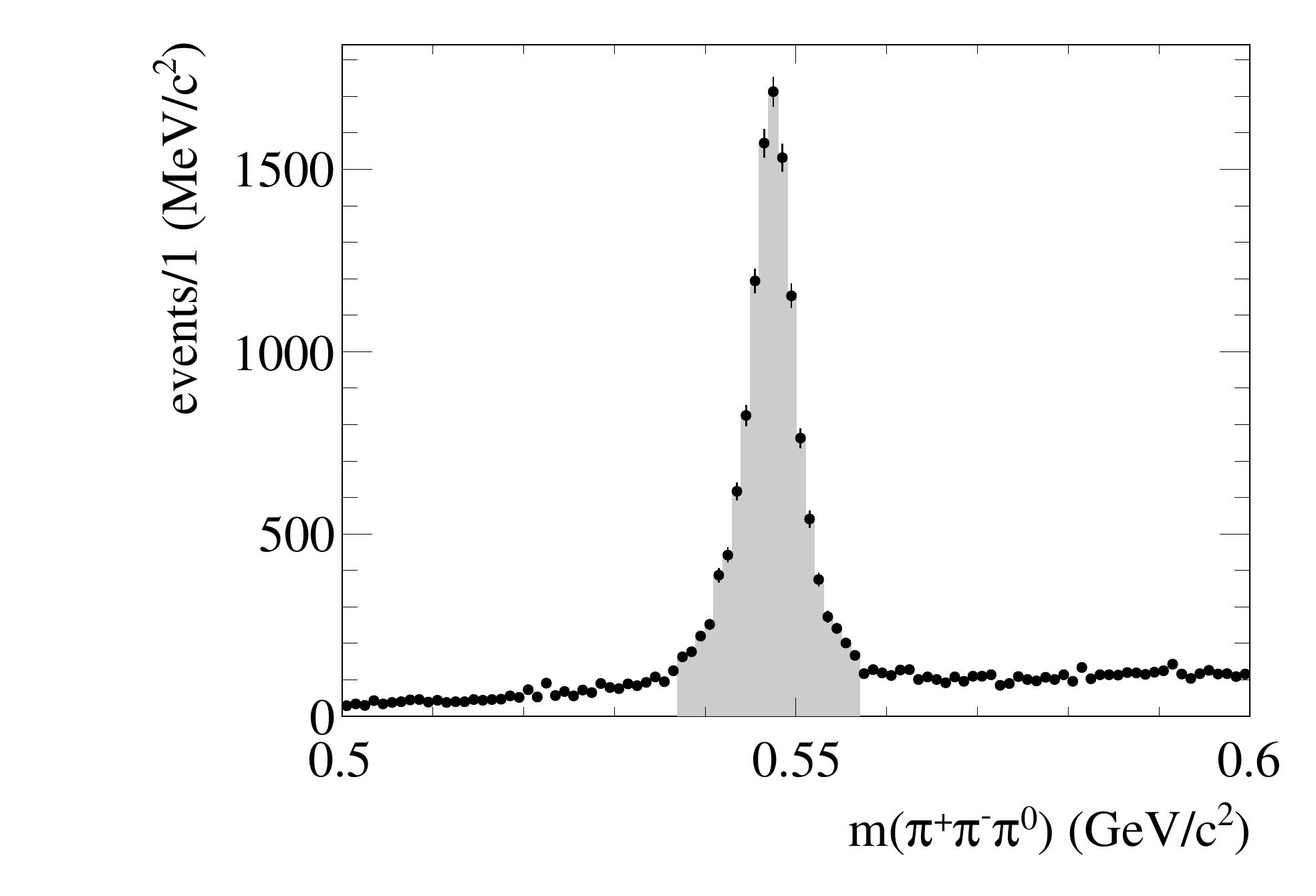}
\caption{Distribution of the reconstructed $\pip\pim\piz$  mass for selected $\gamma \gamma \to \eta \pip \pim$ candidate events. 
The shaded area indicates the $\eta$ selection region.}
\label{fig:fig5}
\end{center}
\end{figure}

The $\eta$ candidates are reconstructed by combining every pair of oppositely charged tracks with each of the \piz\ candidates in the event.
The resulting $\pip \pim \piz$ invariant-mass spectrum is shown in Fig.~\ref{fig:fig5}. 
A clean $\eta$ signal can be seen; we select candidates in the mass region $538<m(\pip\pim\piz)<557~\mevcc$. 
The $\eta$ is then reconstructed by adding the momentum three-vectors of the three pions and computing the $\eta$ energy using its nominal PDG mass.

\begin{figure*}
\begin{center}
\includegraphics[width=8.5cm]{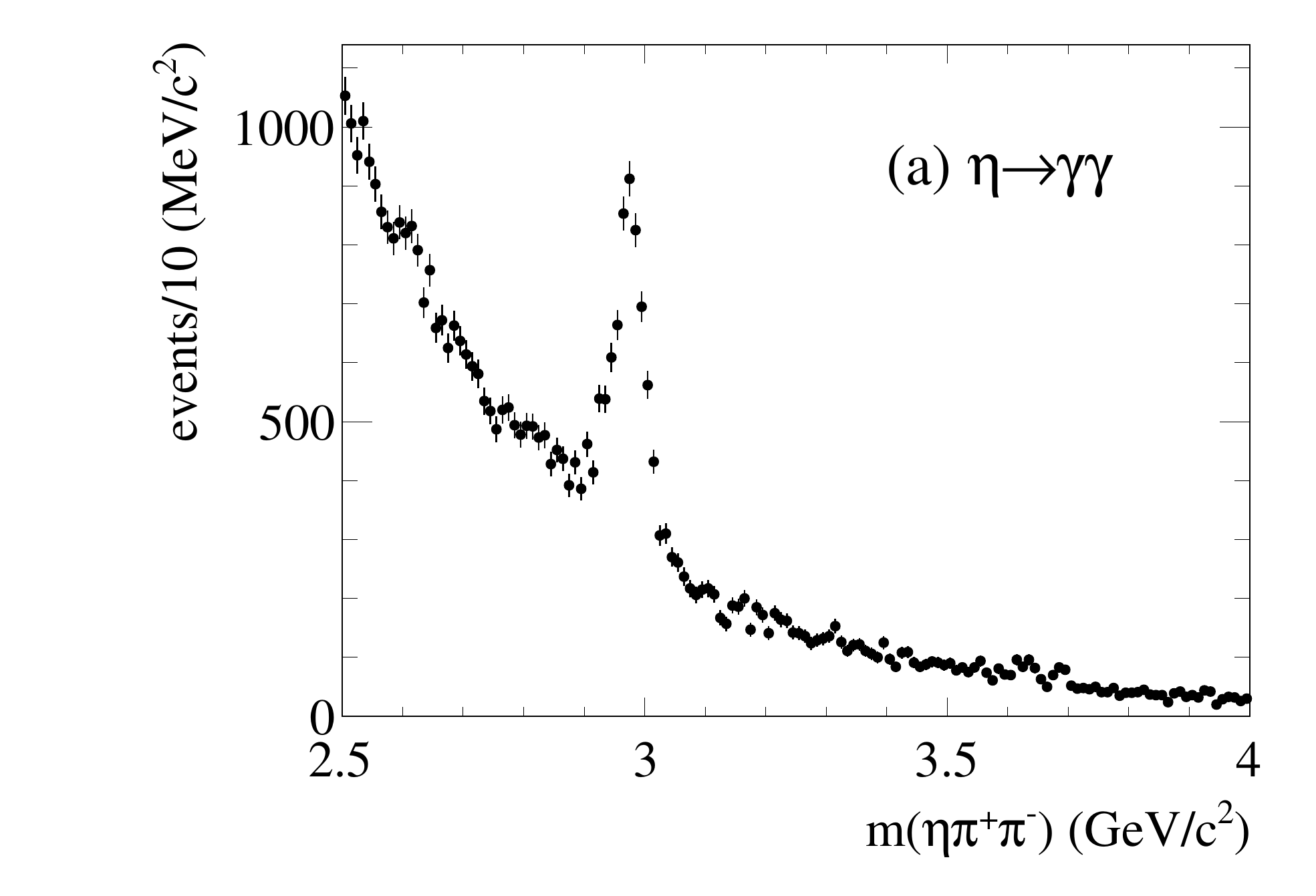}
\includegraphics[width=8.5cm]{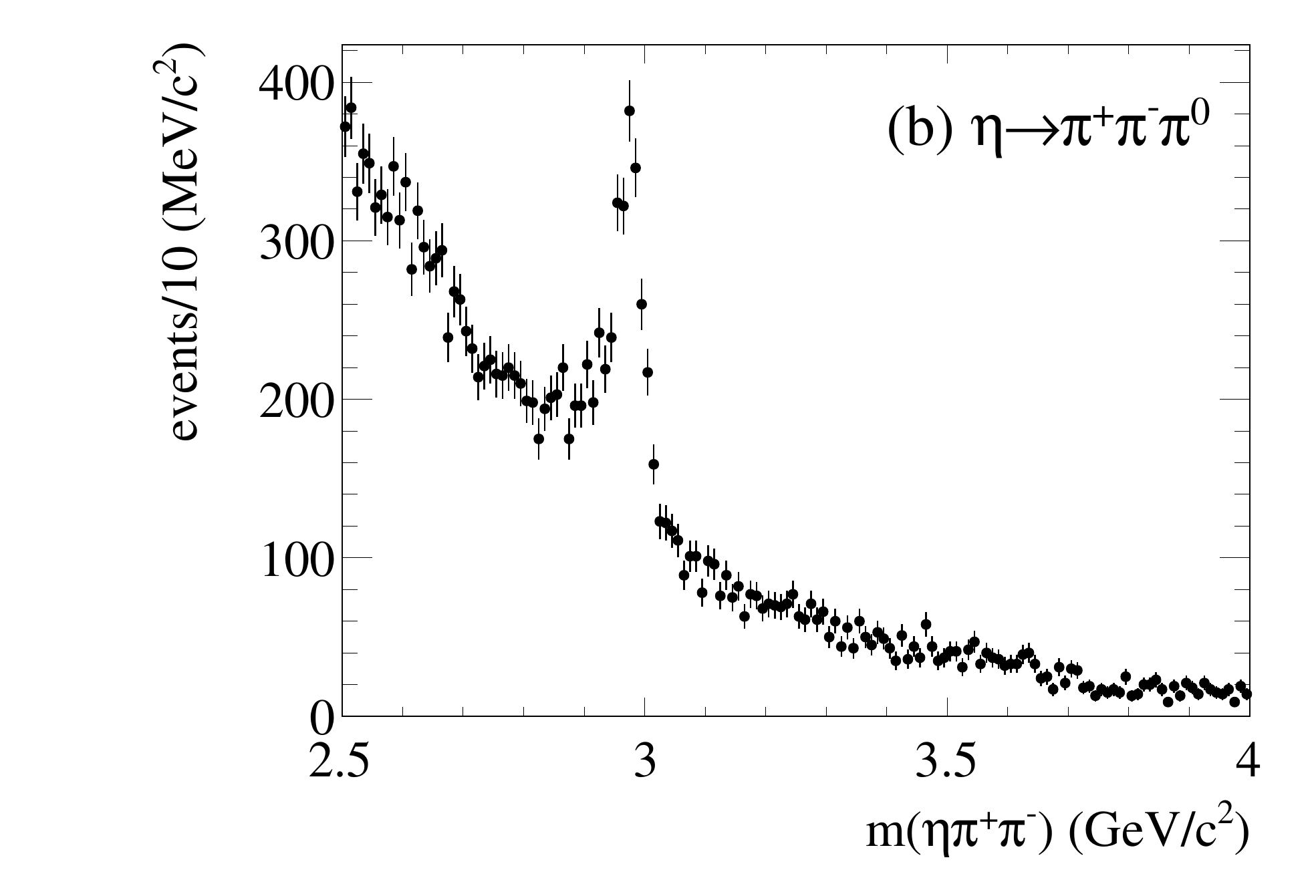}
\caption{The \etapipi invariant-mass spectra for selected events with (a) $\eta \to \gamma \gamma$ and (b) $\eta \to \pip \pim \piz$.}
\label{fig:fig6}
\end{center}
\end{figure*}

The \pt distribution for such events in the charmonium mass region is compared with \etac signal MC simulation in Fig.~\ref{fig:fig4}(b), where a clear signal of the two-photon reaction is observed.
In this case, a maximum of the $PS$ figure of merit leads to the requirement $\pt < 0.1$ \gevc.
The resulting $\eta \pip \pim$ invariant-mass spectrum is shown in Fig.~\ref{fig:fig6}(b), where
the \etac signal can be observed together with some weak activity in the \etactwo mass region.

\section{{ \boldmath Efficiency and \etac invariant-mass resolution}}
\label{sec:effy}

To compute the reconstruction and selection efficiency, MC signal events are generated using a detailed detector simulation~\cite{geant,BabarZ} in which the \etac\ mesons decay uniformly in phase space.
These simulated events are reconstructed and analyzed in the same manner as data. 
We define the helicity angle $\theta_H$ as the angle formed by the $h^+$ (where $h=\pi,K$), in the $h^+ h^-$ rest frame, and the \etapr ($\eta$) direction in the $h^+ h^- \etapr$ ($h^+ h^- \eta$) rest frame. 
For each final state, we compute the raw efficiency in 50$\times$50 intervals of the invariant-mass, $m(h^+ h^-)$, and $\cos \theta_H$, as the ratio of reconstructed to generated events in that interval. 

To smoothen statistical fluctuations, the efficiency maps are parameterized as follows. 
We first fit the efficiency as a function of $\cos \theta_H$ in each of the 100 \mevcc\ wide intervals of $m(h^+ h^-)$, using Legendre polynomials up to $L=12$:
\begin{equation}
\epsilon(\cos\theta_H) = \sum_{L=0}^{12} a_L(m) Y^0_L(\cos\theta_H),
\end{equation}
where $m$ denotes the $h^+ h^-$ invariant-mass.
For a given value of $m(h^+ h^-)$, the efficiency is interpolated linearly between adjacent mass intervals.

\begin{figure*}
\begin{center}
  \includegraphics[width=16cm]{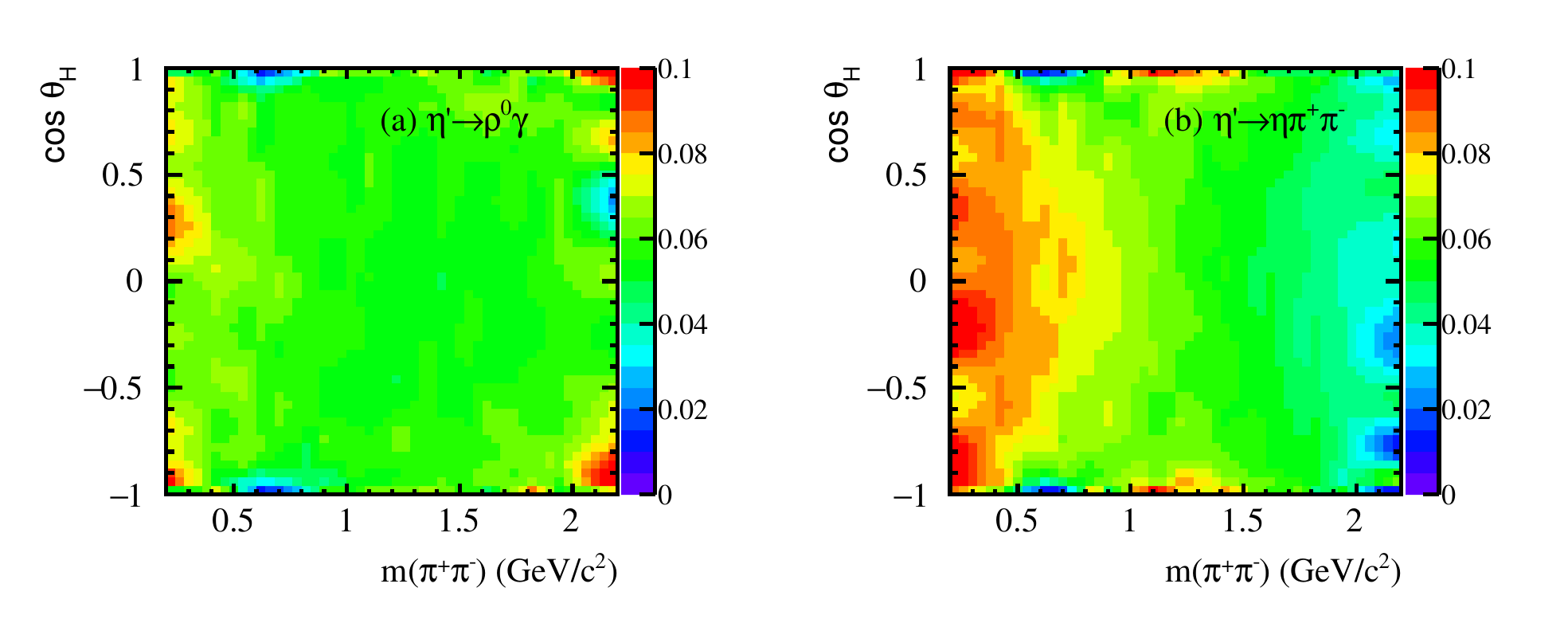}
  \includegraphics[width=16cm]{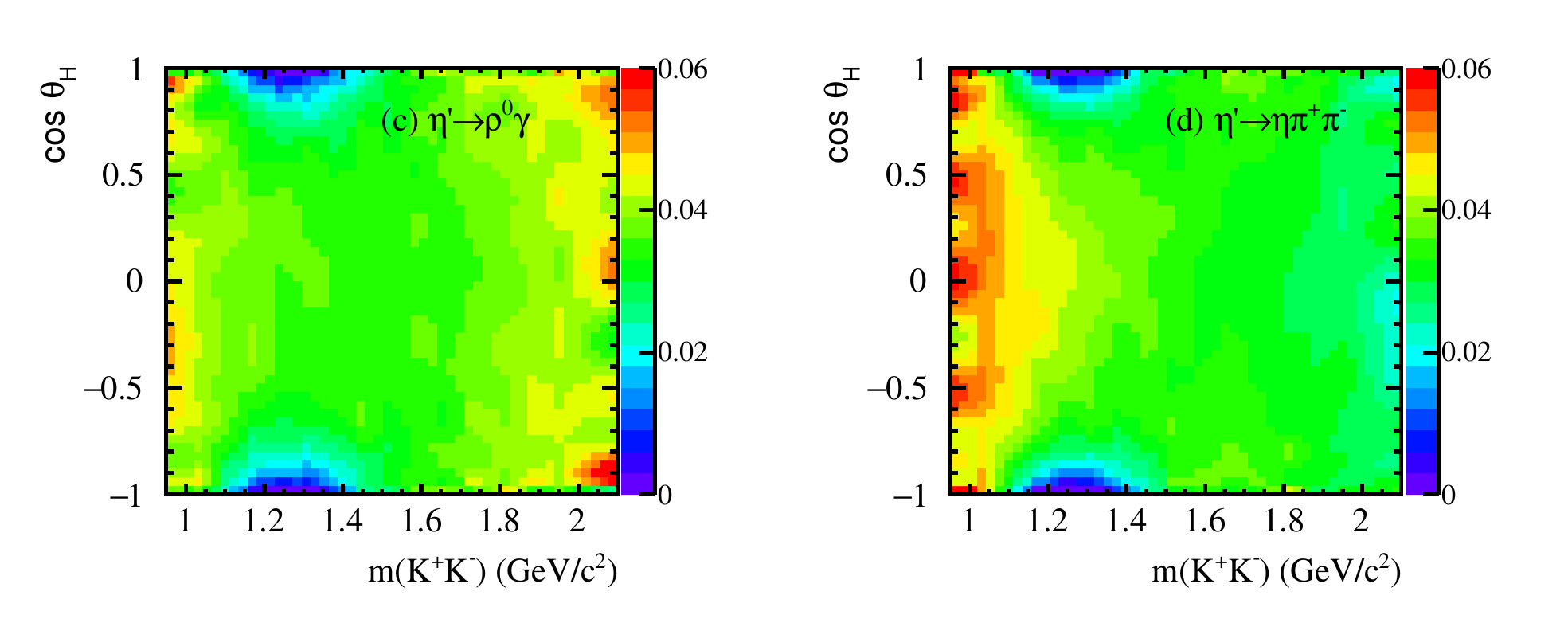}
\caption{Parametrized detection efficiencies in the $\cos \theta_H \  vs. \ m(h^+ h^-)$ plane for simulated 
(a) $\etac \to \etaprpipi$, $\etapr \to \rho^0 \gamma$, 
(b) $\etac \to \etaprpipi$, $\etapr \to \eta \pip \pim$, 
(c) $\etac \to \etaprkk$, $\etapr \to \rho^0 \gamma$, and 
(d) $\etac \to \etaprkk$, $\etapr \to \eta \pip \pim$ events.
The average value of the efficiency is shown in each interval.}
\label{fig:fig7}
\end{center}
\end{figure*}

\begin{figure}
\begin{center}
  \includegraphics[width=8.5cm]{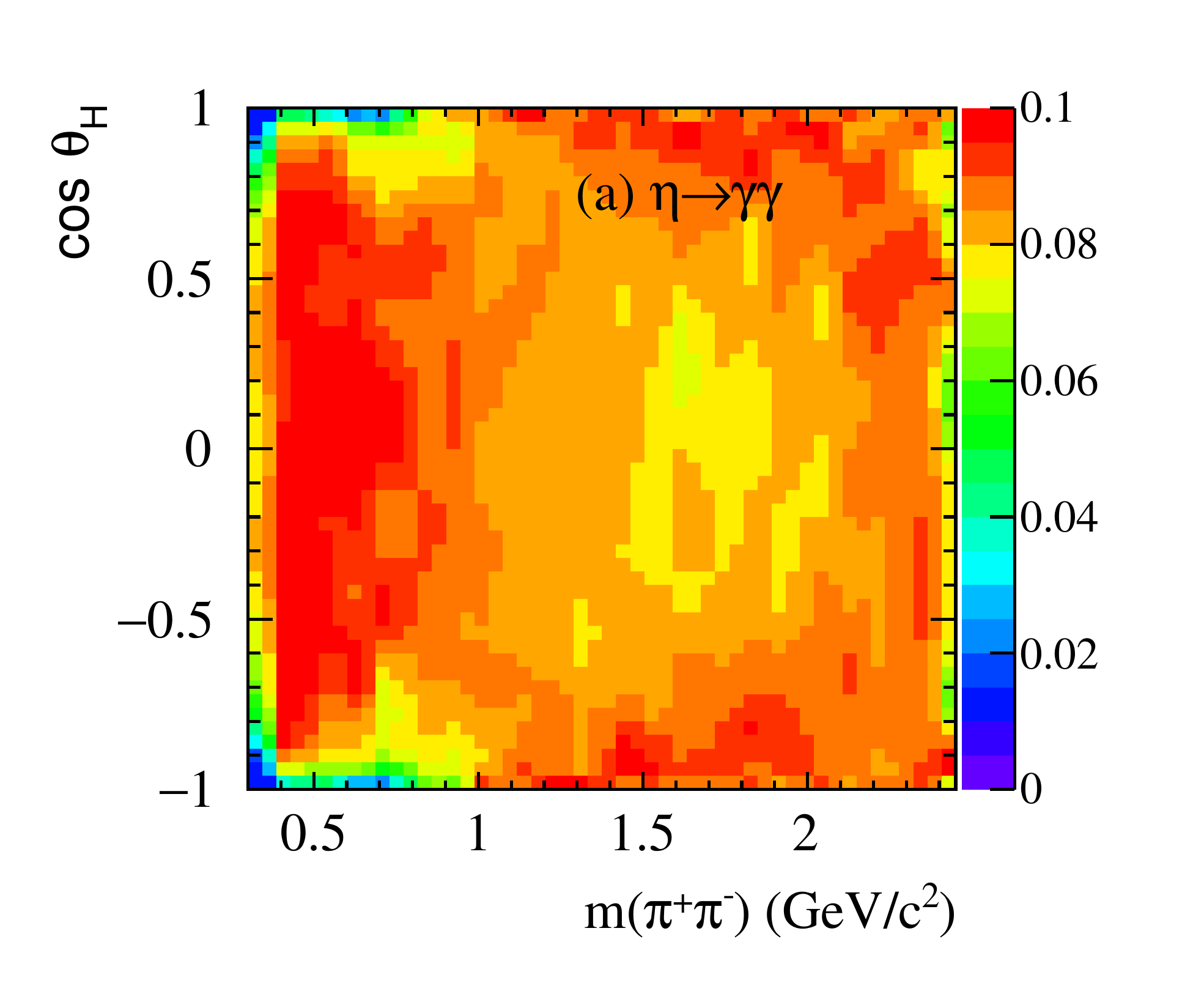}
  \includegraphics[width=8.5cm]{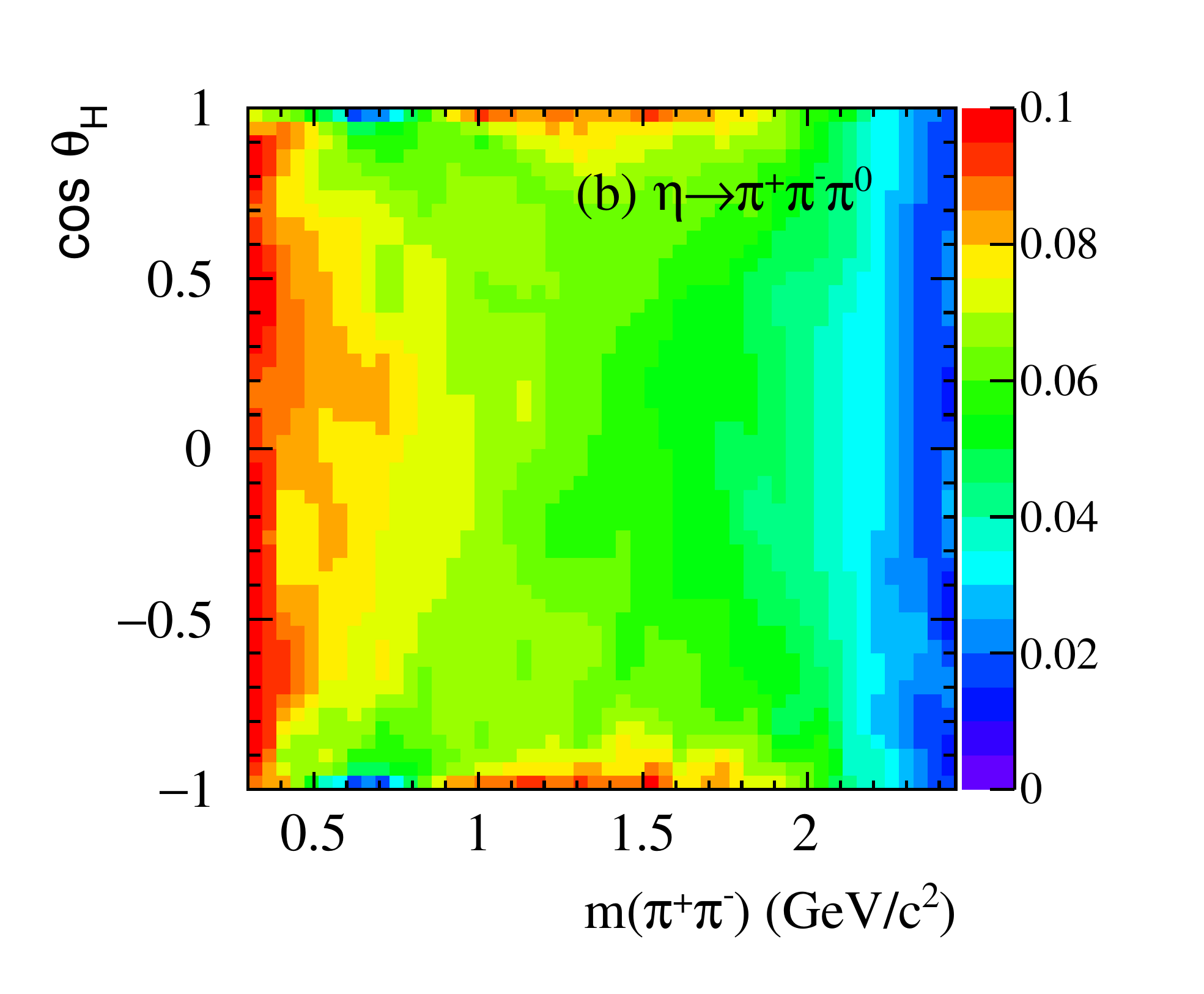}
\caption{Parametrized detection efficiencies in the $\cos \theta_H \  vs. \ m(\pip \pim)$ plane for simulated $\etac\to\eta\pip\pim$ events with (a) $\eta \to \gamma \gamma$ and (b) $\eta \to \pip \pim \piz$. 
The average value of the efficiency is shown in each interval.}
\label{fig:fig8}
\end{center}
\end{figure}

Figure~\ref{fig:fig7} shows the resulting efficiency maps $\epsilon(m,\cos \theta_H)$ for the four $\etapr h^+ h^-$ final states, and
Fig.~\ref{fig:fig8} shows the maps for the two $\eta \pip \pim$ final states.
The small regions of very low efficiency near $|\cos\theta_H| \sim 1$
are the result of the difficulty of reconstructing $\Kpm$ mesons with
laboratory momentum less than $\approx$ 200~\mevc, and \pipm\ mesons
with laboratory momentum less than $\approx$100~\mevc, due to energy loss in the beam pipe and inner-detector material.

The mass resolution is determined from the distribution of the difference ($\Delta m$) between the generated and reconstructed $\etapr h^+ h^-$  or $\eta \pip \pim$ invariant-mass values.
The $\Delta m$ distributions are parameterized by the sum of a Crystal Ball~\cite{cb} and a Gaussian function, which describe well
the distributions, and have root-mean-squared values of:
11.5~\mevcc\ for \etaprpipi, $\etapr \to \rho^0 \gamma$;
13.9~\mevcc\ for \etaprpipi, $\etapr \to \eta \pip \pim$;
8.2~\mevcc\ for \etaprkk, $\etapr \to \rho^0 \gamma$; 
12.2~\mevcc\ for \etaprkk, $\etapr \to \eta \pip \pim$; 
15.9 \mevcc\ for \etapipi, $\eta \to \gamma \gamma$; and
13.8 \mevcc\ for \etapipi, $\eta \to \pip \pim \piz$.

\section{Yields and branching fractions}
\label{sec:fits}

In this section, we fit the invariant-mass distributions to obtain the numbers of selected \etac\ events, $N_{\etapr K^+ K^-}$,  $N_{\etapr \pi^+ \pi^-}$, and $N_{\eta \pi^+ \pi^-}$, for each \etapr\ or $\eta$ decay mode. 
We then use the \etaprkk and \etaprpipi yields to compute the ratio of branching fractions for \etac to the \etaprkk and \etaprpipi final states.  This ratio is computed as
\begin{equation}
\begin{split}
\calR = &\frac{\BR(\etac \to \etapr \Kp \Km)}{\BR(\etac \to \etapr \pip \pim)} \\
  = &\frac{N_{\etapr K^+ K^-}}{N_{\etapr \pi^+ \pi^-}}\frac{\epsilon_{\etapr \pi^+ \pi^-}}{\epsilon_{\etapr K^+ K^-}}
\end{split}
\end{equation}
for each \etapr decay mode, 
where $\epsilon_{\etapr K^+ K^-}$ and $\epsilon_{\etapr \pi^+ \pi^-}$ are the corresponding weighted efficiencies described in the following Sec.~\ref{sec:bf}.

\subsection{Fits to the invariant-mass spectra} 

\begin{figure*}
  \begin{center}
  \includegraphics[width=8.5cm]{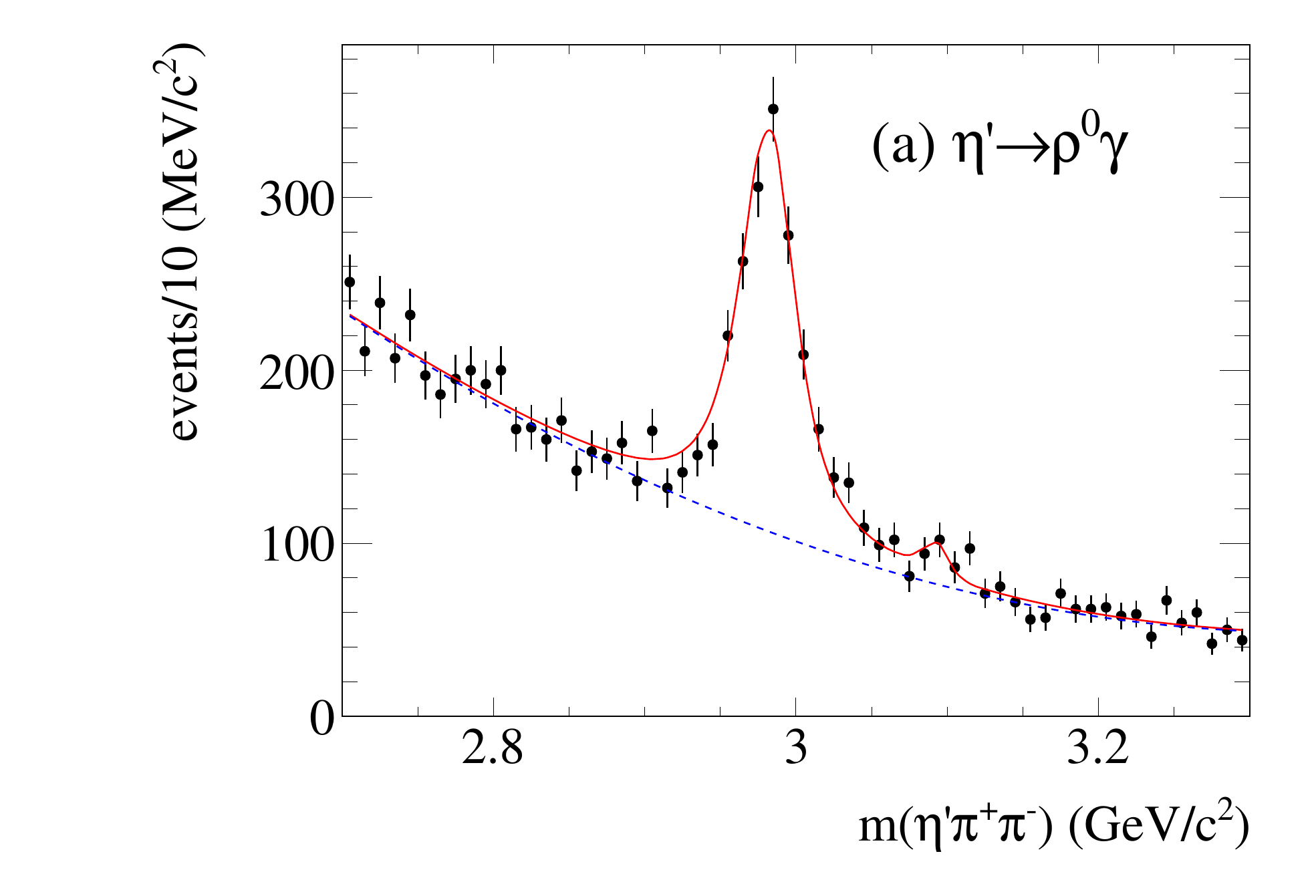}
  \includegraphics[width=8.5cm]{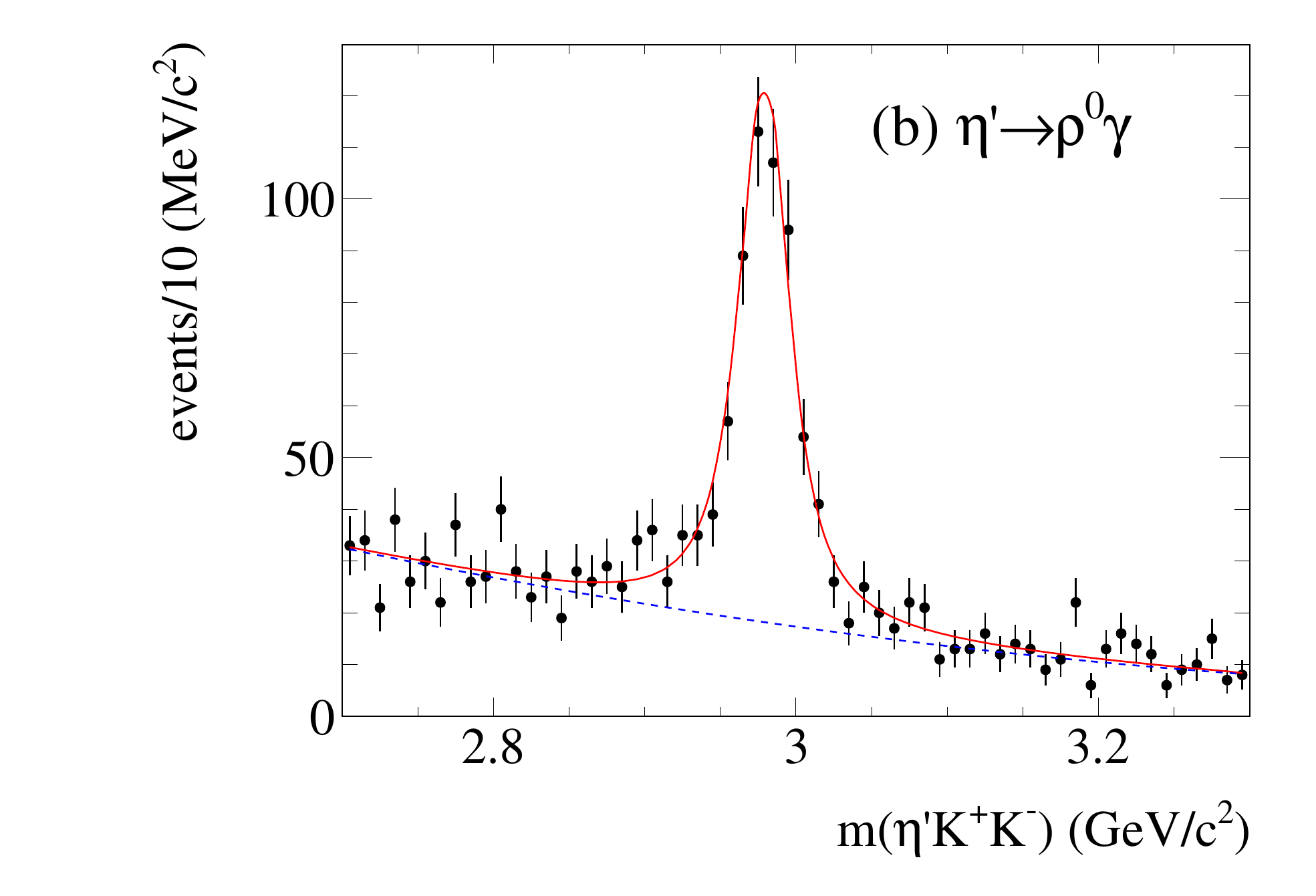}    
  \includegraphics[width=8.5cm]{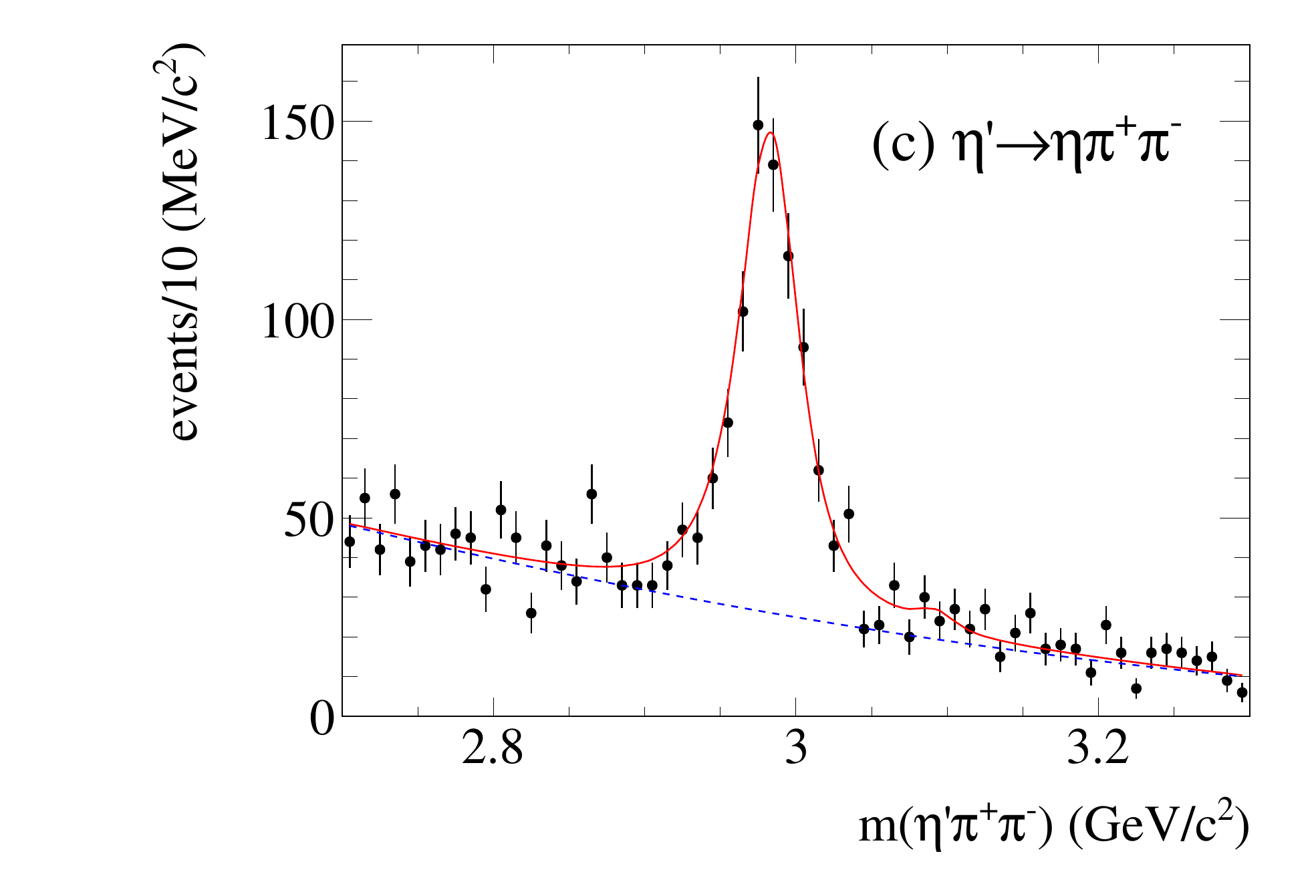}
  \includegraphics[width=8.5cm]{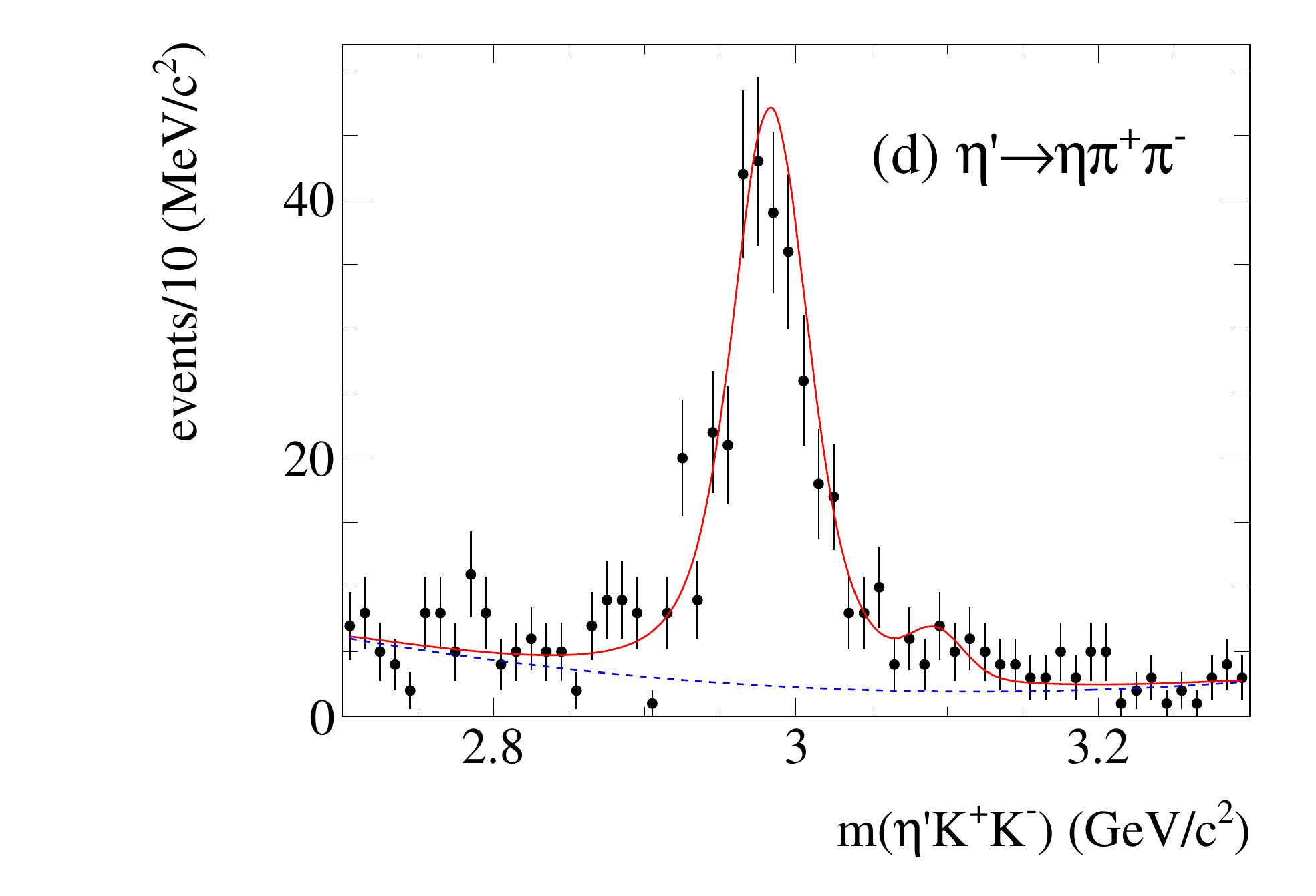}
\caption{Invariant-mass distributions of selected (left) \etaprpipi and (right) \etaprkk candidates for (top) $\etapr \to \rho^0 \gamma$ and (bottom) $\etapr \to \eta \pip \pim$.
The lines are the results from the fits described in the text.}
\label{fig:fig9}
\end{center}
\end{figure*}

We determine $N_{K^+K^-\etapr}$ and $N_{\pi^+ \pi^- \etapr}$ from \etac decays by performing binned $\chi^2$ fits to the \etaprkk and \etaprpipi invariant-mass spectra, in the 2.7-3.3 \gevcc\ mass region, separately for the two \etapr decay modes. In these fits, the \etac signal contribution is described by a simple Breit-Wigner (BW) function convolved with a fixed resolution function described above, with \etac parameters fixed to PDG values~\cite{PDG}.
An additional BW function is used to describe the residual background from ISR $J/\psi$ events, 
and the remaining background is parameterized by a $2^{nd}$ order polynomial.
The fitted $\etapr h^+ h^-$ invariant-mass spectra are shown in Fig.~\ref{fig:fig9}.
The fits generally describe the data well, although the fit to the $\etapr \Kp \Km$ invariant-mass spectrum for $\etapr \to \eta \pip \pim$ (Fig.~\ref{fig:fig9}(d)), which has low statistics, appears to the eye to have a somewhat distorted lineshape. 
For this fit, we add two additional parameters by leaving free the parameters of the Gaussian component of the resolution function.
To minimize the dependence of the $N$'s on the fit quality, the \etac signal yields are obtained by integrating the data over the \etac signal region after subtracting the fitted backgrounds.

Statistical errors on the \etac yields are evaluated by generating,
from each invariant-mass spectrum, 500 new spectra by random Poisson fluctuations of the content of each bin. 
The generated mass spectra are fitted using the same model as for the
original one and the resulting distributions of the \etac subtracted
yields are fitted using a Gaussian function, whose $\sigma$ is taken
as the statistical uncertainty. 
The resulting yields and $\chi^2$ per degree of freedom for the fits, $\chi^2/{\rm ndf}$ are reported in Table~\ref{tab:tab1}.

     
      We test the fitting procedure by leaving free the \etac parameters and find agreement, within the errors, with world averages.
For the decay $\etac \to \eta \pip \pim$, however, the fits without
interference do not describe the data well for either $\eta$ decay mode.
Leaving free the \etac parameters, the fits return masses shifted down by $\approx10$ \mevcc\ with respect to PDG averages.
We test the possibility of interference effects of the \etac with each non-resonant two-photon process~\cite{Zhang:2012tj},
modifying the fitting function by defining
\begin{equation}
      f(m) = |A_{\rm nres}|^2+|A_{\etac}|^2+c\cdot 2Re(A_{\rm nres} A^*_{\etac}),
      \label{eq:int}
\end{equation}
where $A_{\rm nres}$ is the non-resonant amplitude with $|A_{\rm nres}|^2$ described by a $2^{nd}$ order polynomial;
the coherence factor $c$ is the fraction of the non-resonant events that are true two-photon production of the same final state;
the resonant contribution is $A_{\etac}=\alpha\cdot BW(m) \cdot \exp(i\phi)$, where $BW(m)$ is a simple Breit-Wigner with parameters fixed to PDG values; and $\alpha$, $\phi$, and $c$ are free parameters.
The sum of $f(m)$ and the \jpsi\ contribution is convolved with the experimental resolution.

\begin{figure}
  \begin{center}
    \includegraphics[width=8.5cm]{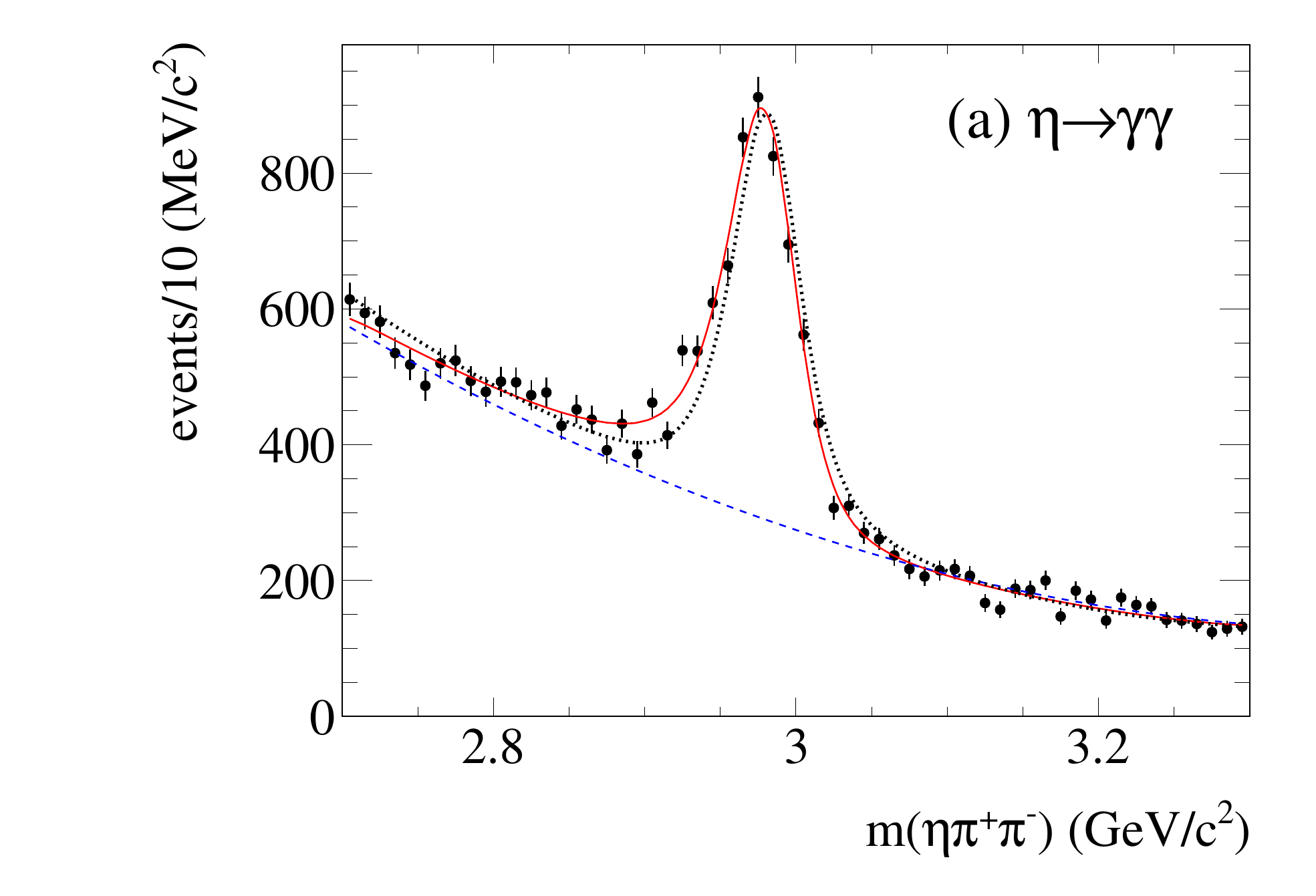}
    \includegraphics[width=8.5cm]{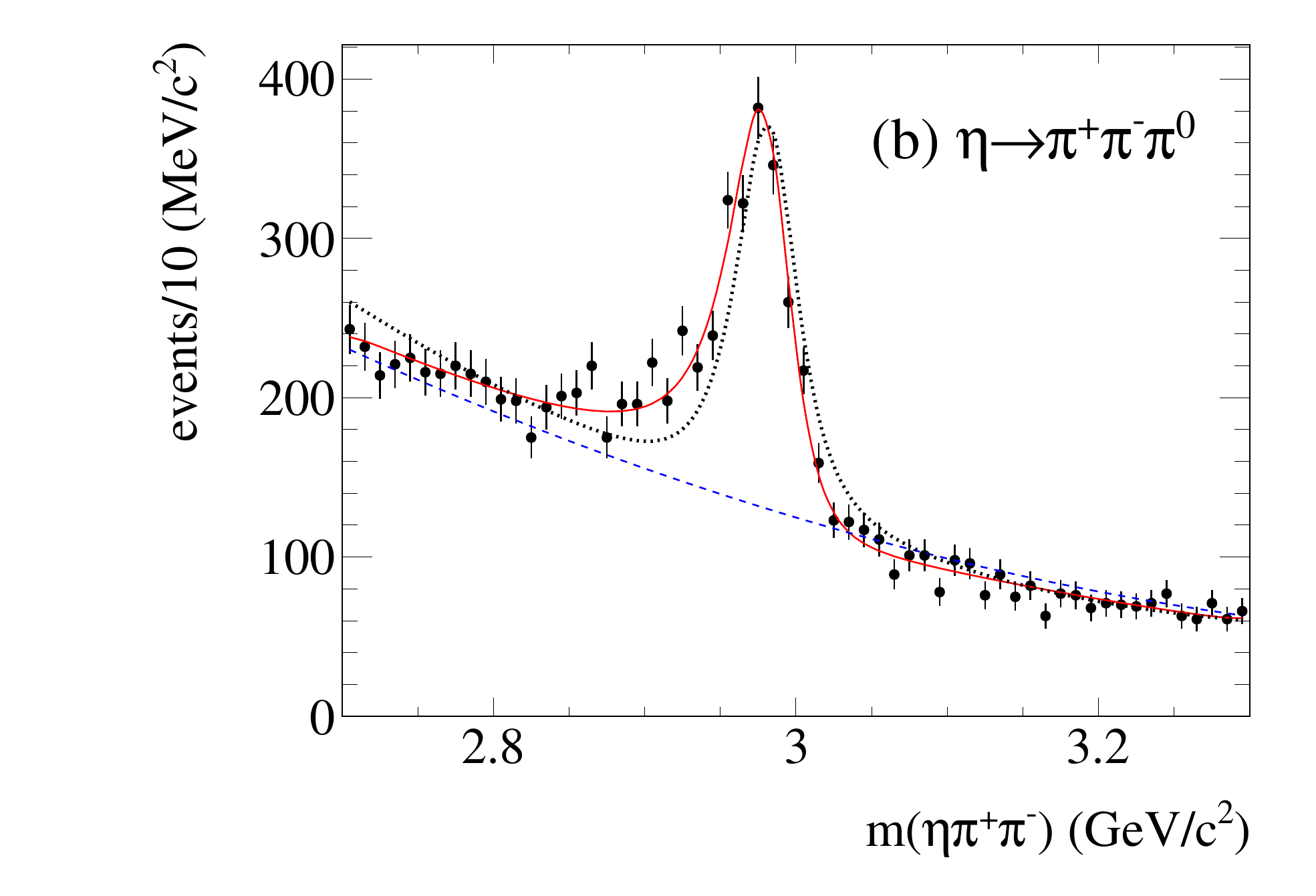}
    \caption{ Invariant-mass spectra for selected $\eta \pip \pim$ candidate events with (a) $\eta \to \gamma \gamma$ and (b) $\eta \to \pip \pim \piz$.
    The solid (red) lines represent the fits including interference described in the text. The dashed (blue) line represents the fitted non-resonant components. The dotted lines represent the fits without interference.}
\label{fig:fig10}
\end{center}
\end{figure}

Fits with interference and fixed PDG parameters give values of $\chi^2/{\rm ndf}=77/54$ ($p$-value=2.2\%) and $\chi^2/{\rm ndf}=46/54$ ($p$-value=77\%) for $\eta\to \gamma \gamma$ and $\eta \to \pip \pim \piz$ decay modes, respectively. 
The fitted relative phases are $\phi = 1.41 \pm 0.02_{\rm stat} \pm 0.02_{\rm sys}$ rad and $\phi = 1.26 \pm 0.03_{\rm stat} \pm 0.02_{\rm sys}$ rad. Systematic uncertainties are related to the use of \etac fixed parameters and on errors in the background shape. The fits, on the other hand, show little sensitivity to the $c$ parameter.
The fitted invariant-mass spectra are shown in Fig.~\ref{fig:fig10}, where reasonable descriptions of the data are evident.
As a comparison we also fit the two mass spectra with no interference and fixed \etac parameters and
obtain the dotted lines distributions shown in Fig.~\ref{fig:fig10} with corresponding $\chi^2/{\rm ndf}=160/55$ and $\chi^2/{\rm ndf}=139/55$, respectively.

We find that the interference model does not produce significant
improvements in the description of the data for final states that include an \etapr . 
As a cross check, we reanalyze the data reported in Ref.~\cite{Lees:2014iua}, and find no evidence for such interference effects also for the $\etac \to  \eta \Kp \Km$ decay mode.

Systematic uncertainties on the yields due to the fitting procedure are estimated by varying the \etac parameters according to the PDG uncertainties. 
An additional uncertainty of 4\% is assigned to the yield for $\etac \to \etapr \Kp \Km$ with $\etapr \to \eta \pip \pim$ due to the variation of the resolution function.
We also take the integral of each full function used to describe the
\etac as an estimate of the yield, and take the difference as the systematic uncertainty.
The quadratic sums of these uncertainties are given in Table~\ref{tab:tab1}.

\subsection{Branching fractions}
\label{sec:bf}

We estimate $\epsilon_{\etapr K^+ K^-}$ and $\epsilon_{\etapr \pi^+ \pi^-}$ for the \etac signals using the \mbox{2-D} raw efficiency functions described in Sec.~\ref{sec:effy}.
Each event is first weighted by $1/\epsilon(m,\cos \theta_H)$. 
Since the backgrounds below the \etac signals have different distributions in the Dalitz plot, 
we perform a sideband subtraction by assigning an additional weight of $+1$ to events in the \etac signal region, defined as the (2.93-3.03) \gevcc\ mass region, and a
weight $-1$ to events in the sideband regions, (2.77-2.87)~\gevcc\ and (3.09-3.19)~\gevcc. 
The weights in the sideband regions are scaled by a small amount to
match the fitted \etac signal/background ratio, and added to those in
the signal region, to produce the weighted yields shown in Table~\ref{tab:tab1}.

Systematic uncertainties on the efficiencies have been evaluated as follows.
The uncertainty due to the limited MC statistics is computed by generating 500 new efficiency tables, obtained from the original tables by random variation, according to a Poisson distribution, of the generated and reconstructed MC yields in each cell.
The distributions of the resulting weights are fitted using a Gaussian function whose $\sigma$ values are taken as systematic uncertainties and are listed in Table~\ref{tab:tab1}.
To estimate an uncertainty on the method of sideband subtraction, we use the average weights in the signal region, and take the difference as an uncertainty.
The quadratic sums of these uncertainties are given in Table~\ref{tab:tab1}.

      \begin{table*} [htb]
        \caption{\small Information for the evaluation of the branching fractions. The reported yields are obtained from the integration of the \etac signal after background subtraction in the \etac signal region.
          The first error is statistical, the second systematic.
        }
        \centering
        {\small
  \begin{tabular}{lcccc}
    \hline \\ [-2.3ex]
    Final state & yield & weight & weighted yields & $\chi^2/{\rm ndf}$ \cr
\hline \\ [-2.3ex]
$\eta_c \to \etapr \pip \pim$ ($\etapr \to \rho^0 \gamma$) & $1160 \pm 57 \pm 47$ & $17.37 \pm 0.28$ & $20149 \pm 990 \pm 878$ & 51/55\cr
\hline \\ [-2.3ex]
$\eta_c \to \etapr \Kp \Km$ ($\etapr \to \rho^0 \gamma$) & $473 \pm 29 \pm 3$ & $26.79 \pm 0.35$ & $12672 \pm 777 \pm 184$ & 58/55\cr
\hline \\ [-2.3ex]
$\eta_c \to \etapr \pip \pim$ ($\etapr \to \pip \pim \eta$) & \al $619 \pm 35 \pm 11$ & $18.42 \pm 0.18$ & $11401 \pm 645 \pm 231$ & 72/55 \cr
\hline \\ [-2.3ex]
$\eta_c \to \etapr \Kp \Km$ ($\etapr \to \pip \pim \eta$) & \al $249 \pm 20 \pm 11$ & $30.77 \pm 0.40$ & \al $7662 \pm 615 \pm 353 $ & 90/53 \cr
\hline \\ [-2.3ex]
  \end{tabular}
  }
  \label{tab:tab1}  
      \end{table*}
      
We label with $\calR_1(\rho^0 \gamma)$ and $\calR_2(\eta \pip \pim)$  the measurements of the branching fraction for the two \etapr decay modes.
In each case, the numerator and denominator involve the same number of charged tracks and $\gamma$'s, so the systematic uncertainties on their reconstruction efficiencies cancel in the ratio. 
The only difference is the presence of two kaons in the numerator and two pions in the denominator.
The uncertainties in the particle identification efficiencies are correlated; we assign a systematic uncertainty of 1\% to the identification of each kaon and 0.5\% to each pion.
Table~\ref{tab:tab2} summarizes the largest systematic uncertainties on the branching fraction, which arise from MC statistics, the use of the full fitting function in extracting the yield (labelled full-BW), the sideband subtraction in the efficiencies (labelled no-sideband), and the kaon/pion identification (labelled PID).

\begin{table}[htb]
    \caption{\small Summary of the systematic uncertainties on the branching fraction.
        }
    \centering
  \begin{tabular}{lccccc}
    \hline \\ [-2.3ex]
$\calR$ & MC stat. & full-BW & no-sideband & PID & Total \cr
\hline \\ [-2.3ex]
$\calR_1(\rho^0 \gamma)$ & 0.029 & 0.014 & 0.003 & 0.014 & 0.035 \cr
\hline \\ [-2.3ex]
$\calR_2(\eta \pip \pim)$ & 0.034 & 0.066 & 0.019 & 0.015 & 0.078 \cr
\hline \\ [-2.3ex]
 \end{tabular}
  \label{tab:tab2}  
       \end{table}

Adding the systematic uncertainties in quadrature, we obtain the following values of the branching ratios:
      \begin{equation}
        \calR_1(\rho^0 \gamma) = 0.629 \pm 0.049_{\rm stat} \pm 0.035_{\rm sys},
      \end{equation}
      \begin{equation}
         \calR_2(\eta \pip \pim) = 0.672 \pm 0.066_{\rm stat} \pm 0.078_{\rm sys},
      \end{equation}
      and an average value of
      \begin{equation}
        \frac{\calB(\etac \to \etapr \Kp \Km)}{\calB(\etac \to \etapr \pip \pim)} = 0.644 \pm 0.039_{\rm stat} \pm 0.032_{\rm sys}.
        \label{eq:br_etapr}
      \end{equation}
      
\section{Dalitz plot analyses}
\label{sec:daly}

We perform Dalitz plot analyses of the \etaprpipi, \etaprkk, and \etapipi systems in the \etac mass region using unbinned maximum likelihood fits.
The likelihood function is written as
\begin{eqnarray}
\mathcal{L} = \nonumber\\
 \prod_{n=1}^N&\bigg[&f_{\rm sig} \cdot \epsilon(x'_n,y'_n)\frac{\sum_{i,j} c_i c_j^* A_i(x_n,y_n) A_j^*(x_n,y_n)}{\sum_{i,j} c_i c_j^* I_{A_i A_j^*}} \nonumber\\
& &+(1-f_{\rm sig})\frac{\sum_{i} k_iB_i(x_n,y_n)}{\sum_{i} k_iI_{B_i}}\bigg],
\end{eqnarray}
\noindent where:
\begin{itemize}
\item $N$ is the number of events in the signal region;
\item $f_{\rm sig}$ is the fraction of those events attributed to \etac\ decays;  
\item for the $n$-th event, $x_n=m^2(\eta/\etapr h^+)$, $y_n=m^2(\eta/\etapr h^-)$, and
\item $\epsilon(x'_n,y'_n)$ is the efficiency, parameterized as a function of $x'_n=m(h^+ h^-)$ and $y'_n=\cos \theta_H$ (see Sec.~\ref{sec:effy});
\item $c_i$ is the complex amplitude of the $i-$th signal component; the $c_i$ are free parameters of the fit;
\item for the $n$-th event, $A_i(x_n,y_n)$ describe the $i-th$ complex signal-amplitude contribution;
\item $k_i$ is the magnitude of the $i-$th background component; the $k_i$ parameters are obtained by fitting the sideband regions;  
\item for the $n$-th event, $B_i(x_n,y_n)$ is the probability-density function of the $i$-th background contribution; we assume that interference between signal and background amplitudes can be ignored;
\item $I_{A_i A_j^*}=\int A_i (x,y)A_j^*(x,y) \epsilon(m(h^+ h^-),\cos \theta_H)\ {\rm d}x{\rm d}y$ and 
$I_{B_i}~=~\int B_i(x,y) {\rm d}x{\rm d}y$ are normalization
 integrals; numerical integration is performed on phase-space generated events.
\end{itemize}
Amplitudes are parameterized as described in Refs.~\cite{Asner:2003gh} and~\cite{ds}.
They include a relativistic Breit-Wigner function having a variable width modulated by the Blatt-Weisskopf~\cite{blatt} spin form factors
and the relevant spin-angular information. Note that these factors are both one for scalar resonances.

The efficiency-corrected fractional contribution $f_i$ due to resonant or non-resonant contribution (NR) is defined as follows:
\begin{equation}
f_i = \frac {|c_i|^2 \int |A_i(x_n,y_n)|^2 {\rm d}x {\rm d}y}
{\int |\sum_j c_j A_j(x,y)|^2 {\rm d}x {\rm d}y}.
\end{equation}
The $f_i$ do not necessarily sum to 100\% because of interference effects. The uncertainty for each $f_i$ is evaluated by propagating the full covariance matrix obtained from the fit.

The search for the amplitudes contributing to the signal or background is performed by starting with the largest resonance observed in the mass projections, which is taken as the reference amplitude with $c_1=1$ and phase zero. We then add, one by one, possible processes that could contribute to the decay, testing for an increase in the likelihood value.
Amplitudes are discarded if no
      significant improvement in the likelihood ($\Delta(-2\log\calL)>2$) is obtained. Each excluded resonance is reiterated many times
      in combination with other possible resonant contributions.
      Where possible, resonance parameters are left free, for comparison with existing values;  otherwise, they are fixed to PDG values.

Table~\ref{tab:tab3} summarizes the information on the structure of the samples used in the Dalitz analyses.
Yields and purities are computed in the $\eta_c$ signal region, defined as the mass ranges (2.93-3.03) \gevcc\ for $\etapr h^+h^-$ and
(2.92-3.02) \gevcc\ for $\eta \pip \pim$.

      \begin{table} [htb]
        \caption{\small Information for the Dalitz analysis.
        }
        \centering
  \begin{tabular}{llrcc}
\hline \\ [-2.3ex]
Final state & Decay mode & Yield & Fraction & Purity  (\%)\cr
\hline \\ [-2.3ex]

$\eta_c \to \etapr \Kp \Km$ &$\etapr \to \rho^0 \gamma$ & 656 & 0.705  & $69.7 \pm 1.7$\cr
\hline \\ [-2.3ex]
$\eta_c \to \etapr \Kp \Km$ &$\etapr \to \pip \pim \eta$& 274 & 0.295  & $85.7 \pm 2.0$\cr
\hline \\ [-2.3ex]
$\eta_c \to \etapr \pip \pim$ &$\etapr \to \rho^0 \gamma$& 2239 & 0.717 & $51.8 \pm 1.1$ \cr
\hline \\ [-2.3ex]
$\eta_c \to \etapr \pip \pim$ &$\etapr \to \pip \pim \eta$ & 883 & 0.283 & $69.0 \pm 1.6$ \cr
 \hline \\ [-2.3ex]
$\eta_c \to \eta \pip \pim$ &$\eta \to \gamma \gamma$ & 6512 & 0.700 & $58.0 \pm 0.6$ \cr
\hline \\ [-2.3ex]
$\eta_c \to \eta \pip \pim$ &$\eta \to \pip \pim \piz$ & 2791 & 0.300 & $52.7 \pm 1.0$ \cr
\hline \\ [-2.3ex]
 \end{tabular}
\label{tab:tab3}
      \end{table}
      The widths of the resonances contributing to the \etac decays are much larger than the experimental resolution, and therefore resolution effects
      are ignored. The only exception is the $\phi(1020)$ resonance, which contributes to the background to $\etac \to \etapr \Kp \Km$.
      We obtain an enhanced $\phi(1020)$ signal by relaxing the selection criteria and in particular the \pt selection. The resulting
      $\Kp \Km$ mass distribution shows a prominent $\phi(1020)$ signal, which is fitted with a $P$-wave relativistic BW function yielding a width $6.1 \pm 0.3$ \mevcc. The fitted BW function is used to describe this contribution to the background.

      Each Dalitz plot analysis deals with two sets of data contributing to the given \etac final state, with different efficiencies and purities: $\etapr \to \rho^0 \gamma$ and  $\etapr \to \eta \pip \pim$ for
      $\etac \to \etapr h^+h^-$, $\eta \to \gamma \gamma$ and  $\eta \to \pip \pim \piz$ for
      $\etac \to \eta \pip \pim$.
 Therefore we use the sum of two different likelihood functions, which share the free parameters and fitting model. 
 Due to the lack of statistics we do not separate the contributing backgrounds for the two sets of data. 

      \section{Dalitz plot analysis of $\etac \to \etapr \Kp \Km$}
      \label{sec:daly1}
      
\begin{figure}
  \begin{center}
  \includegraphics[width=8.5cm]{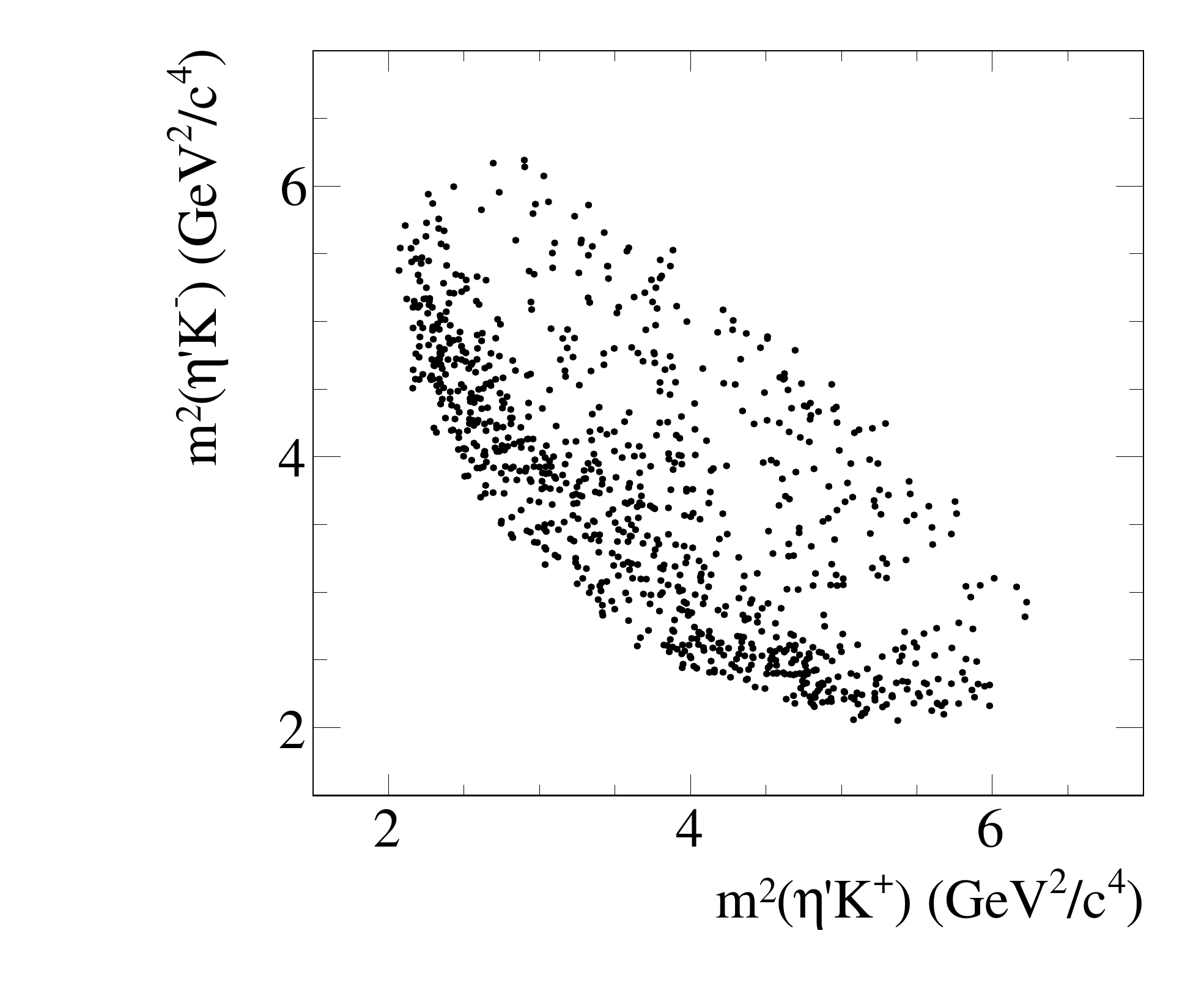}
\caption{Dalitz plot for selected $\etac \to \etapr \Kp \Km$ candidates in the \etac\ signal region, summed over the two \etapr decay modes.}
\label{fig:fig11}
\end{center}
\end{figure}

\begin{figure*}
  \begin{center}
    \includegraphics[width=16cm]{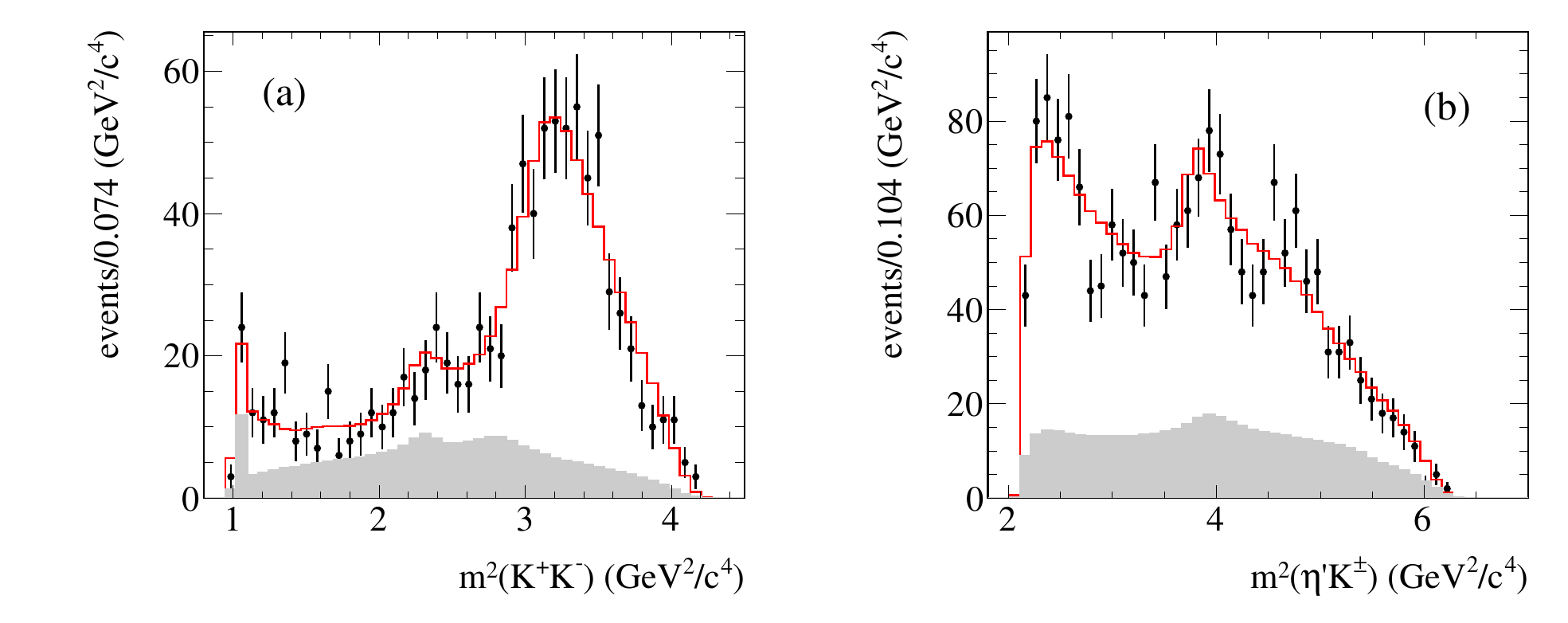}
    \includegraphics[width=16cm]{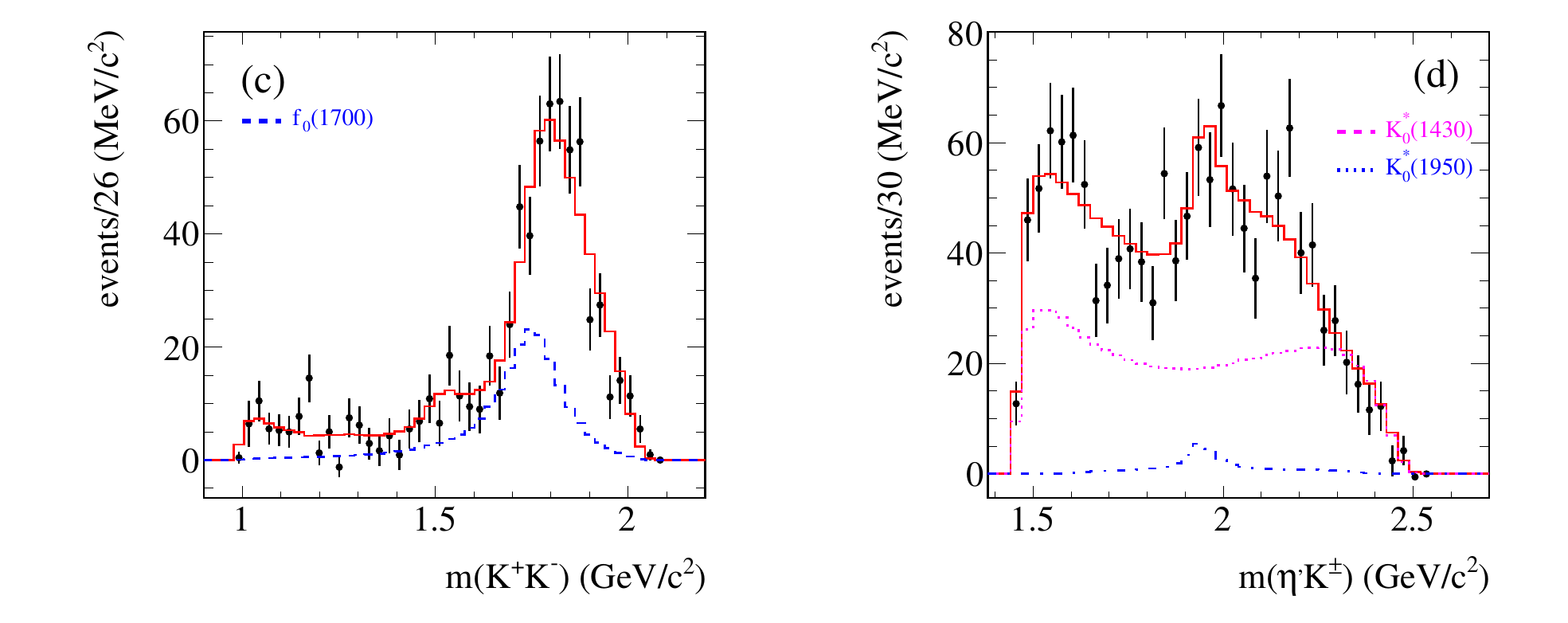}
    \caption{Squared-mass projections (a) $m^2(\Kp \Km)$ and (b) $m^2(\etapr K^{\pm})$ of the measured $\etac \to \etapr \Kp \Km$ Dalitz plot.
The shaded (gray) histograms are the background interpolated from fits to the two \etac\ sidebands.
Linear-scale mass projections (c) $m(\Kp \Km)$ and (d) $m(\etapr K^{\pm})$, after subtraction of the background.
The solid (red) histograms represent the results of the fit described in the text (solution (A)).
The other histograms display the contributions from each of the listed components.      
  The $\etapr K^{\pm}$ mass projections have two entries per event.}
\label{fig:fig12}
\end{center}
\end{figure*}

      Figure~\ref{fig:fig11} shows the Dalitz plot for the selected $\etac \to \etapr \Kp \Km$ candidates in the data, for the two \etapr decay modes combined.
      Figure~\ref{fig:fig12}(a)-(b) shows the two squared mass projections. 
      
We observe that this \etac decay mode is dominated by a diagonal band
on the low mass side of the Dalitz plot. The $m^2(\Kp \Km)$ spectrum shows a large structure in the region of the $f_0(1710)$ resonance. The combined
$m^2(\etapr K^{\pm})$ invariant-mass spectrum shows a structure at threshold due to the $K^*_0(1430)$ accompanied by weaker resonant structures.

 We first fit the two \etac\ sidebands separately, using an incoherent sum of amplitudes, which includes contributions from the $\phi(1020)$, $\phi(1680)$, $f_2'(1525)$, $K^*_0(1430)$, and $K^*_0(1950)$ resonances.  
To model the background composition in the \etac\ signal region, we take a weighted average of the two fitted fractional contributions, and normalize using the results from the fit to the \etaprkk invariant-mass spectrum. 
The estimated background contributions are indicated by the shaded regions in Figs.~\ref{fig:fig12}(a)-(b), and
we show the corresponding background-subtracted invariant-mass spectra in Figs.~\ref{fig:fig12}(c)-(d).

\begin{figure*}
  \begin{center}
    \includegraphics[width=16cm]{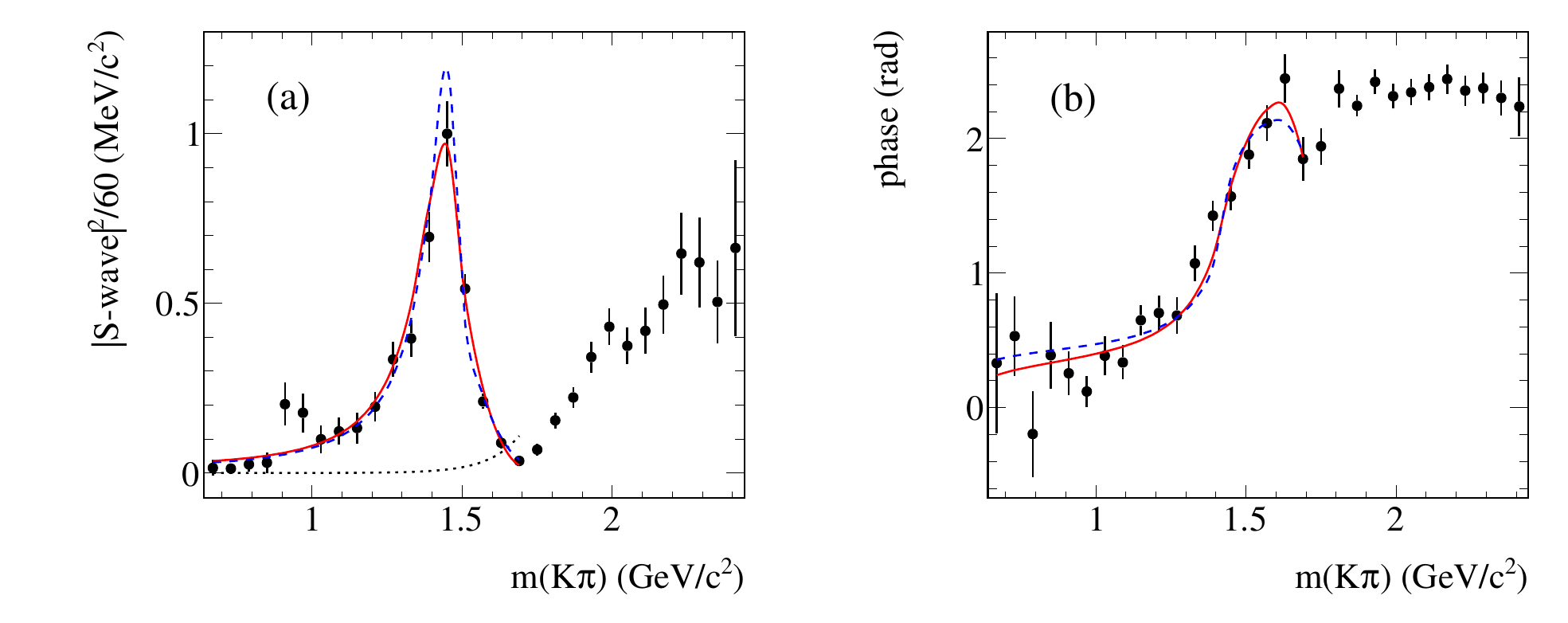}
  \caption{The (a) squared modulus and (b) phase of the $K \pi$ $S$-wave averaged over the $\eta_c \to \KS K \pi$ and
    $\eta_c \to \Kp \Km \piz$ from the \babar~\cite{Lees:2015zzr} QMI analysis. Statistical uncertainties only are shown.
    The full (red) lines represent the result from the fit with free $g^2_{K\etapr}$ and $g^2_{K\pi}$ parameters. The dashed (blue) lines
    represent the result from the fit with a fixed $g^2_{K \etapr}/g^2_{K \pi}$ ratio.
    The dotted (black) line in (a) represents the empirical background contribution.
}
\label{fig:fig13}
\end{center}
\end{figure*}

The $K^*_0(1430)$ is a relatively broad resonance decaying to $K \pi$,
$K \eta$, and $K \etapr$. The measured $K \eta$ relative branching fraction is $\frac{\BR(K^*_0(1430) \to K \eta)}{\BR(K^*_0(1430) \to K \pi)} = 0.092 \pm 0.025^{+0.010}_{-0.025}$~\cite{Lees:2014iua}, while the $K \etapr$ has only been observed in Ref.~\cite{Ablikim:2014tww}. To describe the $K^*_0(1430)$ lineshape in the $K \etapr$ projection, we model it using a simplified coupled-channel Breit-Wigner function, which ignores the small $K \eta$ contribution. We parameterize the $K^*_0(1430)$ signal as

\begin{equation}
        BW(m) = \frac{1}{m_0^2 - m^2 - i(\rho_1(m)g^2_{K \pi} + \rho_2(m)g^2_{K \etapr})},
        \label{eq:ch}
\end{equation}
\noindent
where $m_0$ is the resonance mass, $g_{K \pi}$ and $g_{K \etapr}$ are the couplings to the $K \pi$ and $K \etapr$ final states, and \mbox{$\rho_j(m)=2P/m$} are the respective Lorentz-invariant
phase-space factors, with $P$ the decay particle momentum in the $K^*_0(1430)$ rest frame. 
The $\rho_2(m)$ function becomes imaginary below the $K \etapr$ threshold.
The values of $m_0$ and the $g_{Kj}$ couplings cannot be derived from the $K\etapr$ system only,
and therefore we make use of the $K \pi$ $S$-wave measurement from \babar~\cite{Lees:2015zzr}. 
We average the reported quasi model-independent (QMI) measurements of the $K\pi$ $S$-wave from $\eta_c \to \KS K \pi$ and $\eta_c \to \Kp \Km \piz$ decays, and obtain the modulus squared of the amplitude and the phase shown in Fig.~\ref{fig:fig13}.

We perform a simultaneous binned $\chi^2$ fit to the $K \pi$ $S$-wave amplitude and phase from threshold up to 1.72~\gevcc. 
Above this mass, other resonant contributions are present, which make the amplitude and phase more complicated.
We model the $K \pi$ $S$-wave in this region as:
\begin{equation}
  S{\text{-}\rm wave}(m) = B(m) + c \cdot BW_{K \pi}(m) e^{i \phi},
\end{equation}
 where $BW_{K \pi}(m)$ is given by Eq.(~\ref{eq:ch}), $B(m)$ is an empirical background term, parameterized as
      \begin{equation}
        B(m) = \rho_1(m)e^{-\alpha m},
      \end{equation}
      and $c$, $\phi$, and $\alpha$ are free parameters.
      The results of the fit are shown in Fig.~\ref{fig:fig13} as the solid (red) lines.
      We obtain a $\chi^2/{\rm ndf}=55/31$ ($\chi^2/{\rm ndf}=25/31$ with included systematic uncertainties) and the $K^*_0(1430)$ parameters listed in Table~\ref{tab:tab4}.      
      \begin{table*}[h] 
        \caption{\small Resonance parameters from the Dalitz plot analyses of $\eta_c \to \etapr \Kp \Km$, $\eta_c \to \etapr \pip \pim$, and $\eta_c \to \eta \pip \pim$. In the case of the $K^*_0(1430)$, the first two rows report results from fits to the $K \pi$ $S$-wave with free $K^*_0(1430)$ parameters
          and fixed $\frac{g^2_{\etapr K}}{g^2_{\pi K}}$ ratio, respectively. 
          When two errors are listed the first is statistical, the second systematic.
          The calculated significances do not include systematic uncertainties.
          }
        \centering
  \begin{tabular}{lcccc}
    \hline \\ [-2.3ex]
Resonance & Mass (\mevcc) & $g_{K \pi}^2$ (\gevccc) &  $g_{K \etapr}^2$ (\gevccc) & \cr
\hline \\ [-2.3ex]
& &  $\eta_c \to \etapr \Kp \Km$   & &\cr
\hline \\ [-2.3ex]
\hline \\ [-2.3ex]
$K^*_0(1430)$ & & & \cr
\hline \\ [-2.3ex]
$\etac \to K \bar K \pi$ & \almm\almm\alm\aln$1447 \pm 8$ &  \alm\alm$0.414 \pm 0.026$ & $0.197 \pm 0.105$ &\cr
fixed $\frac{g^2_{\etapr K}}{g^2_{\pi K}}$ & \almm\almm\aln$1453 \pm 22$ &  \alm\alm$0.462 \pm 0.036$ &   &\cr
\\ [-2.3ex]
\hline \\ [-2.3ex]
Resonance & Mass (\mevcc) & $\Gamma$ (\mev) &   & significance (n$\sigma)$\cr
\hline \\ [-2.3ex]
$f_0(1710)$ & $1757 \pm 24 \pm 9$ & $175 \pm 23 \pm 4$ & & 11.4 \cr
$K^*_0(1950)$ &  $1942 \pm 22 \pm 5$ & \al $80 \pm 32 \pm 20$ & & 3.3\cr
\hline \\ [-2.3ex]
\hline \\ [-2.3ex]
& &  $\eta_c \to \etapr \pip \pim$  & & \cr
\hline \\ [-2.3ex]
\hline \\ [-2.3ex]
$f_0(500)$ &  \alm\alm$953 \pm 90$ &  \almm\alm\alm\aln$335 \pm 81$ & &\cr
$f_2(1430)$ & $1440 \pm 11 \pm 3$ &  $46 \pm 15 \pm 5$ & & 4.4 \cr
$f_0(2100)$ & \al$2116 \pm 27 \pm 17$ & $289 \pm 34 \pm 15$ & & 10\cr
\hline \\ [-2.3ex]
\hline \\ [-2.3ex]
&  & $\eta_c \to \eta \pip \pim$  & &\cr
\hline \\ [-2.3ex]
\hline \\ [-2.3ex]
$a_0(1700)$ & \alm$1704 \pm 5 \pm 2$ & $110 \pm 15 \pm 11$ & & 8\cr
\hline \\ [-2.3ex]
\hline \\ [-2.3ex]
  \end{tabular}
\label{tab:tab4}
      \end{table*}
      We note a large statistical error on $g^2_{K\etapr}$ that is expected because of the weak sensitivity of the $K \pi$ $S$-wave to the opening of the $K \etapr$ threshold. We also note the presence of a very small background term.
      We attempt to replace the background term with a BW function with parameters fixed to the PDG averages for the $\kappa/K^*_0(700)$ resonance, but
      obtain a poor description of the data. 
      For comparison, the
  $K^*_0(1430)$ parameters used by BESIII in the Dalitz plot analysis of $\chi_{c1} \to \etapr \Kp \Km$~\cite{Ablikim:2014tww} are those measured by the CLEO $D^+ \to \Km \pip \pip$ Dalitz plot analysis~\cite{Bonvicini:2008jw}, $m = 1471.2\ \mevcc$, $g^2_{K\pi} = 0.299$~\gevccc,
      and $g^2_{K \etapr} = 0.0529$ \gevccc.
      
      We perform a Dalitz plot analysis of the $\etac \to \etapr \Kp \Km$ decay channel by using the $\etapr f_0(1710)$
      intermediate state as the reference amplitude. 
If there are regions of the phase space not well described by the fit, we add postulated $\Kp K^{*-}_{0}$, $\etapr f_{0,2}$, or $\etapr a_0$ intermediate states, and accept them if $\Delta(-2\log\calL)>2$.
At each stage, we test for the presence of a non-resonant contribution.

    We describe the $K^*_0(1430)$ according to Eq.~(\ref{eq:ch}) first with $m_0$ and $g^2_{K\pi}$
      parameters fixed to the values from the fit to the $K \pi$ $S$-wave and $g^2_{K\etapr}$ free.
      We observe little sensitivity to the $g^2_{K\etapr}$ parameter, expressed by the large error, and therefore we also fix the value of this parameter to that from the fit to the $K \pi$ $S$-wave. 

The projections of the fit result are shown in Fig.~\ref{fig:fig12}, along with the largest signal components. 
To test the fit quality, we generate a large number of phase-space MC-simulated events, which are weighted by the likelihood function obtained by the fit. 
These MC-simulated events are then normalized to the observed yield and are superimposed to the data.
To test the fit quality we also project the fit on the ($m(\Kp \Km),\cos \theta_H)$ plane and compare data and simulation in each cell of the plane.
Labelling with ${\rm ndf}=N_{\rm cells}-N_{\rm par}$, where $N_{\rm cells}$ is the number of cells having at least two expected events and $N_{\rm par}$ the number of free parameters in the Dalitz analysis, we obtain $\chi^2/{\rm ndf}=285/264=1.1$ corresponding to a $p$-value of 18\%.
 
 The intermediate states retained by this procedure are listed in the left half of Table~\ref{tab:tab5}, together with their fitted fractions and relative phases. 
 We label this fit as solution (A).
 The non-resonant contribution is consistent with zero.      
      
      \begin{table*}[htb] 
        \caption{\small Fractions and relative phases from the Dalitz plot analysis of $\eta_c \to \etapr \Kp \Km$.
         The first errors are statistical, the second systematic. 
        }
        \centering
  \begin{tabular}{lcccc}
\hline \\ [-2.3ex]
Intermediate state & fraction (\%) & phase (rad)& fraction (\%) & phase (rad)\cr
\hline \\ [-2.3ex]
& Solution (A) & &  Solution (B)\cr
\hline \\ [-2.3ex]
$f_0(1710) \etapr$ & $29.5 \pm 4.7 \pm 1.6$ & 0. & $29.4 \pm 4.5 \pm 1.6$ & 0.\cr
$K^*_0(1430)^+ K^-$ & $53.9 \pm 7.2 \pm 2.0$   & \al$0.61 \pm 0.13 \pm 0.45$ & $61.4 \pm 8.1 \pm 2.6$   & \al$0.79 \pm 0.12 \pm 0.59$ \cr
$K^*_0(1950)^+ K^-$ & \al $2.4 \pm 1.2 \pm 0.4$ & \al$0.46 \pm 0.29 \pm 0.50$ & \al$2.6 \pm 1.2 \pm 0.5$ & \al$0.21 \pm 0.28 \pm 1.10$  \cr
$f_0(1500) \etapr$ & \al$0.8 \pm 1.0 \pm 0.3$ & \al$0.32 \pm 0.54 \pm 0.10$ & \al$0.9 \pm 1.0 \pm 0.3$ & \al$0.24 \pm 0.52 \pm 0.10$  \cr
$f_0(980) \etapr$ & \al$4.7 \pm 2.7 \pm 0.4$ & \aln$-0.74 \pm 0.55 \pm 0.05$ & \al$5.8 \pm 3.0 \pm 0.5$ & \aln$-1.01 \pm 0.46 \pm 0.05$\cr
$f_2(1270) \etapr$ & \al\all$2.9 \pm 1.5 \pm 0.1$ & \al\al$2.9 \pm 0.38 \pm 0.09$ & \al$2.6 \pm 1.6 \pm 0.2$ & \al$2.73 \pm 0.39 \pm 0.09$  \cr
\hline \\ [-2.3ex]
sum & $94.3 \pm  9.3 \pm 2.6$ & & $102.6 \pm 10.0  \pm 3.2$ & \cr
\hline \\ [-2.3ex]
$\chi^2/{\rm ndf}$ & 285/264=1.1 & &281/260=1.1 &  \cr
$p$-value & 18\% & & 18\% & \cr
\hline \\ [-2.3ex]
 \end{tabular}
\label{tab:tab5}
      \end{table*}
     
We measure the $f_0(1710)$ parameters, listed in Table~\ref{tab:tab4}.
In addition to the strong $f_0(1710)\etapr$ and $K^*_0(1430)^+ \Km$ contributions there is evidence for a signal of the $K^*_0(1950)^+ \Km$ decay mode. We measure the parameters of the $K^*_0(1950)$ (see Table~\ref{tab:tab4}) for which there is only one previous measurement from the LASS collaboration~\cite{Aston:1987ir}.
There are smaller contributions from $f_0(980)\etapr$, $f_2(1270)\etapr$, and $f_0(1510)\etapr$.  
The latter is indistinguishable from an $f_2'(1525)\etapr$ contribution, but for simplicity, we report only the $f_0(1510)\etapr$, which gives a slighly larger likelihood improvement.

Statistical significances of resonances contributing to the decay are evaluated using the Wilks' theorem~\cite{wilks} from the difference in log likelihood between fits with and without the specific signal component, taking into account the difference of two free parameters.
For $f_0(1710)\etapr$ and $K^*_0(1950)^+\Km$ we obtain $\Delta(-2\log \calL)=135.9$ and $\Delta(-2\log \calL)=15.3$, respectively. 
The corresponding significances are listed in Table~\ref{tab:tab4}.

We evaluate systematic uncertainties on the fitted fractions, phases, and resonance parameters. 
For resonances having parameters fixed to PDG values, we vary these parameters according to their PDG uncertainties. 
We modify the purity of the \etac signal according to its statistical uncertainty. 
We replace the fitted efficiency with the raw efficiency, defined in Sec.~\ref{sec:effy}. 
The Blatt-Weisskopf~\cite{blatt} form factor present in the relativistic BW functions, nominally fixed at 1.5 GeV$^{-1}$, is varied between 0 and 3.0 GeV$^{-1}$. 
The background description is modified by varying each resonant fraction by its statistical uncertainties in the fits to the sidebands. 
All the contributions are added in quadrature.

An inspection of Fig.~\ref{fig:fig12}(b)-(d) suggests an additional enhancement in the $m^{(2)}(\etapr K^{\pm})$ around a mass of $\approx 2100$ \mevcc.
We explore this possibility adding, in the Dalitz plot analysis, an additional scalar resonance in this mass region with free parameters.
The presence of this additional resonance also affects the parameters of the $K^*_0(1950)$ which are also left free in the fit.
The fit returns the following values of the parameters of these resonances
\begin{equation}
  \nonumber
  \begin{split}
    m(K^*_0(1950)) &= 1979 \pm 26_{\rm stat} \pm 3_{\rm sys}\ \mevcc; \\
    \Gamma(K^*_0(1950)) &= 144 \pm 44_{\rm stat} \pm 21_{\rm sys}\ \mevcc,
     \end{split}
\end{equation}
and
\begin{equation}
  \nonumber
  \begin{split}
    m(K^*_0(2130)) &= 2128 \pm 31_{\rm stat} \pm 9_{\rm sys}\ \mevcc; \\
    \Gamma(K^*_0(2130)) &= 95 \pm 42_{\rm stat} \pm 76_{\rm sys}\ \mevcc.
     \end{split}
\end{equation}

\begin{figure}
  \begin{center}
    \includegraphics[width=8cm]{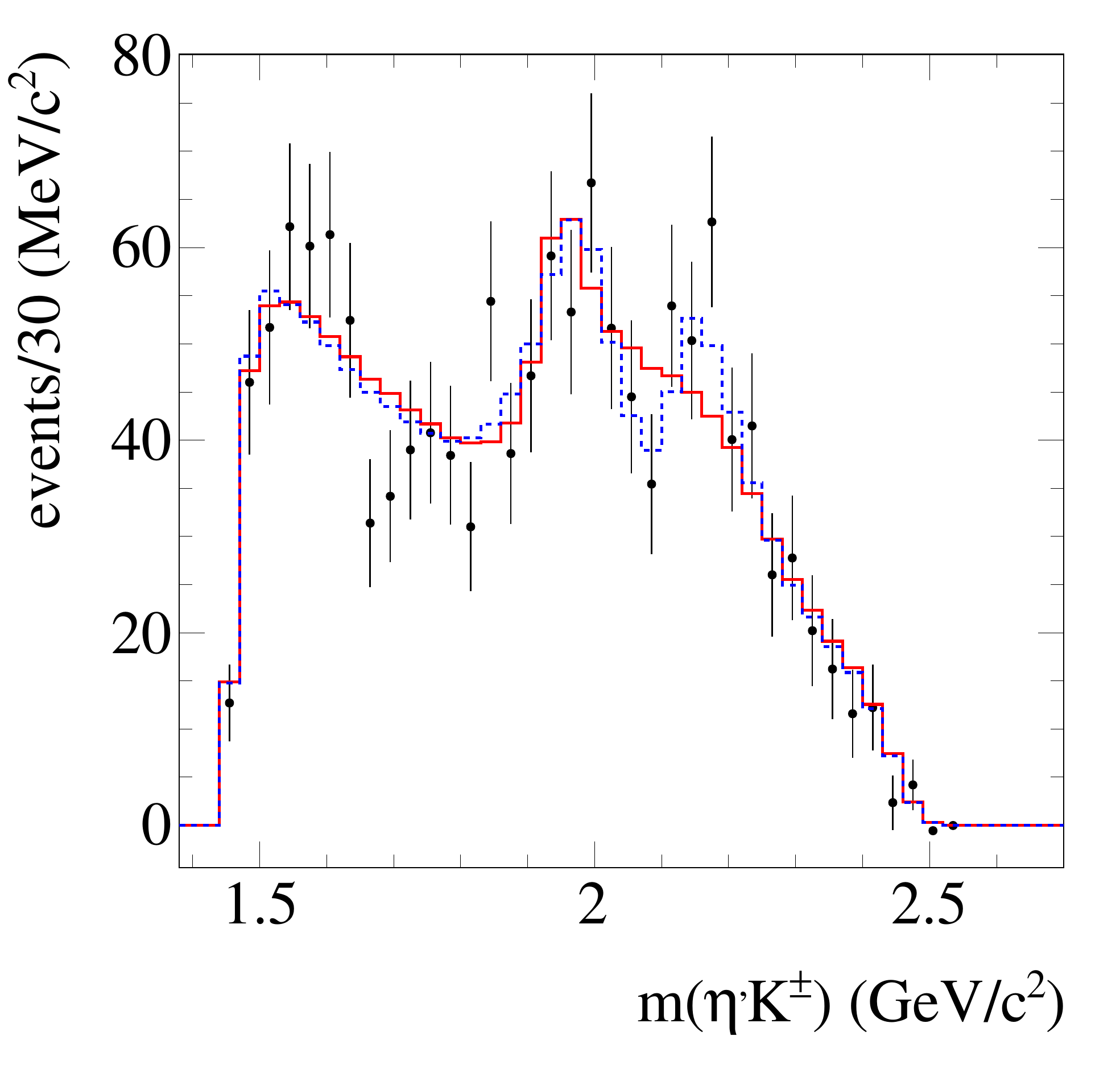}
    \caption{
Linear-scale mass projection $m(\etapr K^{\pm})$, after subtraction of the background.
The solid (red) histogram represent the results of the fit described in the text (solution (A)).
The dashed (blue) histogram represent results of the fit (solution (A)) allowing the presence of an additional $K^*_0(2130)$ resonance.
  The $\etapr K^{\pm}$ mass projection has two entries per event.}
\label{fig:fig14}
\end{center}
\end{figure}

A comparison between the two fits on the $m(\etapr K^{\pm})$ projection is shown in Fig.~\ref{fig:fig14}.
This new hypothesis gives an overall improvement of the likelihood by a factor $\Delta(-2\log \calL)=8.3$. However, an application of the Wilks theorem
for the individual significances of the $K^*_0(1950$ and $K^*_0(2130)$ in this new fit, obtain values of 4.3$\sigma$ and 2.7$\sigma$, respectively.
Since the local significance of the $K_0^*(2130)$ is less than
$3\sigma$, we do not consider anymore in the following the presence of this contribution.

\subsection{Measurement of the relative $K^*_0(1430) \to K \etapr$ coupling}

We make use of previous measurements of \etac decays, combined with the results of the present analysis, to obtain a
measurement of the $K^*_0(1430)$ couplings to the $K \etapr$ and $K \pi$ final states.
The product of the \etac two-photon width and its branching fraction to $\etapr\pip \pim$, $\Gamma_{\gamma\gamma}\calB(\etac \to \etapr \pip \pim)=65.4 \pm 2.6_{\rm stat} \pm 7.8_{\rm sys}$ eV, has been measured
by the Belle experiment~\cite{Xu:2018uye}, while 
$\Gamma_{\gamma\gamma}\calB(\etac \to K \bar K \pi)=386 \pm 0.008_{\rm stat} \pm 0.021_{\rm sys}$ eV has been measured by the \babar\ experiment~\cite{delAmoSanchez:2011bt}.
The isospin decomposition of the \etac decay to $K \bar K \pi$ includes decays to $\bar K^0 \Kp \pim$, $K^0 \Km \pip$, $K^0 \bar K^0 \piz$, and $\Kp \Km \piz$, where the latter contributes with a factor 1/6.
Dividing the \babar\ result by a factor 6 to obtain the $\etac \to \piz \Kp \Km$ component, we have

\begin{equation}
  \frac{\calB(\etac \to \etapr \pip \pim)}{\calB(\etac \to \piz \Kp \Km)}  = 1.016 \pm 0.040_{\rm stat} \pm 0.121_{\rm sys}.
  \label{eq:bab_bell}
\end{equation}

Combined with the $\calB(\etac \to \etapr \Kp \Km)/\calB(\etac \to \etapr \pip \pim)$, given above, Eq.~(\ref{eq:br_etapr}), this gives

\begin{equation}
  \frac{\calB(\etac \to \etapr \Kp \Km)}{\calB(\etac \to \piz \Kp \Km)} = 0.655 \pm  0.047_{\rm stat} \pm 0.085_{\rm sys}.
  \label{eq:brr}
\end{equation}

The \babar\ Dalitz plot analysis of $\etac \to \piz \Kp \Km$ measured
the fraction $\calB(\etac \to K^- K^*_0(1430)^+(\to \Kp \piz)) =(33.8 \pm 1.9_{\rm stat} \pm 0.4_{\rm sys})\%$~\cite{Lees:2014iua}. The present analysis measures $\calB(\etac \to K^-K^*_0(1430)^+(\to \Kp \etapr)) = (53.9 \pm 7.2_{\rm stat} \pm 2.0_{\rm sys})\%$ (left section of Table~\ref{tab:tab5}).
Combining these, and applying a factor 3 due to the isospin related unseen decay modes, we obtain the ratio

\begin{equation}
  \calB = \frac{\calB(K^*_0(1430) \to K \etapr)}{\calB(K^*_0(1430) \to K \pi)}=0.348 \pm 0.056_{\rm stat} \pm 0.047_{\rm sys}.
  \label{eq:kstar}
\end{equation}

  This ratio can be written as

\begin{equation}
  \calB = \frac{g^2_{K \etapr}}{g^2_{K \pi}}\frac{I_{K \etapr}}{I_{K \pi}},
  \label{eq:calb}
\end{equation}

\noindent
where $I_{K \etapr}$ and $I_{K \pi}$ are the integrals over the \etac phase space of the coupled-channel Breit-Wigner function
  describing the $K^*_0(1430)$ in the $\etac \to \etapr \Kp \Km$ and $\etac \to \piz \Kp \Km$ decay modes (Eq.~(\ref{eq:ch})).
 Using Eq.~(\ref{eq:calb}), we obtain the ratio of the couplings $\frac{g^2_{K \etapr}}{g^2_{K \pi}}= 1.43\pm 0.23_{\rm stat} \pm 0.22_{\rm sys}$,
 to be compared with the results from the fit to the $K \pi$ $S$-wave, (from the first row in Table~\ref{tab:tab4}), of $\frac{g^2_{K \etapr}}{g^2_{K \pi}}=0.476 \pm 0.254$.

To resolve this discrepancy (of the order of $2.3\sigma$), we perform several fits to the $K \pi$ $S$-wave with  $\frac{g^2_{K \etapr}}{g^2_{K \pi}}$
varying from 0.476 to 1.75, observing a steady increase in $\chi^2$ from 55 to 80.
Using each set of fitted $K^*_0(1430)$ resonance parameters,
we repeat the Dalitz plot analysis to obtain new values of the fractional contributions,
 and recalculate the ratio $\frac{g^2_{K \etapr}}{g^2_{K \pi}}$ according to Eq.~(\ref{eq:ch}). This ratio depends weakly on the resonance
 parameters, varying between 1.40 to 1.67. Therefore, we fix $\frac{g^2_{K \etapr}}{g^2_{K \pi}}=1.43$ in the fit to the $K \pi$ $S$-wave, and show the result as the dashed (blue) lines in Fig.~\ref{fig:fig13}.
This fit has a $\chi^2/{\rm ndf}=70/32$ ($\chi^2/{\rm ndf}=32/32$ when systematic uncertainties are included).
 The fitted  $K^*_0(1430)$
parameters are then used in a new Dalitz plot analysis, which we denote solution (B), the results of which are listed 
in the right half of Table~\ref{tab:tab5}. The fitted  $K^*_0(1430)^+\Km$ contribution increases to $\calB(\etac \to K^- K^*_0(1430)^+(\to \Kp \etapr))=(61.4 \pm 8.1_{\rm stat} \pm 2.6_{\rm sys})\%$ which gives the ratio

\begin{equation}
  \calB = \frac{\calB(K^*_0(1430) \to K \etapr)}{\calB(K^*_0(1430) \to K \pi)} = 0.397 \pm 0.064_{\rm stat} \pm 0.054_{\rm sys}
  \label{eq:b_calb}
\end{equation}

and

\begin{equation}
  \frac{g^2_{K \etapr}}{g^2_{K \pi}} = 1.50 \pm 0.24_{\rm stat} \pm 0.24_{\rm sys}, 
\end{equation}
where we have included the change from solution (A) in the systematic uncertainty, as an estimate of the model uncertainty. Similarly, we use the estimates of the $K^*_0(1430)$ mass and $g^2_{K\pi}$ from solution (B), along with the differences from solution (A)
(see Table~\ref{tab:tab4}),
to obtain 
\begin{equation}
  \begin{split}
    m(K^*_0(1430)) = & 1449 \pm 17_{\rm stat} \pm 2_{\rm sys} \mevcc, \\
    g^2_{K\pi} = & 0.458 \pm 0.032_{\rm stat} \pm 0.044_{\rm sys} \ \gevccc.
  \end{split} 
\end{equation}
The inconsistency between the $\frac{g^2_{K \etapr}}{g^2_{K \pi}}$ values may be associated with an imperfect model describing the $K^*_0(1430)$ shape.
The Dalitz plot fit quality of the solution (B) is similar to that of solution (A) with $\Delta(-2\log\calL)=4.8$ and $\chi^2/N_{\rm cells}=281/260=1.1$.

\section{Dalitz plot analysis of $\etac \to \etapr \pip \pim$}
\label{sec:daly2}
      
\begin{figure}
  \begin{center}
  \includegraphics[width=8.5cm]{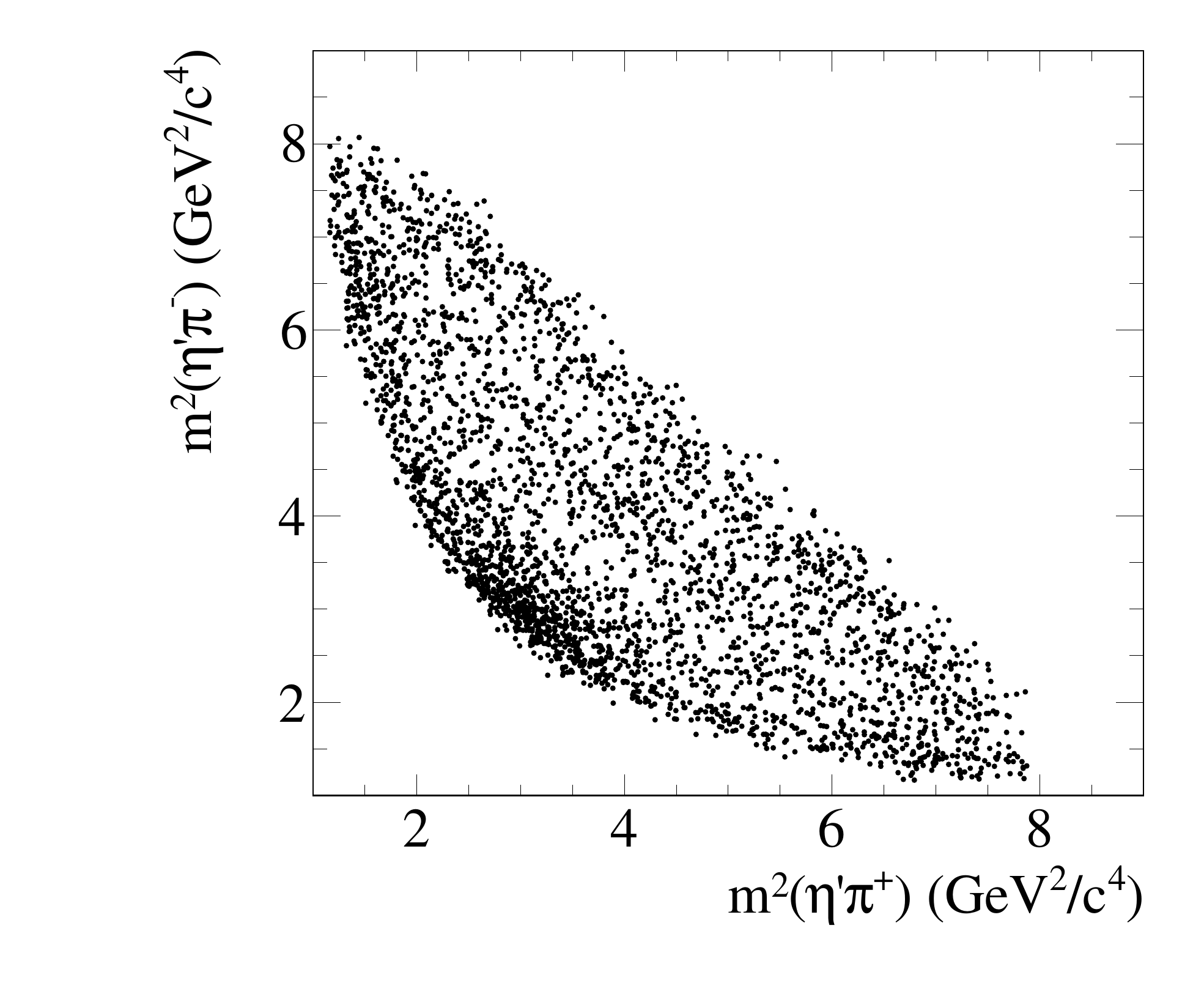}
\caption{Dalitz plot for selected $\etac \to \etapr \pip \pim$ candidates in the \etac\ signal region, summed over the two \etapr decay modes.}
\label{fig:fig15}
\end{center}
\end{figure}

\begin{figure*}
  \begin{center}
    \includegraphics[width=16cm]{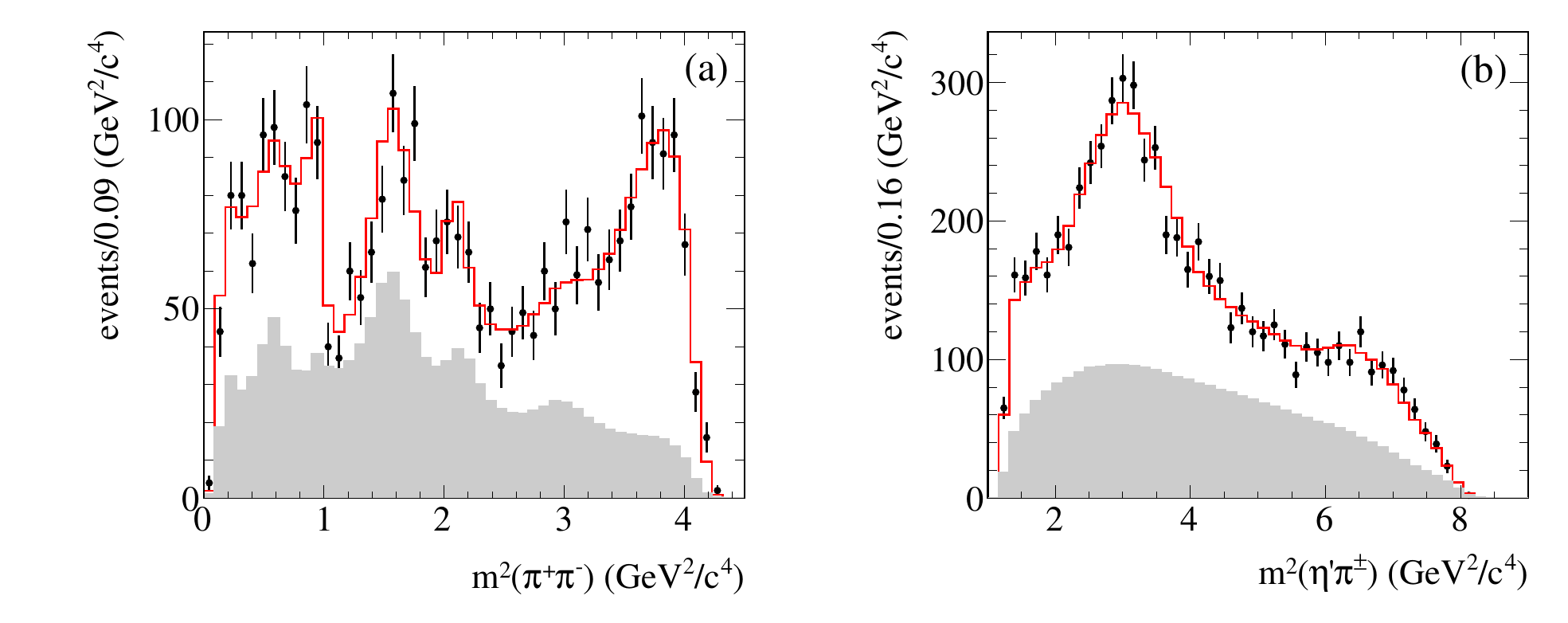}
    \includegraphics[width=16cm]{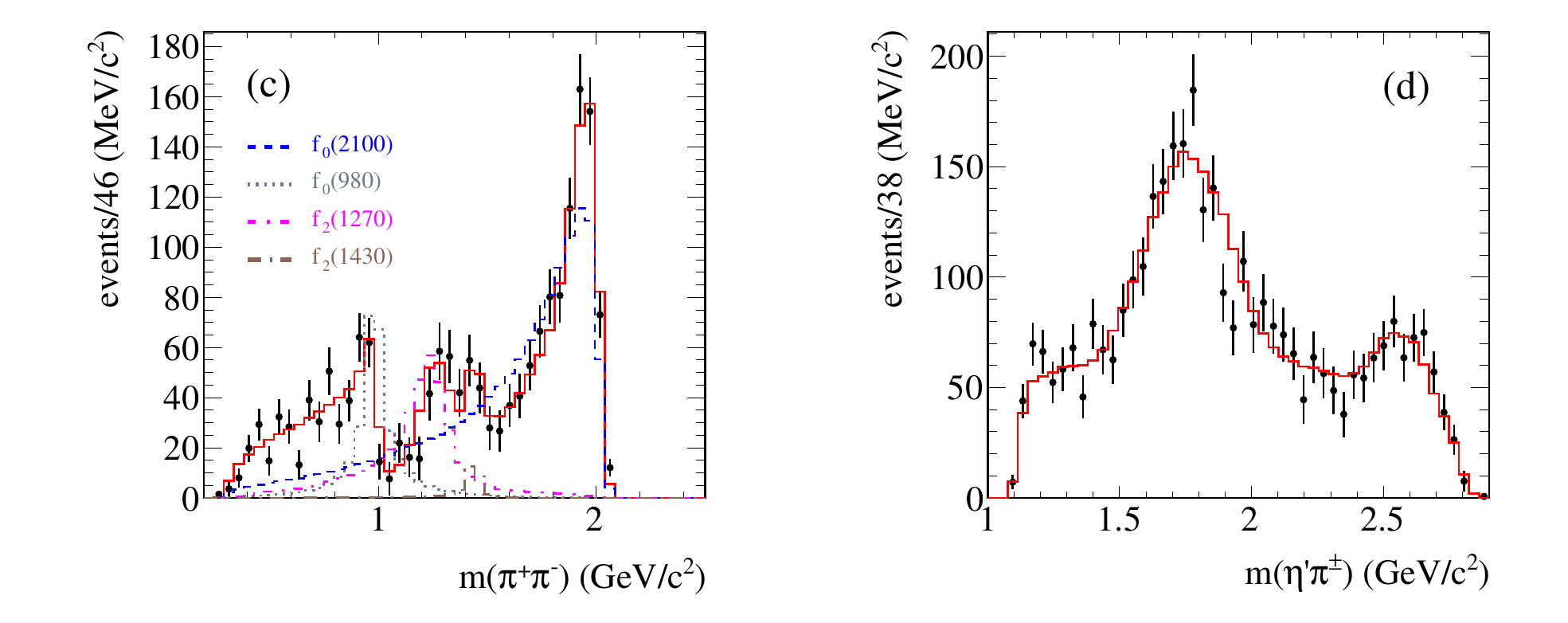}
  \caption{Squared-mass projections (a) $m^2(\pip \pim)$ and (b) $m^2(\etapr \pipm)$ of the measured $\etac \to \etapr \pip \pim$ Dalitz plot.
  The shaded (gray) histograms are the background interpolated from fits to the two \etac sidebands.
  Linear-scale mass projections (c) $m(\pip \pim)$ and (d) $m(\etapr\pipm)$, after subtraction of the background.
  The solid (red) histograms represent the results of the fit described in the text, and the other histograms display the contributions from each of the listed components.
  The $\etapr \pipm$ projections have two entries per event.}
\label{fig:fig16}
\end{center}
\end{figure*}

Figure~\ref{fig:fig15} shows the Dalitz plot for the selected $\etac \to \etapr \pip \pim$ candidates in the data, in the \etac\ signal region, for the two \etapr decay modes combined, and Figs.~\ref{fig:fig16}(a)-(b) show two squared-mass projections. 
We observe several diagonal bands in the Dalitz plot, in particular at the lower-left edge. 
There are corresponding structures in the $m^2(\pip \pim)$ spectrum, including peaks attributable to the $f_0(980)$ and $f_2(1270)$ resonances, and a large structure at high $\pip \pim$ mass. 
In the $m^2(\etapr \pipm)$ spectrum, a large structure is present; 
there is no known resonance decaying to $\etapr \pi$ in this mass region, 
but this could be a reflection of the structure in the high $m^2(\pip \pim)$ region.

We fit the two \etac sidebands using an incoherent sum of amplitudes, which includes contributions from the $\rho^0(770)$, $f_2(1270)$, $f_0(1370)$, and $f_0(2100)$ resonances.
To model the background in the \etac signal region we take a weighted average of the fitted fractional contributions, and normalize using the results from the fit to the \etaprpipi invariant-mass spectrum.
The estimated background contributions are indicated by the shaded regions in Figs.~\ref{fig:fig16}(a)-(b), and
we show the corresponding background-subtracted invariant-mass spectra in Figs.~\ref{fig:fig16}(c)-(d).

A candidate for the large structure in the high $\pip \pim$ mass region is the $f_0(2100)$ resonance, observed in radiative $J/\psi$ decay to $\gamma \eta \eta$~\cite{Ablikim:2013hq}.
We take $f_0(2100)\etapr$ as the reference contribution, and perform a Dalitz plot analysis as described in Sec.~\ref{sec:daly}.
Again, no non-resonant contribution is needed, and the list of the resonances contributing to this \etac decay mode is given in Table~\ref{tab:tab6}, together with their fitted fractions and relative phases.
      \begin{table}
        \caption{\small Fractions and relative phases from the Dalitz plot analysis of $\eta_c \to \etapr \pip \pim$.
          The first errors are statistical, the second systematic.
        }
        \centering
  \begin{tabular}{lcc}
\hline \\ [-2.3ex]
Intermediate state & fraction (\%) & phase (rad)\cr
\hline \\ [-2.3ex]
$f_0(2100) \etapr$ & $74.9 \pm 7.5 \pm 3.6$ & 0.  \cr
$f_0(500) \etapr$ & \al$4.3 \pm 2.3 \pm 0.7$ & $-5.89 \pm 0.24 \pm 0.10$ \cr
$f_0(980) \etapr$ & $16.1 \pm 2.4 \pm 0.5$ & $-5.31 \pm 0.16 \pm 0.04$  \cr
$f_2(1270) \etapr$ & $22.1 \pm 2.9 \pm 2.4$ & $-3.60 \pm 0.16 \pm 0.03$  \cr
$f_2(1430) \etapr$ & \al$1.9 \pm 0.7 \pm 0.1$ & $-2.45 \pm 0.32 \pm 0.11$  \cr
$a_2(1710) \pi$ & \al$3.2 \pm 1.9 \pm 0.5$ & $-0.75 \pm 0.27 \pm 0.11$ \cr
$a_0(1950) \pi$    & \al$2.5 \pm 1.1 \pm 0.1$& $-0.02 \pm 0.32 \pm 0.06$ \cr
$f_2(1800) \etapr$ & \al$5.3 \pm 2.2 \pm 1.4$ & \al\all$0.67 \pm 0.24 \pm 0.08$  \cr
\hline \\ [-2.3ex]
sum & \alm$130.5 \pm 9.5 \pm 4.7$ & \cr
\hline \\ [-2.3ex]
$\chi^2/{\rm ndf}$=409/386=1.1 & \cr
$p$-value & 20\% & \cr
\hline \\ [-2.3ex]
 \end{tabular}
\label{tab:tab6}
      \end{table}
      
The $f_0(2100)$ parameters are first left free in the fit, and we obtain the values listed in Table~\ref{tab:tab4}, which are in agreement with BESIII measurement ($m=2081\pm 13^{+24}_{-36}$ \mevcc, $\Gamma=273^{+27+70}_{-24-23}$) \mev~\cite{Ablikim:2013hq}.
We then fix them to the values listed in the PDG. 
We also leave free the $f_0(500)$ parameters and obtain the values listed in Table~\ref{tab:tab4} which give a good description of the data. Given the low statistics, we do not assign systematic uncertainties to the fitted $f_0(500)$ resonance parameters, which are within the range of other measurements~\cite{PDG}. 
The $f_0(980)$ is parameterized by a coupled-channel Breit-Wigner function with parameters fixed to the measurement from Ref.~\cite{Armstrong}.
To describe the small enhancement around 1.43 \gevcc , we test both spin-2 and spin-0 hypotheses with free resonance parameters; 
we obtain $\Delta(-2\log \calL)=2.4$ in favor of the spin-2 hypothesis, so we attribute this signal to the $f_2(1430)$ resonance, and report the fitted parameter values in Table~\ref{tab:tab4}. 
We test the significance of this signal by removing it from the list of the resonances, obtaining $\Delta(-2\log \calL)=23.8$ and a significance of $4.4 \sigma$.
Replacing the $f_2(1430)$ resonance with $f_0(1500)$ or $f_0(1370)$, we obtain poor fits with fractions from these possible contributions consistent with zero.
The $f_0(2100)$ statistical significance is 10$\sigma$.

The projections of the fit result are compared with the data in Fig.~\ref{fig:fig16}. 
To test the fit quality, we generate a large number of phase-space MC-simulated events, which are weighted by the likelihood function obtained from the fit. 
These MC-simulated events are then normalized to the observed yield and superimposed to the data.
We also project the fit on the ($m(\pip \pim),\cos \theta_H)$ plane and compare data and simulation in each cell, obtaining $\chi^2/{\rm ndf}=409/386=1.1$.
The systematic uncertainties on the fitted fractions, phases and resonance  parameters are evaluated as in the previous section.

\section{Dalitz plot analysis of $\etac \to \eta \pip \pim$}
\label{sec:daly3}

\begin{figure}
  \begin{center}
  \includegraphics[width=8.5cm]{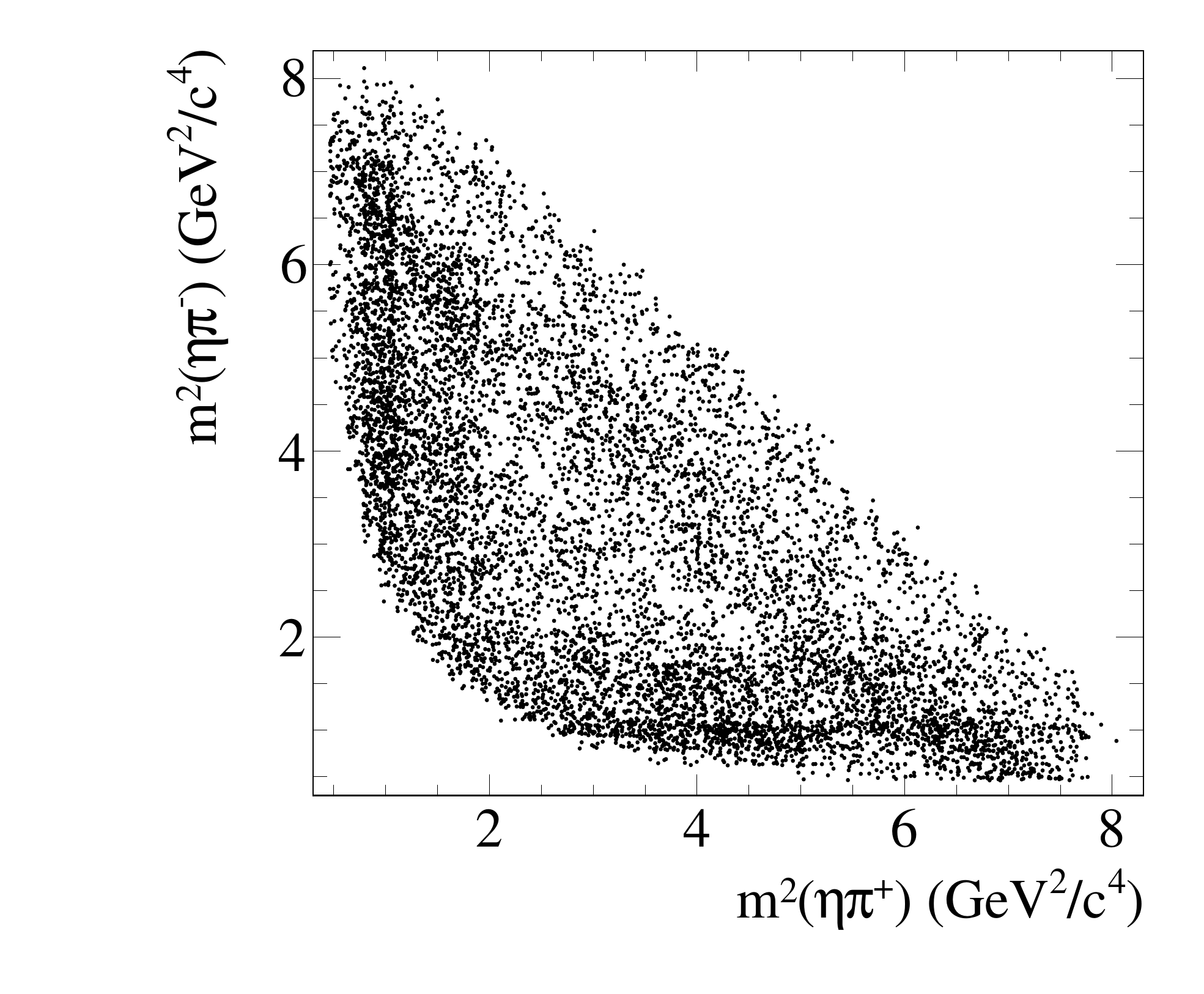}
\caption{Dalitz plot for selected $\etac \to \eta \pip \pim$ candidates in the \etac\ signal region, summed over the two $\eta$ decay modes.}
\label{fig:fig17}
\end{center}
\end{figure}

\begin{figure*}
  \begin{center}
    \includegraphics[width=16cm]{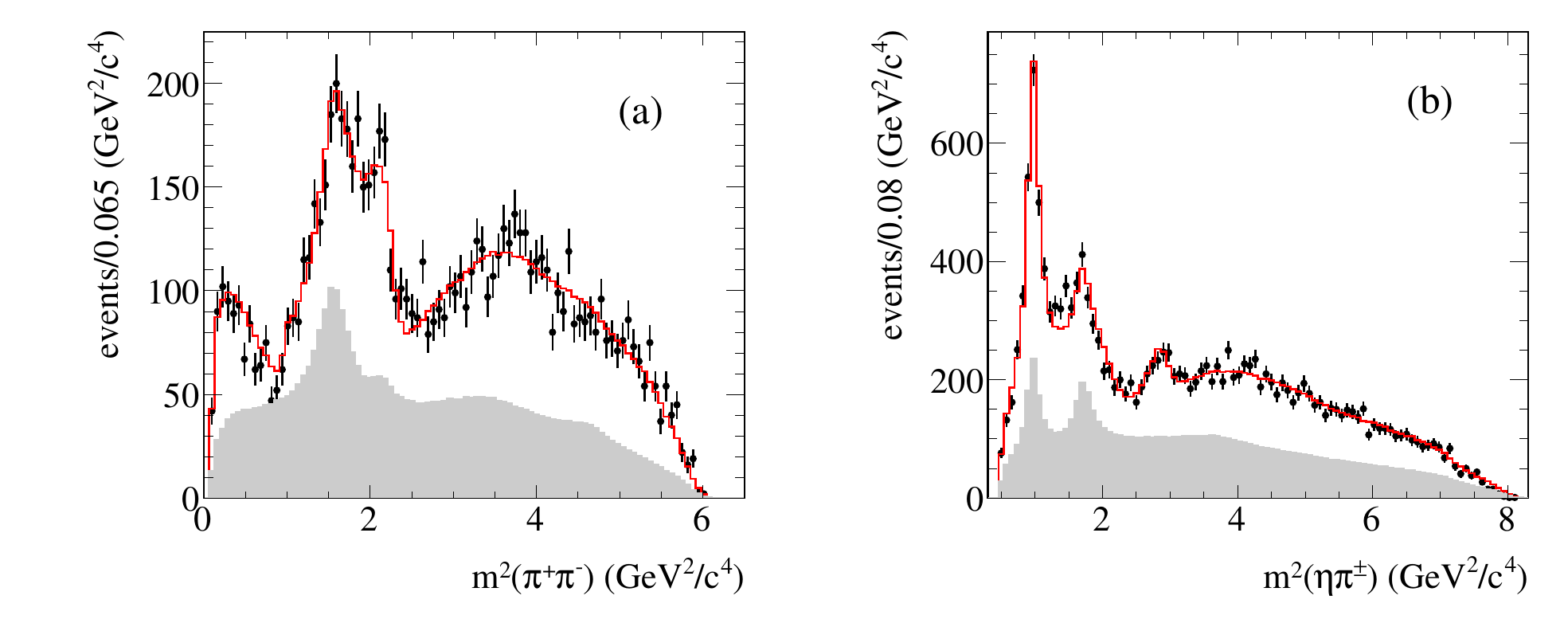}
    \includegraphics[width=16cm]{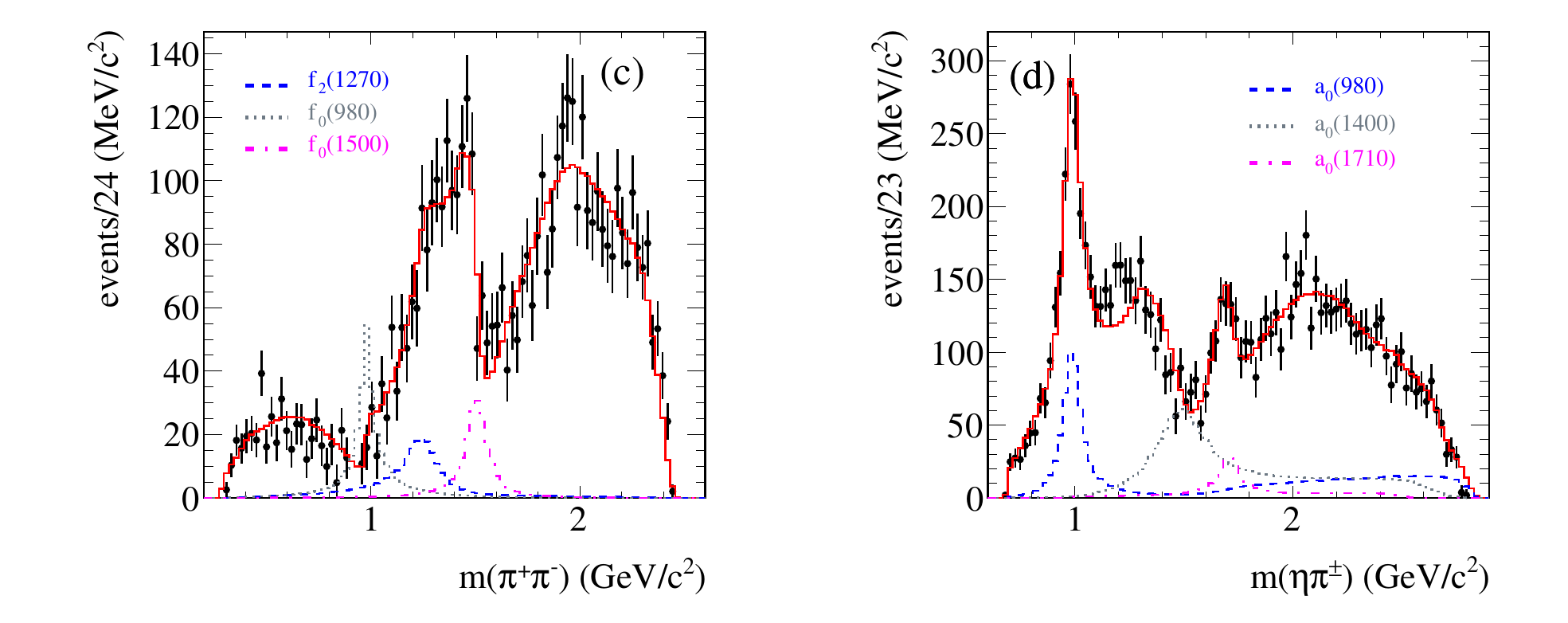}
  \caption{Squared-mass projections (a) $m^2(\pip \pim)$ and (b) $m^2(\eta \pipm)$ of the measured $\etac \to \eta \pip \pim$ Dalitz plot.
  The shaded (gray) histograms represent the background interpolated from fits to the two \etac sidebands.
  Linear-scale mass projections (c) $m(\pip \pim)$ and (d) $m(\eta \pipm)$ after subtraction of the background.
  The solid (red) histograms represent the results of the fit described in the text, and the other histograms display the contributions from each of the listed components. 
  The $\eta \pi^{\pm}$ projections have two entries per event.}
\label{fig:fig18}
\end{center}
\end{figure*}

Figure~\ref{fig:fig17} shows the Dalitz plot for the selected $\etac \to \eta \pip \pim$ candidates in the data, in the \etac\ signal region, for the two $\eta$ decay modes combined, and Figs.~\ref{fig:fig18}(a)-(b) show two squared-mass projections. 
We observe that the Dalitz plot is dominated by horizontal and vertical bands due to the $a_0(980)$ and diagonal bands due to resonances in the $\pip \pim$ final state. 
The squared-mass projections show signals of $f_0(500)$, $f_0(980)$, and $f_2(1270)$.

The \etac sidebands are also rich in resonant structure, and are fitted using an incoherent sum of amplitudes, including contributions from the $a_0(980)$, $f_2(1270)$, $a_2(1310)$, and $f_2(1950)$ resonances.
We take a weighted average of the fitted fractions in the two sidebands, normalized using the results from the fit to the \etapipi  invariant-mass spectrum, to estimate the background in the signal region, shown as the shaded regions in Figs.~\ref{fig:fig18}(a)-(b).

We take $a_0(980)^+ \pim$ as the reference contribution, and perform a Dalitz plot analysis as described above.
The resulting list of contributions to this \etac decay mode is given in Table~\ref{tab:tab7}, together with fitted fractions and relative phases.

\begin{table} 
     \caption{\small  Fractions and relative phases from the Dalitz plot analysis of $\eta_c \to \eta \pip \pim$.
     The first errors are statistical, the second systematic.
        }
        \centering
  \begin{tabular}{lcc}
\hline \\ [-2.3ex]
Intermediate state & fraction (\%) & phase (rad)\cr
\hline \\ [-2.3ex]
$a_0(980)^+ \pim$ & \alm$12.3 \pm 1.2 \pm 2.8$ & 0.  \cr
$a_2(1310)^+ \pim$ & $2.5 \pm 0.7 \pm 0.9$ & $-1.04 \pm 0.13 \pm 0.20$  \cr
$f_0(500) \eta$ & $4.3 \pm 1.3 \pm 1.1$ & \al\all$0.54 \pm 0.14 \pm 0.24$ \cr
$f_2(1270) \eta$ & $4.6 \pm 0.9 \pm 0.8$ & $-1.15 \pm 0.11 \pm 0.05$  \cr
$f_0(980) \eta$ & $5.7 \pm 1.3 \pm 1.5$ & $-2.41 \pm 0.09 \pm 0.07$ \cr
$f_0(1500) \eta$ & $4.2 \pm 0.7 \pm 0.9$ &  \al\all$2.32 \pm 0.13 \pm 0.17$\cr
$a_0(1450)^+ \pim$ & \alm$15.0 \pm 2.4 \pm 3.2$ &  \al\all$2.60 \pm 0.09 \pm 0.11$  \cr
$a_0(1700)^+ \pim$ & $3.5 \pm 0.8 \pm 0.8$ &  \al\all$1.39 \pm 0.15 \pm 0.20$  \cr
$f_2(1950) \eta$    & $4.2 \pm 1.0 \pm 1.0$ & $-1.59 \pm 0.15 \pm 0.21$ \cr
\hline \\ [-2.3ex]
resonant sum & \alm$56.3\pm 3.7 \pm 10.0$  & \cr
\hline \\ [-2.3ex]
$NR$            & \alm$172.7 \pm 8.0 \pm 10.0$ &  \al\all$1.67 \pm 0.07 \pm 0.06$  \cr
\hline \\ [-2.3ex]
sum & \alm$229.0 \pm 8.8 \pm 14.1$ & \cr
\hline \\ [-2.3ex]
$\chi^2/{\rm ndf}$ & 419/382=1.1 & \cr
$p$-value & 9.3\% & \cr
\hline \\ [-2.3ex]
 \end{tabular}
\label{tab:tab7}
\end{table}

We find little sensitivity to the parameters of the $f_0(500)$ resonance, and therefore we use the parameters from the $\etac \to \etapr \pip \pim$ Dalitz plot analysis, listed in Table~\ref{tab:tab4}. 
A new $a_0(1700)$ resonance is observed in the $\eta \pipm$ invariant-mass spectrum, with fitted parameters listed in Table~\ref{tab:tab4}. 
The likelihood change obtained when the resonance is excluded from the fit is $\Delta(-2\log \calL)=72.3$, corresponding to a significance greater than $8\sigma$.
Possible contributions from the $a_2(1710)$ and $f_0(2100)$ resonances have been tested, but both are found to be consistent with zero. 

We note the presence of a very large non-resonant scalar contribution, and in Table~\ref{tab:tab7}, we list both the sum of resonant contributions and the sum including the non-resonant contribution. 
A similar effect has been observed in charmless $B$ decays~\cite{Aubert:2006nu}. 
This effect could be correlated with the interference of the \etac with the two-photon continuum described in Sec.~\ref{sec:fits}.
       
We test the fit quality as described above, with the comparison in the $(m(\pip \pim),\cos \theta_H)$ plane giving $\chi^2/{\rm ndf}=419/382=1.1$.
We evaluate systematic uncertainties as described above but adding an additional uncertainty due to the
possible interference between intermediate resonances from the \etac decay and those present in the
background. To obtain the order of magnitude of the effect we compare the fits to the $\eta \pip \pim$ mass spectra described in Sec. VI.A with and without the interference and obtain an average difference in the \etac yield of the order of 26\%. Multiplying this factor by the sum of all the resonant fractions given
in Table~\ref{tab:tab7}, we obtain an estimate of the uncertainty of the order of 15\% which is added
in quadrature to the other sources of systematic uncertanties.
We also vary the \etac signal region width from 100 \mevcc\ to 60 \mevcc\ and add in quadrature the resulting differences in amplitudes fractions
and phases as an additional source of systematic uncertainties. 

\section{Summary}

We study the processes $\gg\to \etaprkk$, $\gg\to \etaprpipi$, and $\gg\to \etapipi$ using a data sample of 519~\invfb\ recorded with the \babar\ detector operating at the SLAC PEP-II asymmetric-energy \epem\ collider at center-of-mass energies at and near the $\Upsilon(nS)$ ($n = 2,3,4$) resonances.
We observe \etac decays to all the above final states and perform Dalitz plot analyses to measure intermediate resonant fractions and relative phases.
Significant interference effects of the \etac with the two-photon background are observed only for the decay $\etac \to \eta \pip \pim$.

The decay $\etac \to \etapr \Kp \Km$ is observed for the first time and we measure the branching fraction relative to $\etac \to \etapr \pip \pim$
\begin{equation}
  \nonumber
\frac{\BR(\etac \to \etapr \Kp \Km)}{\BR(\etac \to \etapr \pip \pim)} = 0.644 \pm 0.039_{\rm stat} \pm 0.032_{\rm sys}.
\end{equation}

Using published information from the \babar\ and Belle experiments, and this analysis of $\etac \to \etapr \Kp \Km$, we obtain measurements of the $K^*_0(1430)$ resonance parameters:
\begin{equation}
  \nonumber
  \begin{split}
    m(K^*_0(1430)) & =  1449 \pm 17_{\rm stat} \pm 2_{\rm sys}\ \mevcc; \\
    g^2_{K\pi} & = 0.458 \pm 0.032_{\rm stat} \pm 0.044_{\rm sys}\  \gevccc; \\
    \frac{g^2_{\etapr K}}{g^2_{\pi K}} & = 1.50 \pm 0.24_{\rm stat} \pm 0.24_{\rm sys}.
    \end{split}
\end{equation}
 We also measure the ratio of couplings of the $K^*_0(1430)$ resonance to $\etapr K$ and $\pi K$,
\begin{equation}
  \nonumber
  \frac{\calB(K^*_0(1430)^+ \to \etapr K)}{\calB(K^*_0(1430)^+ \to \pi K)} = 0.450 \pm 0.072_{\rm stat} \pm 0.061_{\rm sys}.
\end{equation}

The $\etac \to \etapr \Kp \Km$ decay contains a significant contribution from $\etac \to \etapr f_0(1710)$, and we measure the $f_0(1710)$ resonance parameters:
\begin{equation}
  \nonumber
  \begin{split}
    m(f_0(1710) &= 1757 \pm 24_{\rm stat} \pm 9_{\rm sys}\ \mevcc; \\
    \Gamma(f_0(1710)) &=  175 \pm 23_{\rm stat} \pm 4_{\rm sys}\ \mevcc.
     \end{split}
\end{equation}
Evidence is also found for the $K^*_0(1950)$, whose parameters are measured as:
\begin{equation}
  \nonumber
  \begin{split}
    m(K^*_0(1950)) &= 1942 \pm 22_{\rm stat} \pm 21_{\rm sys}\ \mevcc; \\
    \Gamma(K^*_0(1950)) &= 80 \pm 32_{\rm stat} \pm 20_{\rm sys}\ \mevcc.
     \end{split}
  \end{equation}
We find no evidence for the $\kappa/K^*_0(700)$ in \etac decays.

The $\etac \to \etapr \pip \pim$ decay is found to be dominated by the $f_0(2100)$ resonance, also observed in radiative $J/\psi$ decays, and we measure the resonance parameters:
\begin{equation}
  \nonumber
  \begin{split}
    m(f_0(2100)) &= 2116 \pm 27_{\rm stat} \pm 17_{\rm sys}\ \mevcc; \\
    \Gamma(f_0(2100)) &= 289 \pm 34_{\rm stat} \pm 15_{\rm sys}\ \mevcc.
     \end{split}
\end{equation}
Evidence is also found for the $f_2(1430)$, and we measure the resonance parameters:
\begin{equation}
  \nonumber
  \begin{split}
    m(f_2(1430)) &= 1440 \pm 11_{\rm stat} \pm 3_{\rm sys}\ \mevcc; \\
    \Gamma(f_2(1430)) &=46 \pm 15_{\rm stat} \pm 5_{\rm sys}\ \mevcc.
     \end{split}
\end{equation}

The Dalitz plot analysis of the $\etac \to \eta \pip \pim$ decay shows the presence of a new $a_0(1700)\to \eta \pi$ resonance, for which we measure the following parameters:
\begin{equation}
  \nonumber
  \begin{split}
    m(a_0(1700)) & = 1704 \pm 5_{\rm stat} \pm 2_{\rm sys}\ \mevcc; \\
    \Gamma(a_0(1700)) & =110 \pm 15_{\rm stat} \pm 11_{\rm sys}\ \mevcc.
     \end{split}
\end{equation}

In the framework of the identification of scalar gluonium states, it is interesting to compare the rates of \etac decays into a gluonium candidate state and an $\eta$ or an \etapr meson. 
Table~\ref{tab:tab9} summarizes relevant results from this and our previous analysis.

\begin{table} 
    \caption{\small Fractional contributions to $\etac\to \eta h^+h^-$ and $\etac\to \etapr h^+h^-$ decays of selected scalar mesons, uncorrected for unseen decay modes.
    The first errors are statistical, the second systematic.
        }
        \centering
  \begin{tabular}{lccc}
\hline \\ [-2.3ex]
Final state & $f_0(1500)$(\%) & $f_0(1710)$(\%) & $f_0(2100)$(\%) \cr
\hline \\ [-2.3ex]
$\eta \Kp \Km$ & $23.7 \pm 7.0 \pm 1.8$ & \all$8.9 \pm 0.2 \pm 0.4$ & \cr
$\eta \pip \pim$ & \al$4.2 \pm 0.7 \pm 0.9$ &  & 0. \cr
$\etapr \Kp \Km$ & \al$0.8 \pm 1.0 \pm 0.3$ & \al$29.5 \pm 4.7 \pm 1.6$ & \cr
$\etapr \pip \pim$ & \alm\alm\alm\alm$0.3 \pm 0.2$ & & $74.9 \pm 7.5 \pm 3.5$ \cr
\hline \\ [-2.3ex]
 \end{tabular}
\label{tab:tab9}
      \end{table}

      We observe an enhanced contribution of $f_0(1710)$ in \etac decays to \etapr and an enhanced contribution of $f_0(1500)$ in \etac decays to $\eta$. This effect may point to an enhanced gluonium content in the $f_0(1710)$ meson. A similar conclusion
      is drawn in the study of $J/\psi$ radiative decays~\cite{Ablikim:2013hq}.
In particular, Ref.~\cite{Gui:2012gx} finds that the production rate of the pure gauge scalar
glueball in $\jpsi$ radiative decays predicted by lattice QCD is compatible with the production
rate of \jpsi radiative decays to $f_0(1710)$ and this suggests that $f_0(1710)$ has a larger overlap with the glueball compared
to other glueball candidates (e.g., $f_0(1500)$).
The observation of $f_0(2100)$ in both $J/\psi$ radiative decays and in $\etac \to \etapr \pip\pim$ allows to add this state in the list of the candidates for the scalar glueball.

\section{Acknowledgments}

We are grateful for the 
extraordinary contributions of our \pep2\ colleagues in
achieving the excellent luminosity and machine conditions
that have made this work possible.
The success of this project also relies critically on the 
expertise and dedication of the computing organizations that 
support \babar.
The collaborating institutions wish to thank 
SLAC for its support and the kind hospitality extended to them. 
This work is supported by the
US Department of Energy
and National Science Foundation, the
Natural Sciences and Engineering Research Council (Canada),
the Commissariat \`a l'Energie Atomique and
Institut National de Physique Nucl\'eaire et de Physique des Particules
(France), the
Bundesministerium f\"ur Bildung und Forschung and
Deutsche Forschungsgemeinschaft
(Germany), the
Istituto Nazionale di Fisica Nucleare (Italy),
the Foundation for Fundamental Research on Matter (The Netherlands),
the Research Council of Norway, the
Ministry of Education and Science of the Russian Federation,
Ministerio de Economia y Competitividad (Spain), and the
Science and Technology Facilities Council (United Kingdom).
Individuals have received support from 
the Marie-Curie IEF program (European Union), the A. P. Sloan Foundation (USA) 
and the Binational Science Foundation (USA-Israel).

\renewcommand{\baselinestretch}{1}


\begin{thebibliography}{99}
\bibitem{polosa} G.'t Hooft {\it et al.}, Phys. Lett. B {\bf 662}, 424 (2008);
 W. Ochs, J. Phys. G {\bf 40}, 043001  (2013).
\bibitem{cp} B. Aubert {\it et al.} (\babar\ Collaboration), Phys. Rev. D {\bf 71}, 032005 (2005);
  B. Aubert {\it et al.} (\babar\ Collaboration), Phys. Rev. D {\bf 78}, 034023 (2008);
  B. Aubert {\it et al.} (\babar\ Collaboration), Phys. Rev. D {\bf 78}, 012004 (2008);
  A. Poluektov {\it et al.} (Belle Collaboration), Phys. Rev. D {\bf 81}, 112002 (2010). 
\bibitem{zs} K. Chilikin {\it et al.} (Belle Collaboration), Phys. Rev. D {\bf 88}, 074026 (2013);
  R. Aaij {\it et al.} (LHCb Collaboration), Phys. Rev. Lett. {\bf 112}, 222002 (2014).
\bibitem{bs} R. Aaij {\it et al.} (LHCb Collaboration), Phys. Rev. D {\bf 90}, 072003 (2014).
\bibitem{Lees:2014iua}
J.~P.~Lees \textit{et al.} (\babar\ Collaboration),
Phys. Rev. D \textbf{89}, 112004 (2014).
\bibitem{lass_keta} D. Aston \textit{et al.}(LASS Collaboration), Phys. Lett. B {\bf 201},
169 (1988).
  \bibitem{Lees:2015zzr}
J.~P.~Lees \textit{et al.} (\babar\ Collaboration),
Phys. Rev. D \textbf{93}, 012005 (2016).
\bibitem{Bonvicini:2008jw}
G.~Bonvicini \textit{et al.} (CLEO Collaboration),
Phys. Rev. D \textbf{78}, 052001 (2008).
\bibitem{Ablikim:2014tww}
M.~Ablikim \textit{et al.} (BESIII Collaboration),
Phys. Rev. D \textbf{89}, 074030 (2014)
\bibitem{lattice} Y. Chen \etal\, Phys. Rev. D {\bf 73}, 014516
  (2006).
\bibitem{kopke}  L. K{\"o}pke and N. Wermes, Phys. Rept. {\bf 174}, 67 (1989).
\bibitem{Dobbs} S. Dobbs, A. Tomaradze, T. Xiao, and K.K. Seth, Phys. Rev. D {\bf 91}, 052006 (2015).  
\bibitem{Lees:2018qrk}
J.~P.~Lees \textit{et al.} (\babar\ Collaboration),
Phys. Rev. D \textbf{97}, 112006 (2018).
\bibitem{klempt} E. Klempt and A. Zaitsev, Phys. Rept. {\bf 454}, 1 (2007).
\bibitem{ochs}  W. Ochs, J. Phys. G {\bf40}, 043001 (2013).
\bibitem{mink}  P. Minkowski and W. Ochs, Eur. Phys. J. C {\bf 9}, 283 (1999).
\bibitem{amsler1} C. Amsler and F.E. Close, Phys. Lett. B {\bf 353}, 385 (1995).
\bibitem{amsler2} C. Amsler and F.E. Close, Phys. Rev. D {\bf 53}, 295
  (1996).
\bibitem{gg}  S. Janowski, F. Giacosa and D. H. Rischke,  Phys. Rev. D {\bf 90}, 114005 (2014).
  \bibitem{Ablikim:2013hq}
M.~Ablikim \textit{et al.} (BESIII Collaboration),
Phys. Rev. D \textbf{87}, 092009 (2013).
\bibitem{Gui:2012gx}
L.~C.~Gui \textit{et al.} [CLQCD],
Phys. Rev. Lett. \textbf{110}, 021601 (2013).
 \bibitem{chano} 
   M.~S.~Chanowitz, Phys.\ Rev.\ Lett.\  {\bf 95}, 172001 (2005).
 \bibitem{chao}
   K.~Ta. Chao, X.~G. He and J.~P. Ma, Phys. Rev. Lett. {\bf 98}, 149103 (2007).
   \bibitem{Yang}
     C.~N.~Yang, \pr{77}, 242 (1950).
\bibitem{Harland-Lang:2013ncy}
L.~A.~Harland-Lang, V.~A.~Khoze, M.~G.~Ryskin and W.~J.~Stirling,
Eur. Phys. J. C \textbf{73}, 2429 (2013).
\bibitem{Bass:2018xmz}
S.~D.~Bass and P.~Moskal,
Rev. Mod. Phys. \textbf{91}, 015003 (2019).
\bibitem{Xu:2018uye}
Q.~N.~Xu \textit{et al.} (Belle Collaboration),
Phys. Rev. D \textbf{98}, 072001 (2018).
 \bibitem{BaBar:2013agn}
J.~P.~Lees \textit{et al.} [BaBar],
Nucl. Instrum. Meth. A \textbf{726}, 203-213 (2013) 
  J.~P.~Lees \etal\ (\babar\ Collaboration), Nucl. Instr. Meth. Phys. Res. {\bf 726}, 203 (2013).
\bibitem{BABARNIM} B. Aubert \etal\ (\babar\ Collaboration),
  Nucl. Instr. Meth. Phys. Res. A {\bf 479}, 1 (2002); {\it ibid.} {\bf 729}, 615 (2013).
\bibitem{geant}
  The \babar\ detector Monte Carlo simulation is based on Geant4
  [S. Agostinelli \etal,  Nucl. Instr. Meth. Phys. Res. A {\bf 506}, 250 (2003)] and EvtGen [D.~J.~Lange, Nucl. Instr. Meth. Phys. Res. A {\bf 462}, 152 (2001)].
 \bibitem{BabarZ}
   B.~Aubert \etal\ (\babar\ Collaboration), Phys. Rev. D {\bf 81}, 092003 (2010).
   \bibitem{Rosner} J.~Babcock and J.~L.~Rosner, Phys. Rev. D \textbf{14}, 1286 (1976).
\bibitem{PDG} P.A. Zyla \etal\ (Particle Data Group), Prog. Theor. Exp. Phys. 2020, 083C01 (2020).
 \bibitem{isr} B. Aubert \etal\ (\babar\ Collaboration), Phys. Rev. D {\bf 77}, 092002 (2008).
  \bibitem{BDT} T.G. Dietterich and G. Bakiri, J. Artif. Intell. Res., {\bf 2} 263, (1995). 
     \bibitem{cb}
      M. J. Oreglia, Ph.D. Thesis, SLAC-R-236 (1980);
J. E. Gaiser, Ph.D. Thesis, SLAC-R-255 (1982);
T.  Skwarnicki, Ph.D. Thesis, DESY-F31-86-02 (1986). 
  \bibitem{Zhang:2012tj}
C.~C.~Zhang \textit{et al.} (Belle Collaboration),
Phys. Rev. D \textbf{86}, 052002 (2012).
\bibitem{Asner:2003gh}
D.~Asner,
[arXiv:hep-ex/0410014 [hep-ex]].  
\bibitem{ds} P. del Amo Sanchez \etal\ (\babar\ Collaboration), Phys. Rev. D {\bf 83}, 052001 (2011).
  \bibitem{Aston:1987ir}
D.~Aston \etal\ (LASS Collaboration), 
Nucl. Phys. B \textbf{296}, 493 (1988).
  \bibitem{wilks} S. S. Wilks, Ann. Math. Stat. 9 (1938) 60.
\bibitem{blatt} J. Blatt and V. Weisskopf, Theoretical Nuclear Physics, New York: John Wiley \& Sons (1952).
  \bibitem{delAmoSanchez:2011bt}
P.~del Amo Sanchez \etal\ (\babar\ Collaboration),
Phys. Rev. D \textbf{84}, 012004 (2011).
\bibitem{Armstrong} T.A. Armstrong \etal\ (WA76 Collaboration), Z. Phys. C {\bf 51}, 351 (1991).
 \bibitem{Aubert:2006nu}
B.~Aubert \etal\ (\babar\ Collaboration),
Phys. Rev. D \textbf{74}, 032003 (2006).
\end{thebibliography}
\end{document}